# A Planetary Cooling Hose

Roderick A. Hyde[#]

Geoengineering may become necessary in the near future as a way to temporarily pause global warming, thereby providing time for more permanent solutions, such as clean energy or carbon capture, to become effective. Erection of a high-altitude hose offers an affordable and near-term approach to deliver sulfur-bearing aerosols to the stratosphere, in order to perform geoengineering via solar radiation management (SRM). In this paper, we discuss the design of a hose, extending to an altitude of 20 km, and sized to deliver 100 ktons of sulfur per year. The sulfur, in the form of $H_2S$, is pushed up the hose by a pump on the ground, and then sprayed out at the top, forming $H_2SO_4$ aerosols which scatter enough sunlight to perform geoengineering. Because the hose operates continuously, it only has to deliver about 50 gallons/minute (little more than a garden hose); geoengineering hoses are inherently very modest devices. The flux from a single hose is not sufficient to stop global warming by itself, but is enough to test the effect of the aerosols, and, once replicated to about 20 sites across the planet, can be used to offset all the warming caused by atmospheric $CO_2$.

Two varieties of hose are presented here, one delivering $H_2S$ as a liquid, and the other in gaseous form. Pumping liquid $H_2S$ through a narrow 20 km hose requires a high pressure, and we discuss dealing with this, either by emplacing a number of intermediate pumps along the hose, or by fielding hoses with walls which are sufficiently strong and lightweight. The hose is held in place by a suite of balloons, which may either all be located at its top, or may be distributed along the hose as well. The greatest challenge in suspending such a hose is wind, which pushes strongly on both the balloons and on the hose itself; the resultant deflections can, if sufficiently severe, collapse the hose. The key to surviving strong winds, is to streamline both the balloons and the hose, through using elongated balloons and by enclosing the hose within an airfoil-shaped shroud. A hose sized to deliver gaseous $H_2S$ must be much wider than one delivering liquid, so will suffer even more from the effects of wind; we discuss the challenges involved in fielding an aero-shroud which is large enough to enclose such a hose, yet lightweight enough to be supported by the balloons.

After treating the design of these two hoses, along with a number of their variants, we layout the steps needed to develop and fabricate them, and conclude with some thoughts on how the hoses might be fielded.

[#] Retired : roderickahyde@gmail.com

# Contents

# 1 : Global Warming

We are clearly losing the fight against global warming. Current atmospheric $CO_2$ concentrations and global temperatures, are the highest they've ever been, and both are still rising. These temperature rises, and the increased $CO_2$ levels are already hurting us, and if they keep rising, their damage will grow as well.

The primary cause of global warming stems from hydrocarbon combustion (for energy, transportation, chemical production, etc.), which releases $CO_2$ into the atmosphere. Some of this $CO_2$ then enters the oceans, raising their acidity, while that remaining in the atmosphere is responsible, via the greenhouse effect, for most or all of the temperature rises we're experiencing. It is these temperature rises, which cause global warming's climate and weather changes.

The effects of global warming can be either gradual or abrupt. So far, we've been experiencing the former variety. Some of the leading effects are, disruptions to crop growth and rainfall patterns, rising sea-levels, and, of course, the warmer temperatures themselves. The influence of these changes is growing, but not uniformly across the globe, some places are already suffering significantly, others only slightly, and a few are even seeing benefits. Fortunately, most of these gradual effects are expected to be reversible if/when global warming ends. But, there are a number of potential changes, which, if they occur, would be abrupt and irreversible. Some of the most concerning, are a collapse of the Atlantic Meridional Overturning Circulation, or melting of ice sheets in Greenland or the Antarctic, hypothesized to occur if global temperatures increase beyond "tipping points". The precise value of the tipping points is unknown, but it is worried that global warming will eventually breach them.

## 1.1 : Stopping Global Warming

The world's scientific and political communities are greatly concerned about global warming, and putting a lot of effort into stopping it. Three main strategies have been proposed; abandoning use of hydrocarbons, carbon capture, and geoengineering. Respectively, these act via removing the source of $CO_2$, removing existing $CO_2$, and removing the effects of $CO_2$.

To date, virtually all the effort has gone into the first strategy, transitioning away from hydrocarbon energy to that produced by non-carbon sources. These are typified by so-called "clean energy" sources, wind and solar, as well as other important ones such as nuclear and hydroelectric. A number of approaches are being used to drive this transition, ranging from moral suasion and market incentives, to taxation and laws; to date well over 1 trillion dollars has been expended in this effort[1].

This approach, intended to drive people away from using hydrocarbon-derived energy, amounts to a campaign of "Just Say No" to hydrocarbon usage. Unfortunately, while the educational and public relations elements of this campaign have managed to persuade many people that hydrocarbon usage harms the planet, and that using clean energy is inherently more moral, this has not led to meaningful success against global warming. Over the last 20 years there has indeed been a large increase in wind and solar power usage; however, as seen in Fig. 1.1[2], hydrocarbon usage has also increased, in absolute terms by much more than the wind and solar gains, and it continues to completely dwarf them. The continued rise in hydrocarbon utilization is also reflected in the fact, illustrated in Fig. 1.2[3], that $CO_2$ emissions keep increasing, as do global temperatures, Fig. 1.3[4]. So, despite the effort put into transitioning to clean energy, it is not succeeding; at least not fast enough to successfully deal with global warming.

Figure 1.1 : World Energy Consumption

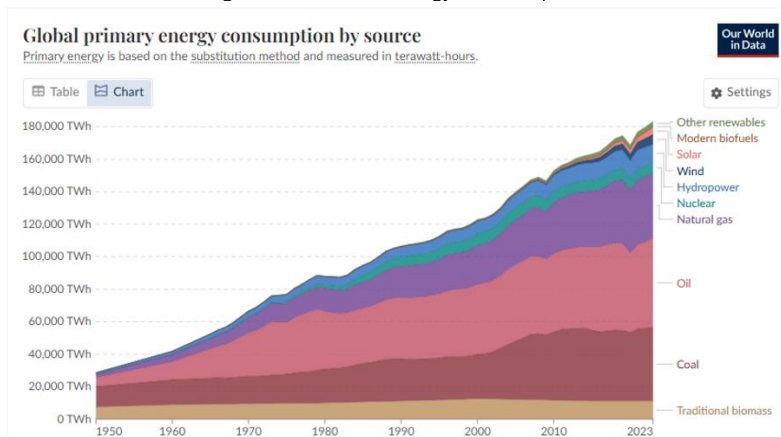

Figure 1.2 : World $CO_2$ Emissions

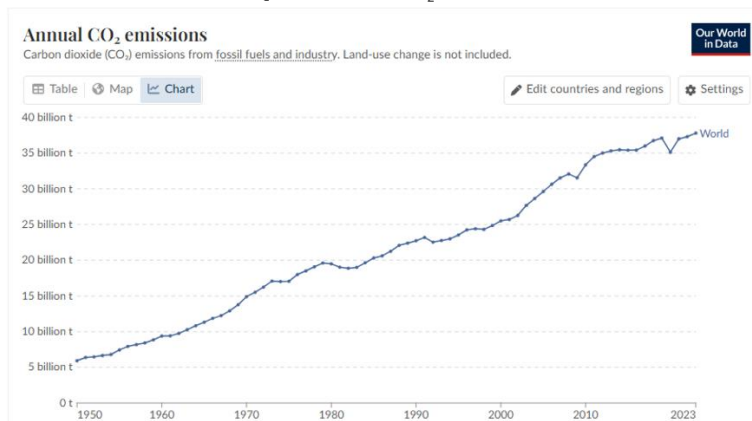

Figure 1.3 : World Temperature Increases

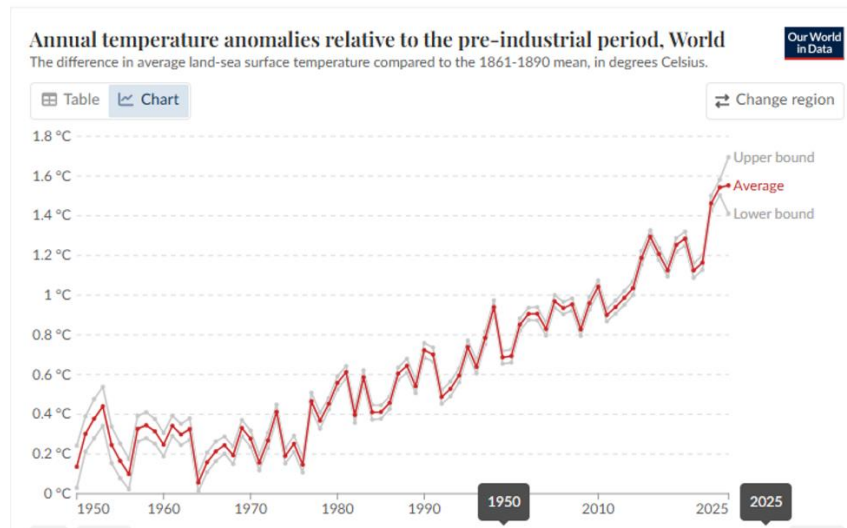

The reasons why the transition to clean energy are so ineffective, are pretty simple: People and countries need energy, and they need it now, not in the indefinite future. Clean energy is not available in the amounts they need, it costs too much, and it doesn't work as well.

A key point, is that while global warming can hurt people, so does lack of energy. Providing people with cheaper and better access to existing energy sources, is more helpful to them than making them wait for clean energy; their countries clearly realize this. Accordingly, while the clean energy transition is making headway in the USA and Western Europe, leading new consumers, such as China and India, pay only lip service to the concept; their actual usage, displayed in Fig. 1.4[5] tells another story.

Figure 1.4 : Greenhouse Gas Emissions

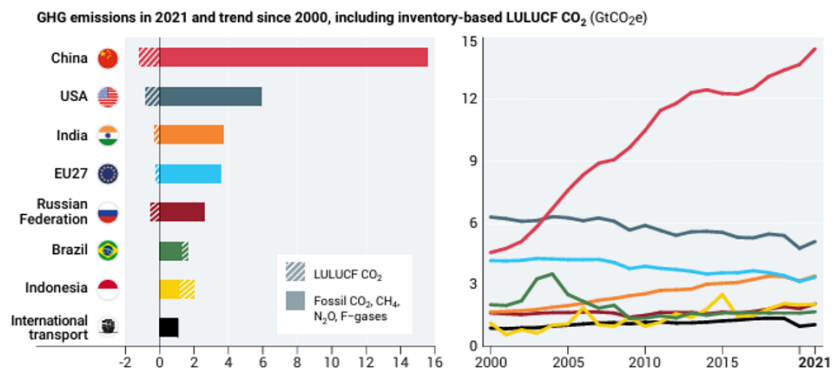

There is no doubt at all that global warming poses an extreme, and ever increasing, danger which must be dealt with. The issue, is just that the current approach of relying solely on a transition from hydrocarbons to clean energy, isn't working quickly enough to solve the problem. While, given time and the progression of technology, a transition away from hydrocarbons may eventually occur; global warming will have gotten a lot more severe before this happens.

As mentioned above, two other ways of attacking global warming have been proposed, carbon capture and geoengineering. Both have strong advantages and disadvantages, but neither is being vigorously pursued; i.e., at even 1% of the effort devoted to the clean energy transition.

Carbon capture[6], particularly direct air capture, has a simple premise; actively remove $CO_2$ from the atmosphere. The potential attraction of this is two-fold. First, note that $CO_2$ is what is responsible for global warming, not hydrocarbon usage per-se. If carbon capture works, then we can still use hydrocarbons to produce cheap and plentiful energy, and can then remove the resultant $CO_2$ before it hurts the climate. Secondly, even if the clean energy transition works, and we stop injecting new $CO_2$ into the environment, that which is already there will continue to drive global warming; carbon capture will be needed to get rid of this, restoring $CO_2$ to previous levels, and finally ending global warming.

The problem with carbon capture, is the same as with the clean energy transition; it won't stop global warming quickly enough. The concerns with cost and scale-up that plague clean energy, are even more severe for carbon capture. From a practical standpoint, the issue is that most carbon capture methods have to collect, and then store, more mass of $CO_2$ than that of the hydrocarbons being burned. Given this, the cost and time needed to reach the enormous scale of the world's hydrocarbon usage, makes it irrelevant for any near-term relief from global warming. Avoiding this conundrum requires a carbon capture method with a "lever arm", in which we can avoid handling nearly so much material. Ocean capture, via fertilization of microalgae[7], is one approach with this potential, but how well it works, and whether it can be scaled-up to global levels, remain open questions.

**1.2 : Geoengineering**

The third approach, geoengineering[8][9][10][11], is very different, and less ambitious, than either carbon capture or transitioning to clean energy. Geoengineering does not attempt to "solve" global warming; it has no effect on the root cause, namely atmospheric $CO_2$. But, what it promises to do, quickly and inexpensively, is to stop (and even roll back) the temperature rises that are the leading threat from global warming.

A wide range of geoengineering approaches have been proposed and analyzed. Their common theme, is to artificially cool the planet, thereby offsetting the warming caused by the increase of atmospheric $CO_2$. The premise behind geoengineering is that, while it's hard to get rid of $CO_2$, it's relatively easy to prevent global temperatures from rising.

Since geoengineering does not stop the growth of atmospheric $CO_2$, it is not a stand-alone policy; efforts to prevent adding more (transitioning to clean energy) and removing that already present (carbon capture) will still be important, and can continue in parallel with geoengineering.

However, given that neither of these two approaches can be scaled up quickly enough to prevent major global warming damage, using geoengineering as a stop-gap to give them time to take effect, should be broadly welcomed. It is not.

Historically, geoengineering has been met with intense opposition by most of the global warming community. While geoengineering (particularly the Solar Radiation Management (SRM) approach emphasized in this report) is the subject of intense academic research, this opposition has until recently prevented all attempts to perform in-the-field experiments. As a result, while climate simulations predict SRM's effectiveness, this has not yet been confirmed with experiments or real-world experience.

Obviously, a big concern for people, and one cause of the opposition, is the question of whether geoengineering is safe.

In terms of health, the approach we'll be discussing (adding small amounts of sulfur to the atmosphere) is clearly safe. The amount needed is truly tiny (about 0.4 parts per billion per year), and will be located up in the stratosphere, far away from air that we breathe. Compare this to $CO_2$ (the contaminant that's driving global warming). This is being added at about 8,000 parts per billion per year, over 20,000 times as much as the proposed sulfur; and the $CO_2$ is down where we breathe it. Furthermore, there is already much more natural sulfur in the atmosphere than we'll be adding, and it also is down where we breathe it.

The broader question, whether geoengineering itself is safe, is more complicated to answer[12][13]. Geoengineering will clearly change the climate (that's the whole point of doing it). The main effect of these climate changes will be to reverse global warming, i.e., to restore the climate. So, it is certainly going to be safer than global warming is. But, while simulations show that both temperatures and precipitation can be controlled[14], the climate restoration is not going to be perfect, everywhere and all the time. Once geoengineering is underway, any adverse weather, whether or not actually due to it, is undoubtably going to be blamed on geoengineering. This political effect will probably be the biggest danger of geoengineering; weather will no longer be viewed as an "Act of God", but rather, as due to human actions, with disputes settled by politics, law, and war.

So, geoengineering is not ideal; it's not something we would normally choose to do. It's simply much safer than global warming, and is the only quick and effective way to stop global warming from continuing, and from getting progressively worse.

**1.3 : Solar Radiation Management (SRM)**

The focus of this paper, is on one way to implement the most studied approach to geoengineering, solar radiation management (SRM) via stratospheric particles.

This geoengineering method works by emplacing many small particles into the stratosphere, whereupon they intercept incoming sunlight. While almost all of the light is not affected, and continues to the surface as usual, a small percentage of light (mostly, invisible ultraviolet light) is reflected back into space, thereby reducing solar heating, and effectively cooling the planet. The amount and type of the particles, are chosen so that this cooling is just enough to counter global warming, while not noticeably dimming the visible light seen by people, animals, and plants.

What makes SRM practical, is that small sub-micron particles are very effective at scattering sunlight. We need to place far less material into the stratosphere, than the hydrocarbons we burn, or the $CO_2$ they emit. This "lever arm" is why geoengineering can be very much cheaper, and hence easier to scale-up, than carbon capture or transitioning to clean energy. And, while humans haven't yet been able to perform experiments to prove that SRM works, nature has done so. Volcanos inject dust and gas into the atmosphere, some of which reach the stratosphere, and then spread across the planet; the Pinatubo eruption of 1991 lowered Northern temperatures by about 0.5°C over the course of a year[15].

Since that event, scientists have recognized the potential for cooling the planet, by artificially emplacing particles into the stratosphere to scatter sunlight. Many different types of particles have been considered[16], although the ones which have received the greatest amount of analysis are $H_2SO_4$ aerosols. Their effect on the climate, can be studied with the same type of climate circulation models (GCMs), as are used to predict the course of global warming itself. The earliest use of GCMs for examining geoengineering[17], done by artificially reducing the solar insolation, was promising, so their use has grown increasingly pervasive and sophisticated. More recent modeling[14][18] applies the aerosols themselves (not just changes to the solar insolation), and also shows that by properly distributing their latitudinal deployment, one can attempt to control both temperature and precipitation profiles. Modeling is not as convincing as experimentation will be, but SRM-based geoengineering is certainly looking promising.

$H_2SO_4$ aerosols are very efficient solar scatterers, so we won't need to inject much material. The actual amount required is uncertain, and will remain so until we're able to start doing stratospheric experiments. But, current estimates[10][19] range from about 1 to 5 Mtons/yr of sulfur; amounts which are only 0.01% of the 40 Gtons/yr of $CO_2$ emitted. This huge difference provides the "lever arm", which makes SRM-based geoengineering so economically attractive.

Another key property of these aerosols, is that they don't stay aloft forever; they eventually, on a few year timescale, fall out of the atmosphere. One effect of this is economic, we have to continually replenish the sulfur; we can't just put it up and leave it. This is actually a good thing, it means that if the aerosols become troublesome, or we have simply put up more than necessary, the situation soon corrects itself, without us having to do anything.

In many ways we can consider the use of this small amount of sulfur, as a fuel additive, released in company with the $CO_2$, in order to prevent most of the harm it would cause. Unlike direct additives (such as those in gasoline), these sulfur aerosols are applied at a distance, up in the stratosphere, but just like direct ones, they are applied in proportion to the fuel used. Nor, is there anything new or unusual about releasing sulfur when burning hydrocarbons. In fact, all of the main hydrocarbon fuels, coal, oil, and natural gas, contain, and release when burned, more sulfur than required for SRM-based geoengineering. The difference, is that this direct sulfur emission occurs low in the atmosphere, where it isn't high enough, doesn't spread far enough, or last long enough, to provide global cooling, yet (unfortunately) interacts far more with people than it would if up in the stratosphere.

SRM requires some way to deliver the scattering particles into the stratosphere. A large number of possibilities have been considered[20][21], including airplanes, balloons, artillery, hoses, and the like. The approach which is generally considered the baseline, is to use a fleet of specialized airplanes[22][23]. A key advantage of this approach, is that it will certainly work. The world knows how to design, build, and operate airplanes, and the need to reach the stratosphere (e.g., 20 km altitude) is not a novel challenge.

However, this approach will be neither cheap nor fast. Designing modern aircraft, even with advanced computational tools, is a complex undertaking, requiring large teams of specialists, and years of work. Actually producing such airplanes, of course, then requires far more people, time, and money. After the airplanes exist, they must be continually operated; pilots may well be replaced by automation, but servicing and maintaining a fleet of airplanes takes manpower. Compared to the scale of global warming, these efforts will be small, but in absolute terms they are quite expensive. Accordingly, this is very much a "big steel" approach to geoengineering, demanding a large political and financial commitment to performing it; a consensus which does not yet exist, and will probably take considerable time to develop. Therein lies the biggest problem with this approach, it's too slow. Bringing it into existence will be a long process; all during this time, global warming will continue to get worse.

**1.4 : Using a Hose**

This paper is focused on an alternative approach to delivering sulfur compounds to the stratosphere, pumping them up a hose.

The idea is simple; place the top end of a hose in the stratosphere, the bottom end at the ground, and pump a sulfur-bearing fluid up the hose, spraying it out into the atmosphere at the top. Because this is a steady and continuous process, the actual flow rate required is quite small, and once the hose is in place, operating it should require very little human involvement. So, it offers the potential of being a simple, cheap way to deliver stratospheric aerosols, and perform SRM.

Over the years, a number of teams have been intrigued by this possibility, and have explored its feasibility; identifying the key challenges, and proposing ways to solve them, so as to turn the concept into a practical reality.

Two, fairly brief, treatments considered the use of a gaseous fluid:

Blackstock's and Koonin's 2009 report[24] on geoengineering, included a short exploration of using a chimney to buoyantly lift a sulfur-donor, within a lighter carrier gas, up to the stratosphere. This facility was sized to deliver 10 Mton/yr flux, so required an extremely massive hose, and a huge balloon to hold it up. They did identify wind as a major concern, and proposed streamlining the hose to reduce its effect.

McClellan et-al$^{\{22\}}$ prepared a 2010 report for Aurora Flight Sciences, on the costs of geoengineering. This was focused on design and costing of an aircraft-based system, but, for comparison purposes, they briefly considered a variety of alternative geoengineering approaches, one of which was an ocean-based hose. They considered two fluid alternatives, one using gas and the other liquid. Their analysis was primarily focused on pumping requirements, and on how this was affected by hose diameter and by the gas/liquid choice. As with Blackstock's report, their system was designed for a large throughput, 1 Mton/yr, requiring a heavy hose with a big, 300 meter, balloon to support it.

There have been two, more detailed, studies performed, both fairly similar in treatment and timing:

The first of the two, was the 2009 Stratoshield proposal$^{\{25\}\{26\}}$ by Intellectual Ventures; I was one of the designers of this concept. This work was focused on delivering liquid $SO_2$ at a considerably smaller rate, 100 kton/yr, than either Blackstock or McClellan, resulting in a system requiring multiple, albeit much smaller scale, facilities. The analysis supporting this report, involved determining the pumping requirements, their tradeoffs with the diameter of the hose, and the possible placement of pumps along the hose to reduce peak pressures. Balloons were, as usual, intended to hold up the hose, and these were envisioned as, either all at the top of the hose, or with some distributed along it. Again, as usual, wind was realized as a major concern, and the advantage of streamlining the hose and balloons was discussed.

The other detailed study, was work done in the United Kingdom, by Davidson and University of Cambridge$^{\{27\}}$, as well as a patent application by Davidson$^{\{28\}}$. This work was similar in thrust to the Stratoshield; it used a hose to pump liquid $SO_2$ (although they did mention other options such as $H_2S$), treated pumping requirements, and analyzed tradeoffs with the hose diameter. One difference from the Stratoshield was size; their system was designed for a 2.5 Mton/yr throughput, requiring both a very heavy hose and a large, 300 meter, balloon. They proposed placing a series of lifting bodies, not simple balloons as in Stratoshield, along the hose to help support it. As with the Stratoshield, they realized the importance of wind, and proposed airfoil-shaped enclosures$^{\{28\}}$ surrounding the hose to help reduce wind forces. Unlike Intellectual Venture's Stratoshield, this work nearly led to an experimental embodiment. The SPICE$^{\{29\}}$ project, intended to field a subscale hose to perform delivery experiments. Unfortunately, it encountered significant political opposition, and was terminated before implementation.

This current report, is a more detailed treatment than, and extension of, the earlier Stratoshield and United Kingdom work. It treats the same basic hose concept as they did, namely transporting a liquid sulfur-bearing compound, using either a single ground-based pump, or a series of smaller ones mounted on the hose. Balloons (either along the hose or only at the top) are used to support the hose, although emphasis is upon a collection of relatively small balloons, rather than a single very large one.

One of the most significant improvements of this work, regards the hose's tension; no longer just using it to support the weight of the hose, but also to avoid pressure-induced buckling, as well as to stiffen the hose against the wind. Another advancement is to self-consistently model the wind-based deflection of the hose. This allows quantitative consideration of the stiffening effects of tension, the benefits of streamlining the hose and balloons, and the cost/benefits of mid-hose balloons.

A couple of other differences in this work, are the selection of $H_2S$, rather than $SO_2$, as the fluid to deliver, and the design of hoses delivering the $H_2S$ as a gas, not just as a liquid; then analyzing and comparing both the gas and liquid options with a uniform level of modeling.

Finally, a utility of this report is simply that it documents the design of a stratospheric hose. One problem with all of the preceding papers, is that they simply report the results of hose designs; they didn't detail the calculations that led to those results. This could leave readers uncertain as to the adequacy of the modeling, and hence the validity of the conclusions. It is the intention of this report, to be more complete and transparent about how the hose was designed and modeled, and, of course, to allow readers to determine where improved modeling is necessary.

Finally, this study, like the Stratoshield[25] effort, is focused on a system delivering a moderately sized throughput, 100 kton/yr, small enough to be affordably developed (in time as well as money), yet large enough to deliver measurable results for testing the SRM approach to geoengineering. This prototype hose can then, if desired, serve as the basis for a global geoengineering system, via simple replication of individual hoses to multiple sites around the planet.

# 2 : Hose – First Glance

In principle, hose delivery of sulfur to the stratosphere is very straightforward. It requires four things, a sulfur-based fluid, a hose, a pump, and a suspension system.

## 2.1 : Fluid Choice

Let's first consider which fluid to use.

We'll focus on sulfur-bearing chemicals for several reasons. One, is that more is known about their stratospheric chemistry and aerosol behavior than other possibilities. Another, is the experimental evidence from volcanos of their effect on global cooling. Because of these reasons, sulfur-based compounds have been the focus of most of the stratospheric aerosol approaches to solar radiation management geoengineering; we'll do the same here.

Four potential choices of liquids to deliver stratospheric sulfur are elemental S, $SO_2$, $H_2S$, and $H_2SO_4$.

| Table 2.1 : Fluid Properties | | | | |
|---|---|---|---|---|
| | Liquid density | Payload penalty ($\Lambda$) | Melting point | Boiling point |
| | gm/cc | | K | K |
| S | 1.82 | 1.00 | 388 | 718 |
| $H_2S$ | 0.92 | 1.06 | 188 | 214 |
| $SO_2$ | 1.46 | 2.00 | 201 | 263 |
| $H_2SO_4$ | 1.83 | 3.06 | 283 | 610 |

Here, we'll define $\Lambda$ as a payload penalty, caused by the mass of extraneous elements other than sulfur in the payload fluid. The need to deliver this extra mass increases the effort (in energy and pressure) required by the hose.

In considering these compounds, note that (barring strong heating or cooling of the hose) wind-mediated convection means that the fluid will be nearly in thermal equilibrium with the atmosphere, so its temperature will vary between about 210K and 300K during its transport.

Accordingly, both sulfur and sulfuric acid will be solids, and hence difficult to deliver by a hose. They could be delivered as suspensions in a carrier liquid, but this will, of course, increase their mass penalty factors. Given that pure sulfuric acid already has a high $\Lambda$ value, it should be ruled out. Elemental sulfur's intrinsic $\Lambda$ is low enough so that some carrier penalty could be acceptable, if not for its high density and the existence of $H_2S$ and $SO_2$ as alternatives.

Previous studies of hose-delivered sulfur baselined SO₂ as the sulfur-bearing compound. But, given that H₂S has a much smaller penalty factor, Λ, and also a lower density (hence less gravitational pressure head), we will consider it as well.

**2.2 : Simple Designs**

Let's perform simple hose designs, using both SO₂ and H₂S, in order to illustrate hose delivery of sulfur to the stratosphere.

We'll size the hose to deliver 100,000 metric tons of sulfur per year. This amount is probably only 5-10% of the total amount required for offsetting the thermal effects of global warming, but is a reasonable unit size to eventually be replicated into a geographically diversified delivery network. Smaller sizes won't save much, if any, of the time or cost needed to develop the hose. And, since the first hose is probably going to be used to study the safety and efficacy of solar radiation management, fielding too small a size will not provide clear, high signal-to-noise, data.

This mass flux works out to be 3.2 kg/sec of sulfur transport through the hose. We can convert this to volumetric terms; for instance, if the liquid is SO₂, the flow is 4.4 liters/sec, or 70 gallons/minute. These are "garden hose" type numbers, and make hose delivery appear simple and compelling.

In practice, extending a garden hose up to the stratosphere is challenging; it will have to be very long (e.g., 20 kilometers), the flow is vertical (imposing a gravity head), and the hose has to be held up somehow (for instance, by balloons). The significance of these effects (all of which were noted in the previous studies) is simple to demonstrate:

The gravitational head is:

$$P_G = \rho g L \qquad (2.1)$$

Thus, a 20 km tall vertical column of liquid SO₂, having a density of about 1.46 gm/cc, will have a gravitational head of 2860 bar; use of H₂S reduces this to 1800 bars.

The walls of the hose will have to be strong enough to contain this pressure. We can use thin-walled pressure vessel equations to estimate the required wall mass:

$$M_w = \frac{\pi D^2}{2} \frac{P}{\sigma_w} \rho_w L \qquad (2.2)$$

For a hose wall employing Kevlar, the strength (neglecting for now any safety-related downrating) is about 3 GPa, and its density, 1.45 gm/cc, is just about the same as that of the SO₂ fluid.

One reason that the distance to the stratosphere is challenging is because we have to maintain the flow rate over a 20 km length. This requires dealing with pipe friction:

$$P_F = \frac{1}{2} f_D \rho v^2 \frac{L}{D} \tag{2.3}$$

Here $v$ is the fluid velocity, $D$ the inner diameter of the hose, and $f_D$ the Darcy friction factor. The velocity and diameter vary inversely with one another, constrained by the need to deliver the required mass rate, $\dot{M}$, of sulfur.

$$\Lambda \dot{M} = \rho v \left(\frac{\pi}{4} D^2\right) \tag{2.4}$$

This $\dot{M}$ constraint means that (not withstanding a weak effect on the friction factor) the frictional pressure scales as $1/D^5$, so will be dominant for small diameters and negligible for large diameters. Unfortunately, both the wall mass, $M_w$, and the overall weight of the fluid, $M_F$, vary in the opposite direction, so that the large diameters needed to avoid big frictional pressures will drive up the overall hose mass $M_H$.

$$M_F = \rho \left(\frac{\pi}{4} D^2\right) L \tag{2.5}$$

$$M_H = M_F + M_w \tag{2.6}$$

The primary penalty for having a heavy hose, is the increased difficulty in suspending it from the stratosphere. A first cut at quantifying this, is to calculate the diameter of a balloon, $B$, needed to support the hose. Below, we'll present a lower bound for this, neglecting the weight of the balloon's gas and skin. Here, $\rho_*$ is the atmospheric density at 20 km, i.e., 88.9 gm/m³.

$$\frac{\pi}{6} B^3 \rho_* = M_H \tag{2.7}$$

We'll illustrate these tradeoffs by comparing hose designs for different values of the flow diameter (2, 3, 4, and 5 cm). To keep things simple, we'll assume a friction factor of 0.03 for all cases (ignoring its weak, $\sim D^{-0.2}$, variation).

| Table 2.2 : Basic Hose Properties using $SO_2$ | | | | |
|---|---|---|---|---|
| $D$ (cm) | 2 | 3 | 4 | 5 |
| $P_G$ (kbar) | 2.86 | 2.86 | 2.86 | 2.86 |
| $P_F$ (kbar) | 41.70 | 5.49 | 1.30 | 0.43 |
| $P_{max}$ (kbar) | 44.56 | 8.35 | 4.16 | 3.29 |
| $M_F$ (ton) | 9.17 | 20.64 | 36.70 | 57.33 |
| $M_w$ (ton) | 27.07 | 11.42 | 10.12 | 12.49 |
| $M_H$ (ton) | 36.24 | 32.06 | 46.82 | 69.82 |
| $B$ (m) | 92.0 | 88.3 | 100.2 | 114.5 |

| Table 2.3 : Basic Hose Properties using $H_2S$ | | | | |
|---|---|---|---|---|
| $D$ (cm) | 2 | 3 | 4 | 5 |
| $P_G$ (kbar) | 1.80 | 1.80 | 1.80 | 1.80 |
| $P_F$ (kbar) | 18.68 | 2.46 | 0.58 | 0.19 |
| $P_{max}$ (kbar) | 20.48 | 4.26 | 2.38 | 1.99 |
| $M_F$ (ton) | 5.78 | 13.01 | 23.12 | 36.13 |
| $M_w$ (ton) | 12.44 | 5.82 | 5.80 | 7.58 |
| $M_H$ (ton) | 18.22 | 18.83 | 28.92 | 43.71 |
| $B$ (m) | 73.1 | 74.0 | 85.3 | 97.9 |

## 2.3 : Preliminary Results

The basic trends are clear:

First, $H_2S$ leads to lower pressures, lower hose masses, and smaller balloons than $SO_2$. Its primary drawback is its low boiling point; $H_2S$ is normally a gas. However, a modest over-pressure, 30 bar, is sufficient to keep it a liquid for the temperatures of interest[30] (i.e., T < 315K). Since the hose needs to operate at kilobar-level pressures to handle gravitational head and flow drag, this requirement is easily met. Accordingly, we will focus on $H_2S$ as the sulfur-bearing fluid in the remainder of this study.

Second, there will be an optimum hose diameter, somewhere in the range of 3 cm. Smaller values tend to reduce the system's mass and balloon size, but at a price of greatly increasing the required pumping pressure; this ultimately will itself drive up the mass and balloon size. For larger diameters, the pressure requirements asymptote, but the greater hose size drives up its mass and the balloon size needed to support it.

And, most importantly, the size and mass values are low enough, so that hose delivery does appear to be an attractive way to deliver large amounts of sulfur up to the stratosphere.

# 3 : Hose – Deeper Looks

While the above scoping analysis demonstrates the potential of a hose to deliver sulfur to the stratosphere, there are aspects which need further consideration. These include a more detailed understanding of H₂S properties, improved structural design of the hose, and the possible placement of pumps and balloons along the length of the hose (rather than just at either end).

## 3.1 : Vertical Support

Before considering these issues, however, it's important to clarify how the hose will be supported.

In the preliminary treatment, we added together the weights of the fluid and hose and used this total to determine the size of balloon necessary to hold up the system. However, in principle it's not clear why this should be correct. The weight of the fluid is actually supported by pressure from below, this is the reason for the 1800 bar gravitational head. Similarly, consider the consequence of the frictional pressure drop; this force pushes downwards on the flow, but its reaction pushes upwards on the walls of the hose. It is possible to choose flow parameters so that this up-force completely supports the weight of the hose walls.

Such an up-force does come with a cost, namely the frictional pressure drop $P_F$ required to induce it. This pressure, in turn, dictates the mass of the hose walls required to contain it; as pressure levels go up, so will the required hose mass, further driving up the frictional pressure necessary to support it.

It is straightforward to show that, with strong enough hose materials, this feedback loop converges, rather than runs away.

$$\left(\frac{\pi}{4}D^2\right) P_F = gM_w = \frac{\pi D^2}{2}\frac{P}{\sigma_w}\rho_w gL \tag{3.1}$$

Cancelling the areas, and adding the gravitational pressure, we get

$$P = \rho gL + 2\frac{P}{\sigma_w}\rho_w gL \tag{3.2}$$

Hence,

$$P = \rho gL\bigg/\left(1 - 2\frac{\rho_w gL}{\sigma_w}\right) \equiv \rho gL/(1 - f_w) \tag{3.3}$$

For the Kevlar properties presented earlier, the $f_w$ feedback term is ~ 0.2. So, the hose could be fully supported from below, requiring a pressure increase of only 25% over that imposed by the gravitational head, i.e., a total of about 2250 bars.

In practice, this support-from-below approach is not feasible. It fails, not for strength reasons, but for stability ones. A long slender pressurized tube, just like a column in compression, is subject to buckling; lateral displacements grow in magnitude rather than being damped. While in principle, such displacements could be controlled by an array of guy wires or thrusters, we will instead baseline a simpler, more reliable, approach; keeping the hose in tension.

As long as the tension force along the hose is greater than the compressive force, it will be stable. Since the maximum pressure in a hose is usually at the ground, this dictates the tension there:

$$T_o \geq \left(\frac{\pi}{4}D^2\right)(P_o) = \left(\frac{\pi}{4}D^2\right)(\rho g L + P_F) = gM_F + \left(\frac{\pi}{4}D^2\right)P_F \tag{3.4}$$

This tension value may either grow or decrease along the length of the hose from that at the base, depending on whether the upward force due to flow friction is greater or less than the downward force due to the weight of the hose walls. The tension at the top of the hose, which is the force which must be supplied by the balloon is:

$$T_T = T_o + \left[gM_w - \left(\frac{\pi}{4}D^2\right)P_F\right] = g(M_F + M_w) = gM_H \tag{3.5}$$

In summary, the hose does need to be under tension, there's a minimum value which this must have at the base, and the tension at the top, supplied by the balloon, is indeed the full weight of the hose.

The above treatment just considers the static stability of the hose. The fact that the fluid is flowing introduces the possibility of dynamical instability[31]. However, because we need to keep the tension high enough to avoid buckling from compressive forces due to the fluid's static pressure, we are even further above the instability threshold caused by its dynamic pressure, since $P_o \gg \rho v^2$.

**3.2 : Hose Configuration**

The tension, which is necessary to support the hose, will result in axial stress within the hose. How should this stress be handled?

One option, is to have the stress carried by the existing hose wall, either by its inherent axial strength, or by the inclusion of additional material (e.g., axial fibers) within the wall. This approach results in a particularly simple hose configuration, and potentially the lowest overall mass. However, it does couple two distinct design factors (pressure containment and hose support) together.

An alternative approach is to separate these two considerations, designing the hose wall for pressure containment (e.g., with maximal circumferential strength), and provide a separate tether running alongside the hose to carry the tension (e.g., with axial fibers). Since this separation of functions will have little, if any, effect on total mass; its primary drawback is the inconvenience of needing two different components. We can estimate the tether's mass by using the overall hose weight to determine the tension that it must carry, and from that its mass:

$$gM_H = T = \sigma_T A_T = \sigma_T \frac{M_T}{\rho_T L} \tag{3.6}$$

Hence,

$$\frac{M_T}{M_H} = \frac{\rho_T g L}{\sigma_T} \equiv f_T \tag{3.7}$$

Since the tether carries only axial load, it can use materials optimized for 1-D strength, likely with better strength to weight than those of the hose wall, e.g., fibers such as Dyneema SK78 (3.6 GPa, 0.97 gm/cc) or Torayca T1100S (7.0 GPa, 1.79 gm/cc). The Dyneema choice gives a tether fraction, $f_T$, of 0.054; a minor penalty.

The key advantage of this approach is that hose and tether implementation become separate, parallel, activities. The hose walls can be designed purely for their primary task, using circumferential strength to optimize pressure containment. Likewise, the tether can be designed purely to carry unidirectional tension, using high-strength-to-weight fibers. Both components can be designed, acquired, and tested separately, with changes to one's requirements not directly affecting the other's. From a programmatic standpoint, this advantage seems compelling, and will be baselined in this study.

For hose designs with particularly high internal pressure or low wall strength, use of the thin-wall approximation (as in equation 2.2) is inappropriate. Accordingly, we will use the corresponding thick-wall version:

$$M_w = \frac{\pi D^2}{4} \left\{ \frac{\sigma_w + P}{\sigma_w - P} - 1 \right\} \rho_w L \tag{3.8}$$

It should be noted that the loads imposed on the hose wall and the tether are not constant, but vary along the length of the hose. We can, therefore, reduce their masses by gradually adjusting their sizes as the loads change. For example, the fluid pressure varies linearly along the hose's length, so the wall thickness can be adjusted to match, reducing its total mass by a factor of two. Similarly, the required tension varies from top to bottom of the hose; by changing the tether's area proportionally, its mass can be reduced. While a continuous variation in wall and tether thickness will be impractical, it's also unnecessary. Most of the potential weight savings can be achieved by dividing the hose and tether into $N_w$ discrete sectors; e.g., 40 sections of 500 meters each. With this approach, properties can be constant within each sector, allowing for easy fabrication, yet adapted to the local load requirements.

### 3.3 : Variation of Fluid Properties

For detailed design, we need to refine our treatment of the $H_2S$ fluid; its properties will vary with ambient temperature[32] along the length of the hose, and also be affected by the large pressures imposed on the fluid. Accordingly, we need to use an equation of state giving $H_2S$'s density as a function of pressure and temperature. Similarly, in order to properly calculate frictional drag, we need to know the $H_2S$ viscosity and its pressure-temperature variation.

For this study, we use a modified SLV-EOS[33] to determine the $H_2S$ density as a function of pressure and temperature. Their combined effect on the density variation along the hose is shown in Fig. 3.1; for a case in which pressures and temperatures vary from 3490 bar, 298K at the base to 30 bar, 216K at the top. As can be seen, the density increases and saturates at high pressure, but its variation is less than 3%, which will have little effect on the behavior of the hose.

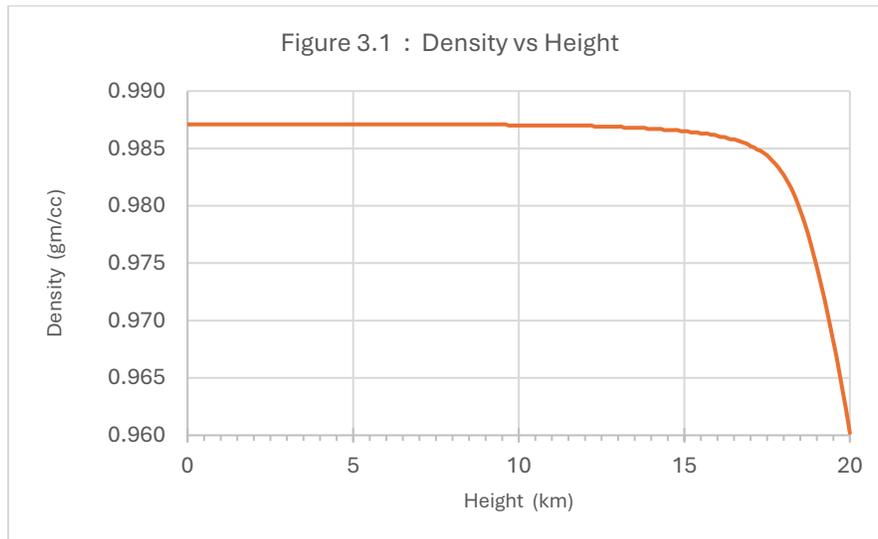

Figure 3.1 : Density vs Height

The $H_2S$ viscosity helps determine fluid friction, and hence the $P_F$ pressure needed to force $H_2S$ to flow through the long hose. We use the Serghides[34] approximation to the Colebrook-White expression in order to calculate the Darcy friction factor. In the absence of a specific viscosity equation-of-state for $H_2S$, we use a pressure-temperature dependent model[35] developed for hydrocarbon fluids. This viscosity is shown below in Fig. 3.2.

While this variation in viscosity is substantial, the Reynold's number is high enough so that this has very little effect in the friction factor, which varies only from 0.0197 to 0.0202 over the length of the hose.

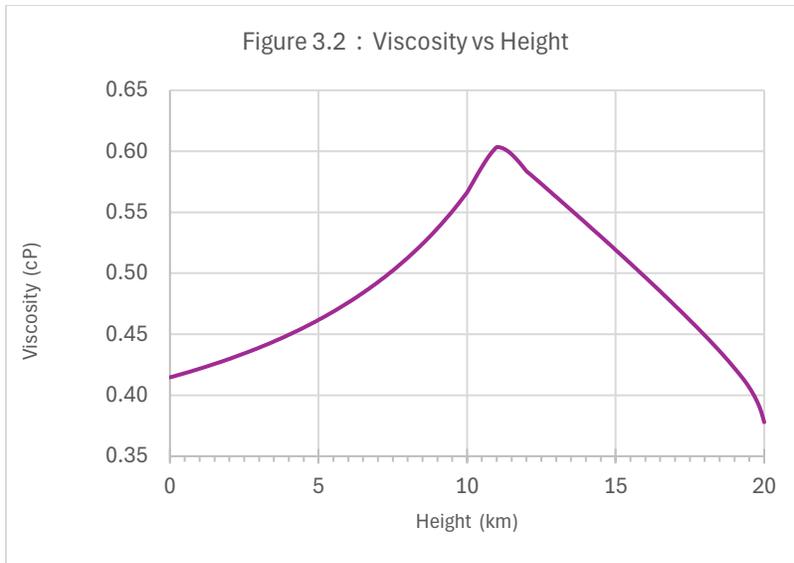

So, while H$_2$S's properties do change during its passage through the hose, these have no appreciable effect on its overall behavior.

**3.4 : Fluid Solidification**

One material property that could become crucial, is the possible solidification of H$_2$S due to either high pressure at the base of the hose or low temperature at the top. Experimental data on H$_2$S's temperature-dependent liquid-solid phase transition is limited, but existing data$^{\{36\}}$ for the melting curve is displayed in Fig. 3.3. To see if solidification will occur, we'll superimpose the variation of H$_2$S's pressure, and that of the hose's temperature, along the length of a 3 cm diameter hose. The abrupt vertical transition in the hose's P(T) curve is due to the near isothermal nature of the atmosphere from 11-20 km, i.e., along the upper half of the hose.

Unfortunately, the fact that the hose's P(T) curve lies fully below the H$_2$S melting curve, is not assurance that H$_2$S will never solidify. The "standard" atmosphere$^{\{32\}}$ used here is a global average over time and latitude, actual temperatures can be hotter or colder than its values. It's clear from Fig. 3.3, that if the stratosphere ever becomes 5K colder (which it will occasionally do), the H$_2$S might solidify in portions of the hose. Or, if we wished to field a longer hose than 20 km, the pressure in the liquid H$_2$S would increase, exceeding the solidification threshold. Just as importantly, the melting curve shown in Fig. 3.3 is based on limited (and fairly old) experimental results; an updated experimental curve may differ enough to make solidification more of a concern, or less of one. Unfortunately, at this point, there is no way to know whether H$_2$S will solidify inside the hose or not. There are, however, steps that can be taken to eliminate all such risk; the most powerful of these will be discussed next.

pg. 18

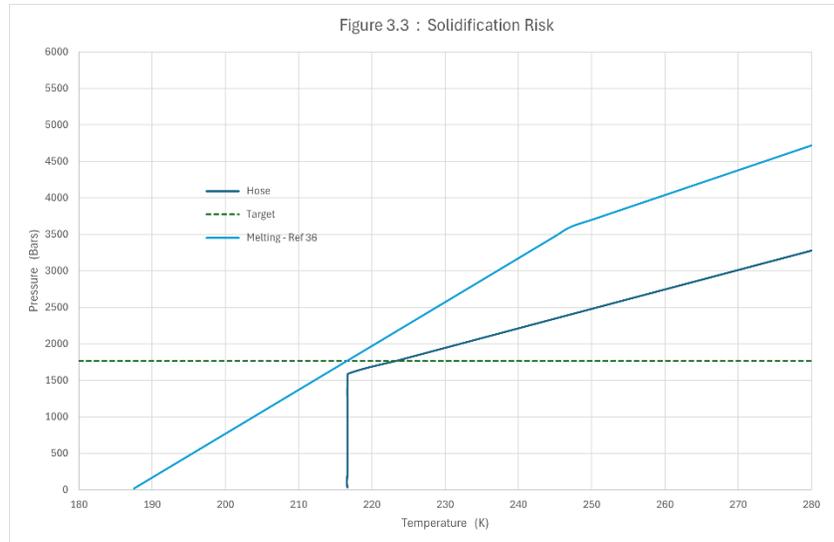

Figure 3.3 : Solidification Risk

### 3.5 : Pumps Along the Hose

There is one way to guarantee that solidification can never occur. This is to lower the pressure imposed on the $H_2S$ by performing some of the pumping along the length of the hose, rather than only at its base. Since both the gravitational and frictional pressure gradients are essentially constant, then pressure drops linearly along the hose. Accordingly, if we use *N* pumps, evenly spaced at *L/N* intervals along the hose (i.e., 1 at the ground and *N-1* mounted on the hose), then the peak pressure drops by a factor of *N*. Replacing the single large pump at the ground with 2 or 3 smaller ones (i.e., 1 or 2 on the hose) should lower the $H_2S$ pressure enough to avoid solidification; if we use 10 or more pumps, then risk of solidification completely disappears.

Placing pumps along the hose represents a major design decision, having several advantages and disadvantages well beyond preventing solidification.

A key advantage is that, because the peak pressures decrease, so does the amount of tension we need to insure lateral stability. Since tension must ultimately be provided by the upward force from balloons, this reduction can lower the balloon size, hence be beneficial. Another advantage to lowering pressures is that the hose walls can be made thinner and lighter. For these reasons, use of multiple pumps should lower the overall mass of the hose system.

There are, however, several major disadvantages to placing pumps along the hose. An obvious factor is simply the weight of these pumps. Their reliability will also be a concern, since it will be hard to service them; we'll probably need to field extra pumps for redundancy. In addition to the pumps themselves, we need to provide electricity to power them. This will require us to run electrical cables up the hose, and choose their cross-sectional area, trading off cable weight against resistive power losses.

We'll also find that most of the tension reduction gained by lowering the peak pressure (and hence the lateral stability requirements) is offset by a shifting of loads from the ground to the hose itself. This results from the reaction force applied to each on-hose pump as it boosts the fluid pressure. In effect, the weight of the most of fluid (except that of the lowest inter-pump segment) is now carried by the hose (and hence balloons), rather than by the ground. Similarly, fluid drag no longer helps support the hose, since the pressure supplying this is reacted into the hose, not the ground. Accordingly, while on-hose pumps should somewhat reduce the balloon requirements, the savings may not be large.

The biggest disadvantage to using on-hose pumps, however, is qualitative; it leads to a more complicated hose system. It's a lot simpler and more reliable to use a single pump on the ground.

**3.6 : Balloon Design**

For scoping purposes, we've so far used an "ideal" balloon; specifying a single large spherical balloon at the top of the hose to hold it up. The balloon itself has been modeled as a sphere, having a buoyancy set by the weight of displaced air, and by ignoring its own mass (both that of its structure and internal gas).

As part of an improved hose design, this balloon treatment should be refined. We will, however, continue to keep the treatment as straightforward as possible. Hence, we will continue to design with spherical, rather than pumpkin-based, shapes, to assume that the balloon is sealed, and that its internal gas is in pressure-temperature equilibrium with the 20 km atmosphere.

One fundamental question in designing the balloon, is whether it should actually be a single large balloon, or whether we should instead use a large number of smaller balloons. From a buoyancy concern, all that matters is the total volume of displaced air; as long as their aggregate volume is the same, both options will provide the same lift. The argument for a single balloon is simplicity and, probably, a lower overall structural mass. The argument for a collection of smaller balloons, is the fact that replicating many small, standardized, balloons is likely cheaper and simpler to implement than fielding a large customized one. It also provides a number of other programmatic advantages; small balloons are easier to test, and their design and acquisition can begin early, in parallel with other hose components, rather than afterwards once the necessary size is known. For these reasons, plus the fact that we can adapt to unanticipated changes in the hose mass simply by changing the number of balloons used, we'll prefer the multi-balloon approach; if its performance penalties are tolerable.

While this decision does not affect the lifting power of the balloon system, it certainly might affect the "cost" represented by the weight of the balloon's gas and structure.

Because the total amount of enclosed gas is the same in both approaches, so will be its mass. The size of this mass is easily calculated: Given pressure-temperature equilibrium, the mass of the internal gas is simply related to that of the external atmosphere by the ratio of their molecular weights. While $H_2$ offers the lowest mass penalty for this fill gas, we'll baseline He. The 2X increase in mass is affordable in exchange for the elimination of safety and operational concerns accompanying $H_2$.

pg. 20

Modern, large volume, balloons are not actually spheres with uniform thickness walls. Instead, they usually adopt a large number of gores, each running pole-to-pole and evenly spaced around the balloon's circumference. Stress is handled asymmetrically; the balloon walls accumulate circumferential pressure loads, transferring them into strong tendons at the gore edges, which provide meridional strength. In principle, the mass of both these structural components will scale with the volume and over-pressure of the balloon's gas; hence will not depend on the number of balloons used. In practice, the balloon wall thickness (particularly for small balloons) is lower-bounded by fabrication, gas leakage, and handling concerns, not by stress. Hence, this portion of the structural mass does scale with surface area, not volume, so will increase as $N_B^{1/3}$.

To be specific, let's consider standardizing on a 20 meter balloon diameter. Using an atmospheric density of 88.9 kg/m³, an "ideal" balloon could lift a mass of 372 kg. For the performance of a "real" balloon, we must subtract the mass of its internal He gas, as well as that of its structure. The gas has a mass, set by the molecular weight ratio with air, of 52 kg. Since tendons can be made from modern high-strength fibers, their mass is much smaller than that of the balloon's skin, which we'll conservatively treat as 20 microns of 1.2 gm/cc plastic.

This, of course, is simply a convenient simplification. In reality, balloons will be pressurized above the value of the surrounding air, and so must be strong enough to withstand a pressure differential. A 20 meter ballon made from a 20 micron film of modern polyethylene would, if operating at a 5 MPa stress, only be able to handle an overpressure of 10 Pa, far below what's necessary. Accordingly, the balloon must actually count on high-strength fibers to carry the primary pressure loads. Whether this should be done by conventional multi-gore layouts, by an overlying fiber web[37], or by use of coated sailcloth for the entire skin, is unclear. But, for mass calculations, we'll continue to model the walls via an areal mass density of 24 gm/m². This leads to the walls of our 20 meter balloon having a mass of 30 kg.

The effect of gas and wall masses is to reduce the lifting power of a "real" balloon from 372 kg, down to 290 kg, i.e., by 22%. If we were to use a single 100 meter balloon instead of 125 of the smaller 20 meter balloons, this penalty fraction would drop to 16%. So, there is indeed an advantage to using a single balloon, but it is probably not compelling enough to offset the programmatic benefits of using multiple smaller balloons.

Temperature variations are another concern that balloons must commonly deal with. Since the density of the gas inside a sealed balloon is fixed, if the temperature changes, so will the pressure inside the balloon. The temperature of the atmosphere itself does not vary much at 20 km, the diurnal variations[38] are around ±1°C, and seasonal ones[39][40] are less than ±5°C. As a result, the largest temperature changes seen by stratospheric balloons don't arise from the atmosphere, but stem from the balloon being out of thermal equilibrium with it; the balloon's skin is heated by the Sun during the day, and radiatively cools at night. A tethered balloon has an advantage here over drifting ones, in that wind-driven convection will help drive the balloon's skin into thermal equilibrium with the surrounding air.

One issue, of course, is where to place the balloons. Obviously, at least some balloons must be located at the top of the hose. But, should they all be placed there, or should some balloons be placed along the length of the hose, providing more localized support?

The advantage of mid-hose balloons, is that they operate within denser air, so provide more buoyancy; we'll need fewer balloons if some are placed along the hose. These balloons also localize lift, closely matching the local weight of the hose; if the balloon's lift is all generated at the top of the hose, then a long tether is needed to transmit this upward force downwards along the hose.

Unfortunately, as will be discussed in Chapter 5, wind will have a major impact in the performance and design of the hose, as well as that of the balloons. Mid-hose balloons will experience stronger wind forces than ones at the top of the hose. And, the lower intra-hose tension allowed by mid-hose balloons, is actually a detriment in strong winds; tension helps stiffen the hose.

Accordingly, we will postpone discussion of mid-hose balloon placement until later, after the effects of wind on the hose have been considered.

In the meantime, we'll base the upcoming discussions of hose design and performance on the assumption that all balloons are located at the top of the hose. Not only may this choice turn out to be both simpler and operationally preferred, it also provides a clearcut and standardized way to compare hose design choices, without conflating these with decisions on where mid-hose balloons should be placed.

**3.7 : Power Requirements**

After the hose is in place, there is an ongoing power requirement to operate it, i.e., to pump $H_2S$ up to the stratosphere.

Fundamentally, the power needed to do this is given by the gravitational potential energy and the flow rate:

$$\mathcal{P}_G = \Lambda \dot{M} g L \qquad (3.9)$$

This can be seen to be equivalent to the $P\Delta V$ work done by the gravitational-head portion of the pressure supplied by the pump(s). The "cost" of delivering 100 kton/yr of $H_2S$ up to a 20 km altitude is 660 kW. In practice, this must be increased somewhat due to pump inefficiencies; assuming 80% efficiency, it becomes 825 kW.

We'll actually need more power than this, both due to resistive losses in cables bringing power to any on-hose pumps, and to overcome flow friction.

Both of these two factors depend on details of the hose design, and are subject to tradeoffs with other factors.

*Resistive Power Loss*

If our design uses on-hose pumps, then power will be lost resistively in the cables transporting it to the pumps. We'll model this by tracking current flow through a three-phase cable network designed to deliver constant power to each pump.

As AC current $I$ travels up each of the 3 cables, a portion $J$ splits off into each pump, while the rest continues upward to the next pump; each path experiencing the same voltage drop $V$.

$$I_k = I_{k+1} + J_k \tag{3.10}$$

$$V_{k-1} = \mathcal{R}_{k-1} J_{k-1} = \mathcal{R} I_k + \mathcal{R}_k J_k \tag{3.11}$$

Here, $\mathcal{R}_k$ models the load resistance for pump $k$ and is adjusted to deliver the proper power to it; $\mathcal{R}$ is the resistance of each inter-pump portion of the cable.

$$\mathcal{P}_p = \frac{1}{\eta} \frac{\mathcal{P}_G + \mathcal{P}_F}{N} = \frac{3}{2} \mathcal{R}_k J_k^2 = \frac{3}{2} \frac{V_k^2}{\mathcal{R}_k} \tag{3.12}$$

$$\mathcal{R} = \rho_e \frac{L/N}{A_c} \tag{3.13}$$

The network is solved recursively, from the top pump, $k=N$, downwards, starting with $I_{N+1} = 0$, and a specified value for the top load resistance $\mathcal{R}_N$. Upon completion, $I_1$ is the current flowing into each phase of the network from the power source, and $J_1$ is that flowing into the lowest, ground-mounted, pump. The total resistively wasted power is $\mathcal{P}_R$:

$$\mathcal{P}_R = \frac{3}{2} \sum_{k=2}^{N} \mathcal{R} I_k^2 \tag{3.14}$$

The primary way to reduce $\mathcal{P}_R$ is to reduce the cable resistance $\mathcal{R}$, by increasing its cross-sectional area $A_c$. The tradeoff, is of course that this increases the hose's weight and hence the number of balloons required to support it.

*Frictional Power*

Overcoming flow friction requires power; this "cost" is just given by that portion of the $P\Delta V$ work. Since the fluid density is nearly constant along the hose, this is given, using Eq. 2.3, by:

$$\mathcal{P}_F = \Lambda \dot{M} \frac{P_F}{\rho} \equiv q_F L \tag{3.15}$$

We can reduce this power by the same steps which cut the frictional pressure drop. By far, the dominant method, is increasing the hose's bore size, $D$. This will invoke the same tradeoff;

increasing $D$ raises the weight of the fluid carried by the hose, and also that of the hose walls needed to contain the pressure.

**3.8 : Heat Dissipation and Fluid Temperature**

It's important to understand what ultimately happens to power delivered to the hose and fluid. Unlike $\mathcal{P}_G$, which does useful work, $\mathcal{P}_F$ winds up as heat in the fluid, the hose walls, and ultimately the atmosphere.

The local frictional heat, $q_F = \mathcal{P}'_F$, will be processed in two ways, by advection within the flowing H$_2$S, and by radial coupling into the atmosphere.

The advection results in a longitudinal temperature gradient:

$$q_{adv} = \Lambda \dot{M} C_p T' \qquad (3.16)$$

If all of the friction heat remained in the H$_2$S as it was pushed up the hose, the resulting temperature increase would be given by:

$$P_F = \rho C_p \Delta T \qquad (3.17)$$

Using a value[41] of 0.5 Cal/gm/K for $C_p$, and a representative $P_F$ value from Table 2.3 (2.5 kbar for a 3 cm bore), we calculate a temperature rise of 125K. If this much temperature rise actually occurred, it would be virtually impossible to keep H$_2$S liquid, preventing its use in the hose.

In reality, however, the frictional heat does not all remain in the H$_2$S, but most is instead transferred to the surrounding atmosphere, via conduction through the hose wall and then wind-mediated convection.

Atmospheric convection can be expressed by a lineal heat transfer rate, $q_{conv}$:

$$q_{conv} = (\pi D_o) h_c (T_W - T_A) \equiv h_A (T_W - T_A) \qquad (3.18)$$

$$h_c = N_u \frac{k_A}{D_o} \quad , so \quad h_A = \pi k_A N_u \qquad (3.19)$$

Here, $k_A$ is the thermal conductivity of air, and $D_o$ is the outer diameter of the hose. The dimensionless Nusselt coefficient, $N_u$, depends on the local atmospheric density, temperature, viscosity, and wind speed. We'll utilize a correlation[42] applicable for high Reynolds number crossflow over cylinders. In Figure 3.4, the $h_A$ value is plotted versus wind speed for atmospheric conditions (density and temperature) corresponding to five different altitudes along the hose.

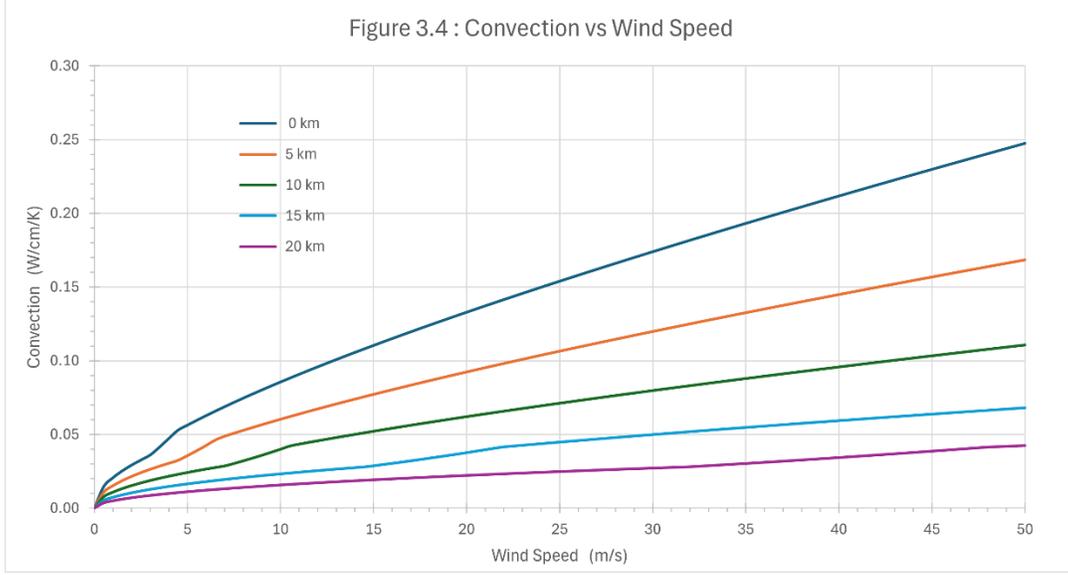

Figure 3.4 : Convection vs Wind Speed

The need to conduct heat through the hose wall, results in a difference between the fluid temperature $T$, and the wall's outer temperature, $T_w$:

$$q_{wall} \equiv h_w(T - T_w) \qquad (3.20)$$

$$h_w = 2\pi k_w / ln(D_o/D) \qquad (3.21)$$

Equating the $q_{wall}$ and $q_{conv}$ heat flows, and calling this $q_{cond}$, we get

$$q_{cond} \equiv h(T - T_A) \quad , \quad h = \frac{h_w h_A}{h_w + h_A} \qquad (3.22)$$

Overall, the fluid temperature can be described by combining conduction and advection:

$$q_F = \left(\frac{\Lambda \dot{M}}{\rho}\right) P'_F = \Lambda \dot{M} C_p T' + h(T - T_A) \qquad (3.23)$$

This can be used to describe how far out of equilibrium, the temperature of the fluid is with that of the air as it moves up the hose.

$$\Lambda \dot{M} C_p \tau' + h\tau = q_F - \Lambda \dot{M} C_p T'_A \qquad (3.24)$$

where
$$\tau \equiv T - T_A$$

As seen by the righthand side of this equation, the thermal mismatch $\tau$ is driven by two separate effects, frictional heating and the atmospheric thermal gradient.

Frictional heating $q_F$ is usually the dominant effect. Its thermal warming tracks the frictional pressure gradient $P'_F$, and hence is most severe for small diameter hoses; increasing the diameter provides a way to greatly reduce this portion of the thermal overshoot.

However, even without any frictional heating, the $H_2S$ temperature will still exceed that of the atmosphere. This is because the atmosphere cools with altitude, and the fluid's thermal inertia prevents its temperature from falling as rapidly.

The actual temperature profile of the fluid depends on the interplay of the heating (from friction and atmospheric cooling), the fluid's thermal inertia, and the strength of thermal coupling to the atmosphere (based on wall conduction and wind-driven convection). We illustrate their effect in Fig. 3.5 by showing temperature profiles for a single-pump hose with different values of the bore diameter (and hence, different amounts of frictional heating). To simplify the comparison, these calculations neglect thermal resistance from the hose walls, and use a simple, fixed wind speed of 20 m/s, avoiding (for now) linking the comparison to the selection of a particular wind-altitude profile.

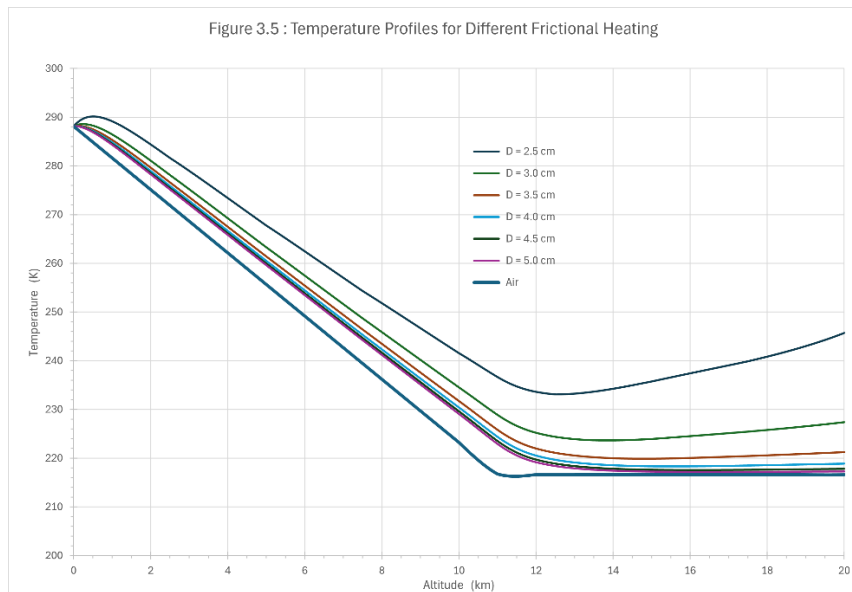

Several things are worth noting. First, as expected, frictional heating has a very strong effect on the fluid temperature. But, a twofold change in diameter can change it from being a dominant influence to a negligible one. Secondly, even when frictional heating is insignificant, the fluid's thermal inertia causes it to become about 5°C warmer than the air. And, note that this inertial warming peaks midway up the hose, where the atmosphere becomes essentially isothermal. This is precisely where the threat of $H_2S$ solidifying peaks; while slight, the extra warmth will help reduce this risk.

For a hose utilizing on-hose pumps, some of the electrical power required by them winds up as waste heat.

One portion of this is lost resistively in the electrical cables bringing power to the pumps, and can be dissipated in a couple of ways, depending on the cable design. In one approach, the cables are held in tight thermal contact with the hose, and cooled by the flowing $H_2S$. In this case, the local gradient in $\mathcal{P}_R$ augments the frictional heating $q_F$, and causes additional warming of the $H_2S$. In another approach, the cables and hose are less integrated, and the cables are cooled directly by atmospheric convection.

Another source of waste heat comes from inefficiencies in the on-hose pumps and their electronics. Given the concentrated nature of this heating, it is unlikely that it can be cooled by local airflow. Instead, the pumps will be cooled by the $H_2S$ fluid flow, causing a local temperature rise downstream of each pump before the heat is eventually convected from the hose into the atmosphere. The base pump also generates such waste heat, but this can be handled separately at the base station, rather than by the hose's $H_2S$ flow.

# 4 : Liquid-based Hoses (Liq-1 and Liq-2)

We have incorporated the design refinements into numerical simulations of the hose. This allows prediction of its capability for delivering $H_2S$ up to the stratosphere, and examination of how this performance varies with changes to key design choices and parameters. These simulations are used to design and model the nominal vertical hose, before deflection by wind forces; these effects will be treated later, in Chapter 5.

**4.1 : Configuration**

The hose system will be designed to deliver a steady flow of $\dot{M}$ kg/s of sulfur (and $\Lambda \dot{M}$ of total mass) to the stratosphere at altitude $H$ km. For now (until consideration of wind-caused deflections) it will be considered vertical, and based at sea-level; so its length $L$ is equal to $H$.

We'll design for $N_P$ evenly spaced pumps. So, they are separated by a distance of $L/N_P$; one pump is on the ground and $N_P$-1 are mounted along the hose.

The hose is composed of several different components; a conduit, a tether, and in some cases electrical cables and pumps. The hose conduit itself has a constant bore of $D$ cm, and contains the pressurized $H_2S$ flow. It has a thin inner wall surrounded by a thicker, fiber-based wall to resist the fluid pressure. The thickness of this wall is adjusted based on the pressure; each region in between pumps is split into $N_W$ sectors, each having a wall thickness chosen to handle the sector's maximum pressure. The tether supports the other hose components; its cross-sectional area is (like the hose wall thickness) constant within each sector, but can vary at each boundary. If we do have on-hose pumps, we'll need electrical cables. In our current design, these cables (containing a conductor and insulation) will have constant properties along the length of the hose. The above components are continuous ones, but there are $N_P$-1 pumps as well; each constitutes a discrete load, both from its own weight as well as the reaction force due to its pumping action.

At the top of the hose, we'll have a sprayer system to disseminate and aerosolize the $H_2S$, as well as the attachment site for a suite of $N_B$ balloons which connect to the tether and ultimately hold up the hose.

**4.2 : Modeling Procedure**

The simulation uses two passes along the length of the hose.

The first pass is downward, from top to bottom. This pass is to determine the pressure along the length of the hose, as well as the $\Delta P$ pressure jumps needed at each pump. The pressure variation is found by starting at the top with enough pressure to prevent the $H_2S$ from boiling, and then integrating the gravitational and flow-friction pressure gradients from equations 2.1 and 2.3 until the next pump site is reached. At this point, pressure is again reduced to the anti-boiling floor, the pump's ΔP value is remembered, and the downward pass continues until the ground is reached.

We next use an upward pass, to determine the variation of hose wall and tether thicknesses, as well as that of the tension carried by the tether, along the length of the hose. This proceeds for each of the $N_P$ pump regions, first across the pump itself, and then through each of the $N_W$ sectors on the way to the next pump. The tension increases abruptly when passing a pump for two reasons; the weight of the pump, and the reaction force due to the ΔP applied by the pump. This ΔP, and hence its reaction force, comes from two components; one from the weight of the $H_2S$ in the region above the pump, and the other from flow friction in the same region. After the pump, we proceed through each of $N_W$ hose sectors. Both the tether and hose wall have constant areas within each sector (as, of course, do the electrical cables); the weight from all three of these increases the tension which the tether must support, while flow friction decreases it. The proper hose wall thickness for the sector is sized by the maximum $H_2S$ pressure within it. Likewise, the proper tether area is sized by the maximum tension within the sector. Since the tether weight itself contributes to this tension, this peak value must be found self-consistently, and can occur either at the top or the bottom of the sector.

Finally, we calculate the electrical power needed to drive any on-hose pumps. Some (generally most) of this is used by the pumps, whereas a fraction is lost within the cables. Since cable sizes are kept uniform in our present design, the resistive losses are highest near the base of the hose. An alternative design would make the cables thicker near the base to balance out the losses, at the cost of more weight there.

As discussed previously, while the $H_2S$ fluid's temperature is usually close to that of the atmosphere, it will be somewhat warmer, the difference depending strongly on the hose diameter, and hence the amount of frictional heating. The actual value of this temperature difference also depends significantly on thermal resistance from the hose walls and on the wind speed. Since wind speed varies with time, altitude, and hose location, we'll simply present the thermal effects of a few representative cases here, and reserve more detailed discussion of wind to Chapter 5, when we treat its role in laterally deflecting the hose.

**4.3 : Hose Properties**

The hose system is characterized by several different parameters; these are tabulated in Table 4.1.

The properties are of different types with different effects on the system's performance. Some of them, like the altitude and flow rate, are set more by external considerations than by the hose itself. Most of the properties act monotonically: For instance, the larger the hose's or tether's specific strength, or the more hose sectors between pumps, the better. Likewise, the lower the mass of the pumps, that of the balloon walls, or the systems at the top, the lighter the hose system will be.

Only a few properties can be truly optimized. The "cleanest" one of these will be the conduit diameter; small values reduce fluid mass, but increase frictional pressure and heating, the reverse is true for larger diameters. In other cases, the tradeoffs are less clearcut: Increasing the number of pumps (i.e., placing more on the hose), will allow the hose to use less balloons, but the tradeoff is a more complex, probably less reliable, overall system. Likewise, if we do use on-hose pumps, then smaller cross-section conductors obviously allow a lighter hose; here the tradeoff, is more resistive losses, so use of more total electricity.

Table 4.1 : Hose Properties

| Property | Nominal | Range | Units | Comments |
|---|---|---|---|---|
| Fluid | $H_2S$ | | | |
| Altitude | 20 | 15 - 25 | km | |
| Flow rate | 100 | 25 - 200 | kg/sec | Sulfur component |
| # Pumps | | 1 - 20 | | One on ground, rest are on-hose |
| Conduit diameter | 3 | 2 - 5 | cm | Constant along hose |
| # Sectors per pump | 10 | 1 - 20 | | Each has constant hose & tether |
| Hose strength | 1.5 | 0.5 – 3.0 | GPa | Derated strength , $\rho$ = 1.44 gm/cc |
| Tether strength | 2.0 | 1.0 – 4.0 | GPa | Derated strength , $\rho$ = 0.97 gm/cc |
| Pump mass | 0.5 | 0.25 – 1.50 | W/gm | Incorporates spares |
| Wind speed | 20 | 0.0 – 50.0 | m/sec | Constant along hose (for now) |
| Hose thermal conductivity | 2.0 | 0.04 – 10.0 | W/(m*K) | |
| Conductor area | 0.15 | 0.05 – 0.25 | $cm^2$ | Al : Constant along hose |
| Pump efficiency | 0.85 | | | |
| Insulator/Conductor | 1.0 | | | Area ratio |
| Stability ratio | 1.25 | | | Minimum tension/compression |
| Balloon size | 20 | | m | |
| Balloon wall | 24 | | $gm/m^2$ | 20 μm @ 1.2 gm/cc |
| Top systems | 200 | | kg | Sprayers, sensors, controls |

**4.4 : Single-Pump Hose (Liq-1)**

The simplest hose would use a single pump, located on the ground at the base of the hose. We've discussed this design earlier, and an updated design (using the nominal parameters in Table 4.1) is presented below, in Fig. 4.1. Here, we plot the number of balloons required to suspend the hose as we vary the conduit diameter. As discussed in section 3.2, one step that can be used to reduce both hose and tether weights is to vary their thicknesses along the length of the hose. The benefits of this are illustrated in Fig. 4.1 by showing the performance for different numbers of sectors along the hose. It's clear that there is a substantial advantage to using multiple sectors, but it's also clear that this effect saturates once we use 10 or more sectors. For clarity, we'll display the number of balloons decimally, rather than rounding upward to the next highest integer.

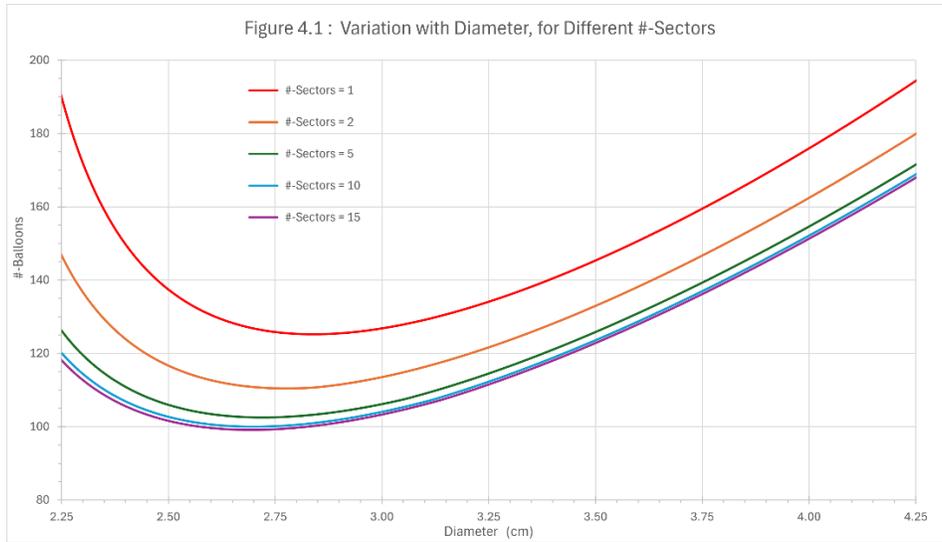

The optimum conduit diameter is about 2.7 cm, but this is fairly broad, extending from about 2.4 to 3.0 cm.

To determine whether some diameter within this range might be preferable to the value which actually minimizes the balloon count, let's examine the effect of conduit diameter on fluid pressure. We expect this variation to be monotonic; increasing diameter will strongly reduce the frictional pressure drop, while having no effect on the gravitational component. This variation is shown in Fig. 4.2; since the number of sectors has no influence on fluid pressure, we only show results from the single-sector design. Two pressure curves are shown, pressure at the base of the hose, and that at 11 km altitude, which (as seen in Fig. 3.3) is where the $H_2S$ is at most risk of solidification. We also display the pressure (also from Fig. 3.3) where solidification is estimated to occur. This figure indicates that it may be possible to avoid $H_2S$ solidification by using conduit sizes which are somewhat above those which minimize the number of required balloons.

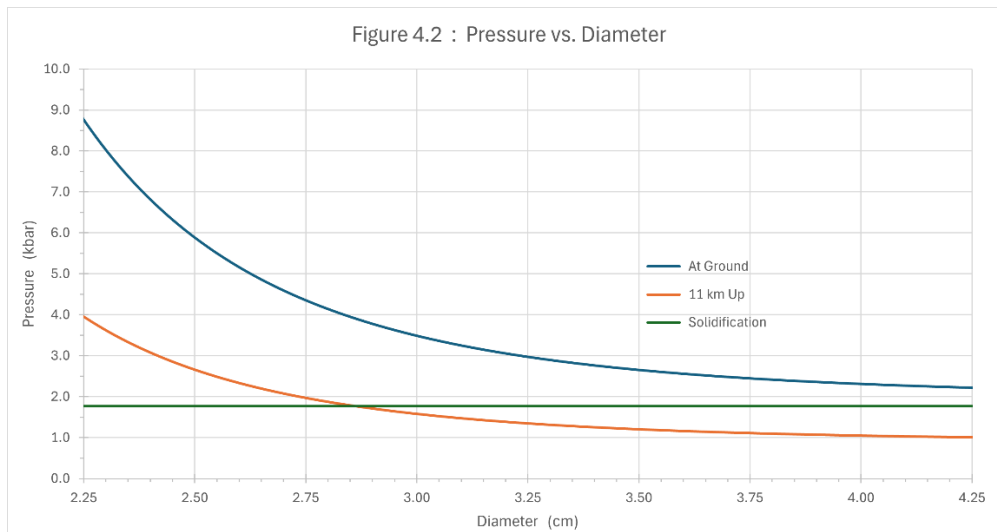

The large frictional pressure at small diameters has a profound effect on many properties of the hose; we'll demonstrate this in the next set of figures for the hose mass, the excess warmth of the $H_2S$, and the power needed to operate the hose.

In Fig. 4.3, we investigate the effect of the conduit diameter on the masses of the hose's components, baselining the use of 10 sectors as determined above. As expected, the fluid mass grows monotonically with the conduit diameter. Since, in a single-pump design, the weight of the fluid is supported from the ground, this increase has no direct consequence. The diameter has a more nuanced effect on the mass of the hose structure; both the walls of the hose and the tether supporting the hose. For very small diameters, overcoming flow friction requires very large pressures. This drives up the wall mass, since the walls need to contain this pressure, and drives up the tether mass, since higher tension values are needed to prevent buckling. Accordingly, the hose mass is high for small diameters, but drops as the diameter increases, reaching a minimum at a value of around 3.0 cm. Beyond this optimum, both portions of the hose mass increase slowly with further growth in the diameter. This occurs because the pressure decrease stops, asymptoting to the gravitational head value, whereupon the growth in hose area causes both the wall area and the tension requirements to increase.

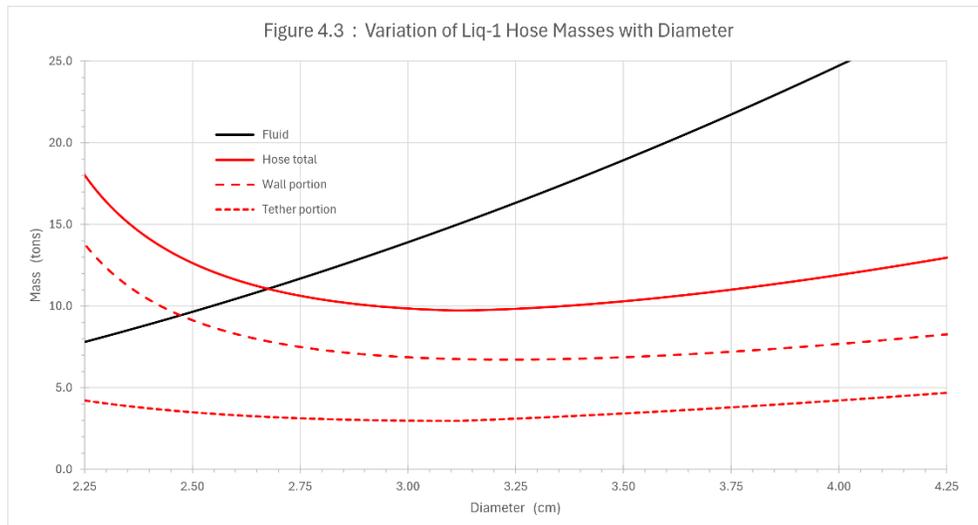

As discussed earlier, the $H_2S$ fluid in the hose will be warmer than the surrounding atmosphere; when frictional heating is strong (i.e., for small conduit diameters) we expect this thermal imbalance to be large. Figure 4.4 quantifies this behavior, for the values of wall conductivity and wind speed listed in Table 4.1. For diameters below about 3.0 cm, the $H_2S$ is much warmer than the atmosphere. Given the high pressure that most of the hose is at, this should not cause the $H_2S$ to vaporize; some warming is helpful as it lowers the risk of $H_2S$ solidification.

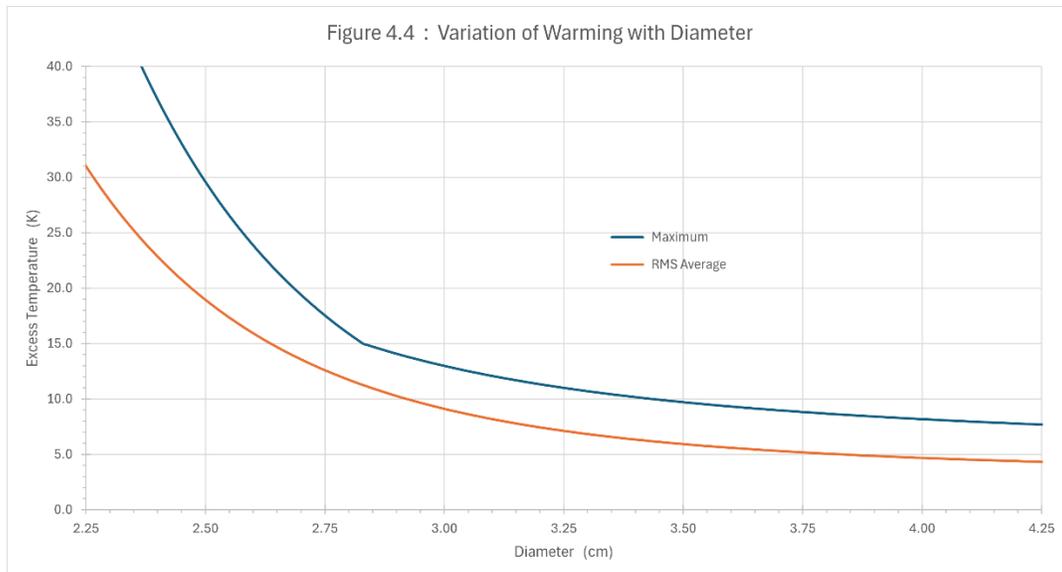

A final effect of small conduit diameters, is on the power needed to operate the hose. Fig. 4.5 illustrates the monotonic decrease in required power with increasing diameter. For small values, the pumping power is large, dominated by the need to overcome flow friction. For larger diameters, frictional pressure drops become small, and the power saturates at that needed to pump fluid up to a height of 20 km.

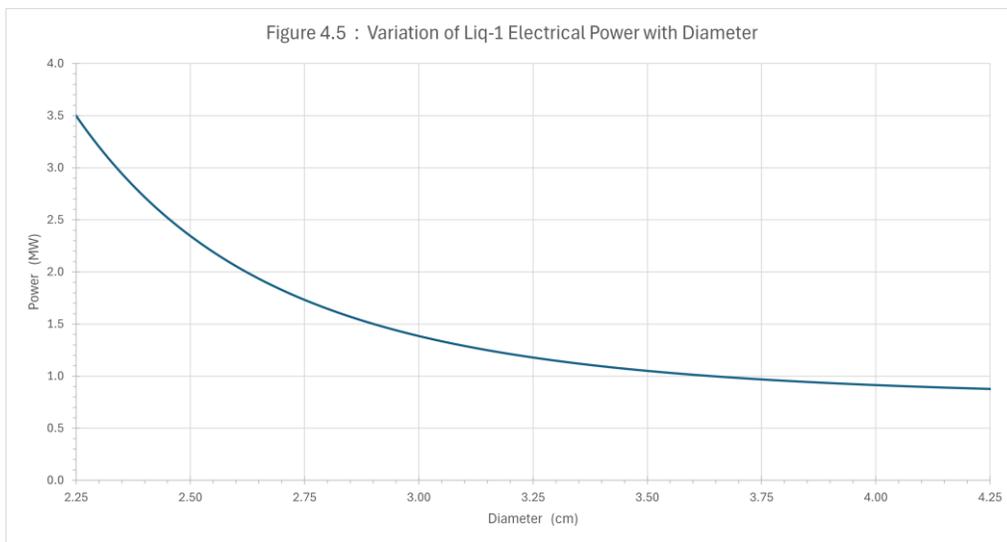

Finally, let's examine the effect of material strength, that of the hose walls in Fig. 4.6 and of the tether in Fig. 4.7. As expected, the stronger these materials are, the lighter the hose becomes, and the less balloons are needed to support it. The nominal values specified in Table 4.1 (and used for most of our subsequent analyses) are derated 2X from those of potential materials, Kevlar for the hose and Dyneema for the tether; the figures show the influence of greater or lesser strengths. By comparing Figs. 4.6 and 4.7, analyzed for a hose with 3.1 cm bore diameter, it's clear that hose strength has more influence than that of the tether. In both cases, however, the penalty for weaker than nominal materials, is larger than the gain from stronger ones.

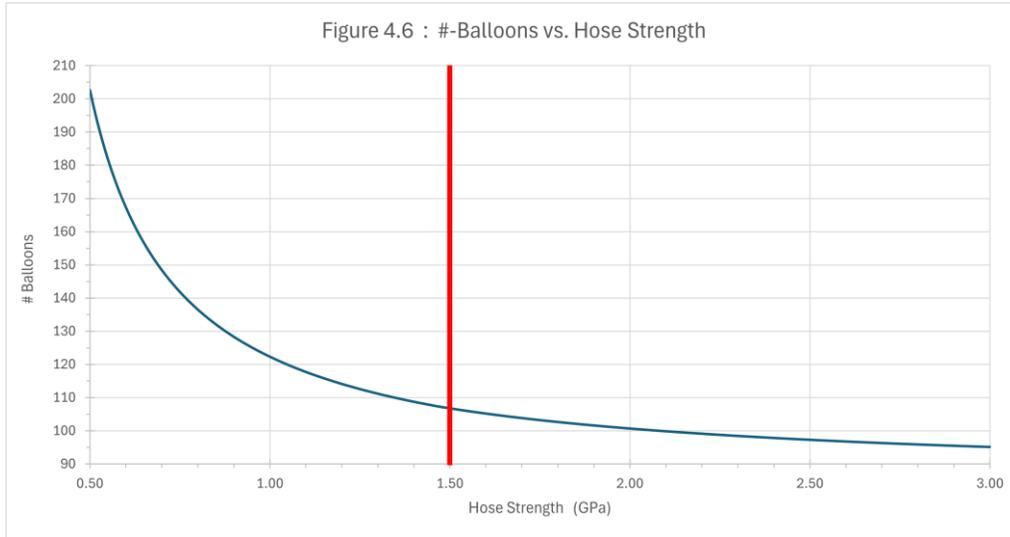

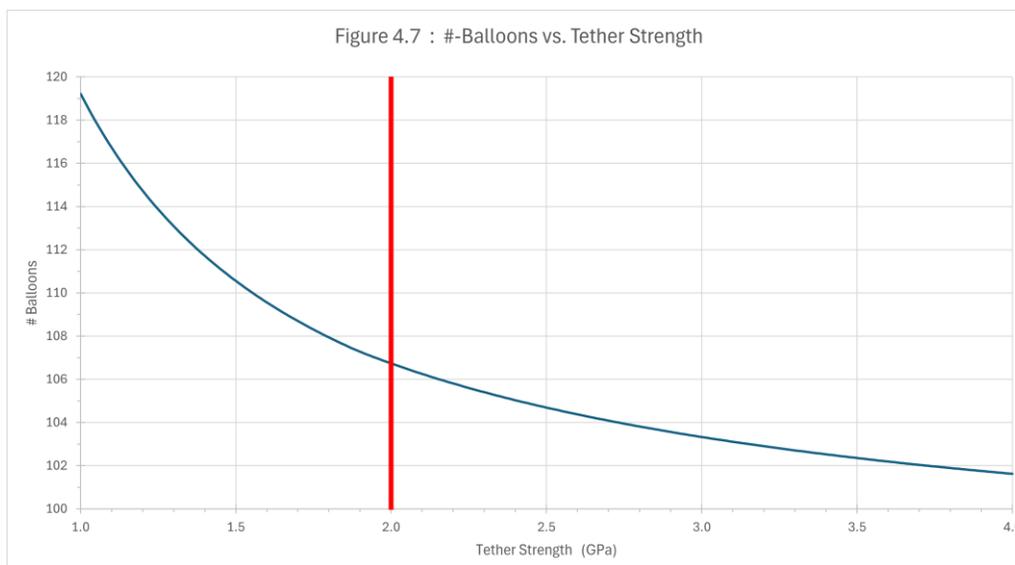

Another parameter, which can, if incorrectly chosen, have a large effect, is the thermal conductivity of the hose walls. These are in series with atmospheric convection, so that if their conductivity is lower than that, they greatly restrict thermal flow. However, for larger values of conductivity, convection dominates, and we no longer benefit from further increases in wall conductivity. A problem, is that high-strength hose materials of interest (e.g., Kevlar) have very low thermal conductivity, around 0.04 W/(m*K). As seen in Fig. 4.8, such conductivities lead to large temperature rises. At the other extreme, metals such as copper, with conductivity of 400 W/(m*K), are too weak and heavy to be used for the hose wall. In practice, of course, it's clear from Fig. 4.8 that we don't need such high values of the conductivity, the geometric mean of 4

W/(m*K) (100X higher than Kevlar, but 100X lower than copper) would work well. Hence, hose designs which incorporate a small amount (a percent or so) of a material such as aluminum would provide enough boost to the conductivity without an appreciable strength or weight penalty. Note, that the hose conductivity, particularly near the midpoint of the hose, represents a useful tool to deal with solidification worries; by lowering this (i.e., not augmenting the wall material's natural conductivity as much) we can increase the local fluid temperature.

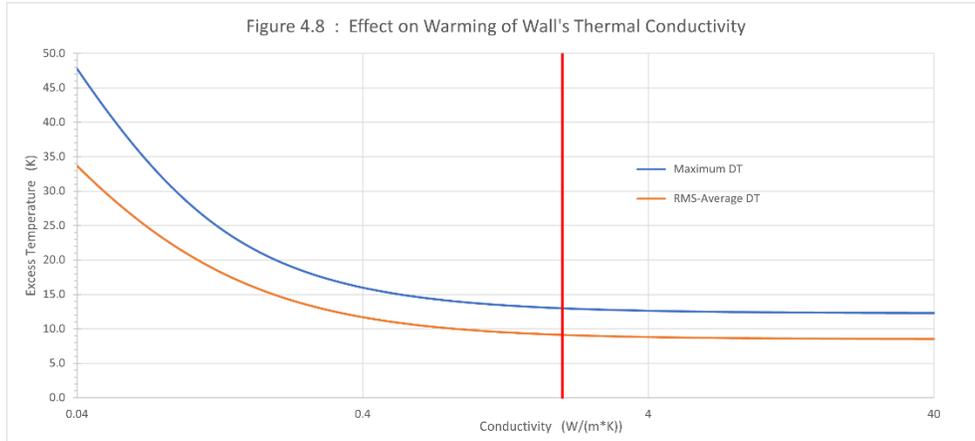

We have focused our hose discussion on a sulfur flowrate of 100 kton/yr. This represents a tradeoff between smaller values (whose cooling effect would probably be too weak to clearly measure) and larger values (too expensive an initial investment). But, before concentrating on the 100 kton/yr value for the rest of this report, it's worthwhile to quickly look at how the properties of the hose change with flowrate.

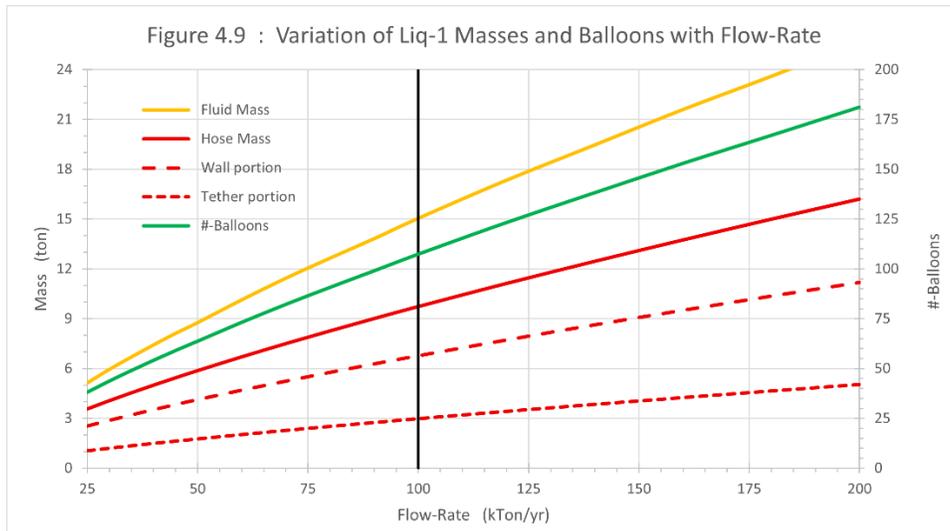

Fig. 4.9 shows the variation in the mass of the hose with flowrate, as well as the resultant number of 20 meter balloons needed to support it. Obviously, as flowrate increases, the hose's diameter should be increased to control the flow resistance; at each flowrate shown in Fig. 4.9, the hose's diameter has been picked to give the smallest hose mass. As expected, larger flowrates require bigger hoses; the mass (the hose's fluid and all its components) increases, as does the number of balloons.

One question, is how high should the hose reach? The advantage of a tall hose is that it releases aerosols higher in the stratosphere, allowing them to stay there longer[19]. But, the downside is, of course, that a taller hose will be harder to field; and, since pressures rise, so will the risk of $H_2S$ solidification. With experience and better modeling, we will be able to properly balance these tradeoffs, but it's too early to do this yet.

What we can do now, is to look at how hose properties change with its length. In Fig. 4.10a, we see that the pressure and the hose mass grow more-or-less linearly with length; This is expected, since the gravitational head grows with height, and the flow-friction and hose masses scale with the length of the hose. The number of balloons needed to hold up the hose, shown in Fig. 4.10b, increases even faster, first because the weight they must support grows, and secondly because the balloon's buoyancy drops with air density.

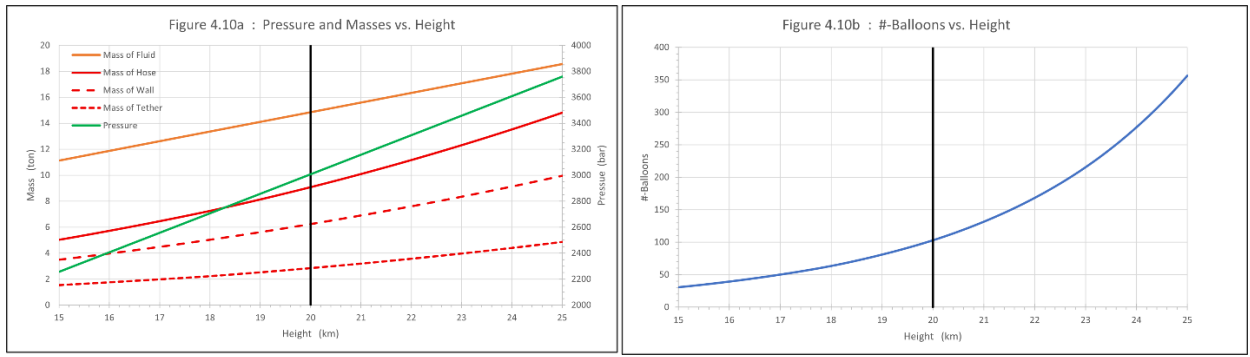

In the remainder of this report, we'll focus on a value of 20 km, both because this is a reasonable a-priori choice, and because adopting a fixed height value makes it easier to examine the effects of other hose design choices.

**4.5 : Multi-Pump Hose (Liq-2)**

While fielding a hose in which all pumping is done from the ground has many practical advantages, its high pressures may cause the $H_2S$ fluid to solidify midway up the hose. While we believe that this can be avoided, it poses a risk that can't be entirely ruled out.

As mentioned earlier, there is one way to guarantee that freezing will never occur; replace the single, large ground-based, pump with multiple smaller ones spaced along the hose, thereby greatly reducing the peak $H_2S$ pressure. While this approach will certainly increase the complexity of the hose, and raise additional reliability and maintenance issues, hose designs employing multiple pumps show that this also can decrease the number of balloons needed to support the hose.

Fig. 4.11 shows the dependence of balloon requirements with the number of pumps used (all simulations using the same bore diameter of 2.7 cm). Most of the gain comes from the first 5 or 6 pumps, after that there is only a small dependence on further pumps, with a weak optimum for designs with 12 pumps.

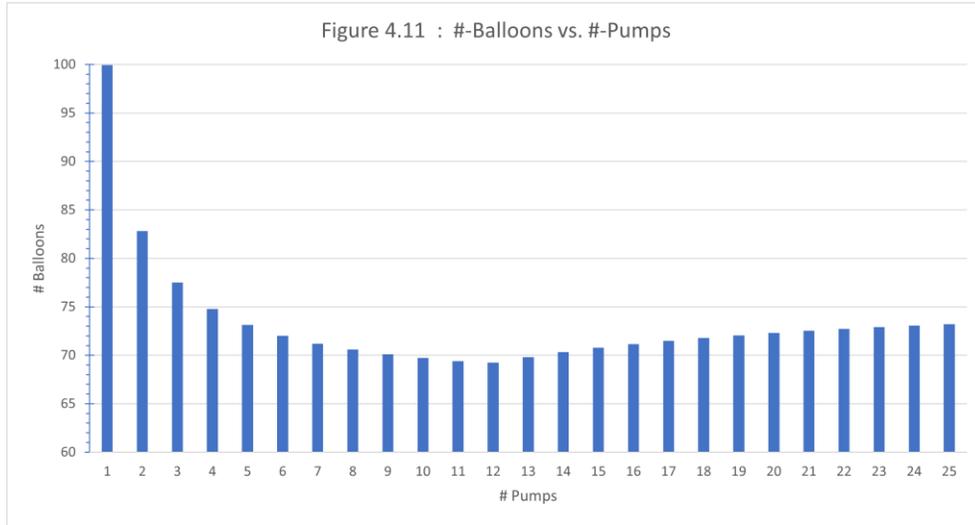

Fig. 4.12 shows that the optimum conduit size for multi-pump designs actually shifts downward to even smaller values, e.g., about 2.5 cm. This difference from the single-pump results stems from the fact that these designs accumulate pressure only between pumps, and hence can tolerate the greater frictional pressure gradients implicit in smaller bore diameters.

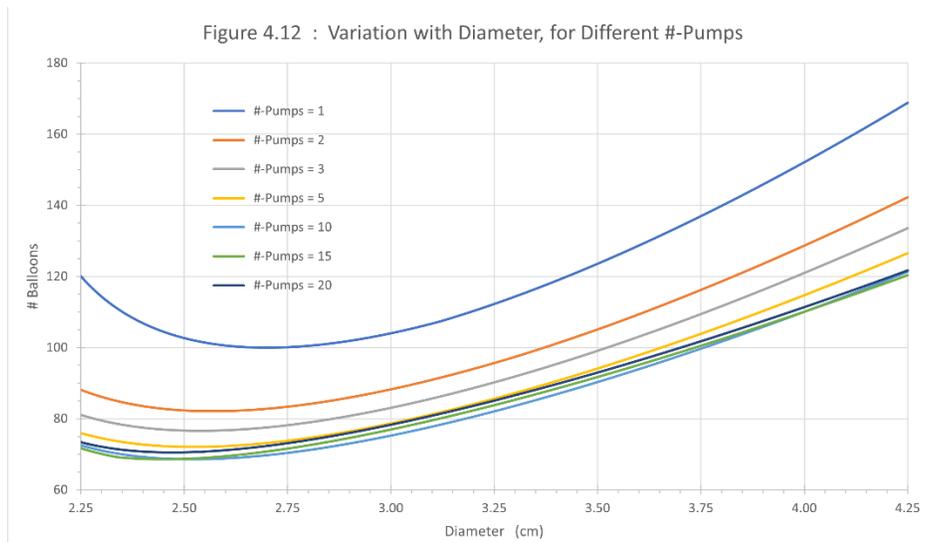

In practice, while small bores require the least number of balloons, larger bores help in other respects. As with the single-pump case, frictional heating is greatly reduced by increasing the hose diameter. This manifests itself in a drop in the hose's power requirements, and in the excess $H_2S$ temperature. The power requirements are shown in Fig. 4.13; they follow the same trend with bore diameter as for the single-pump case displayed in Fig. 4.5. However, while the actual pumping power is similar to the single-pump case, the multi-pump design requires considerably more total power, due to resistive loses in the cables transporting electricity up the hose.

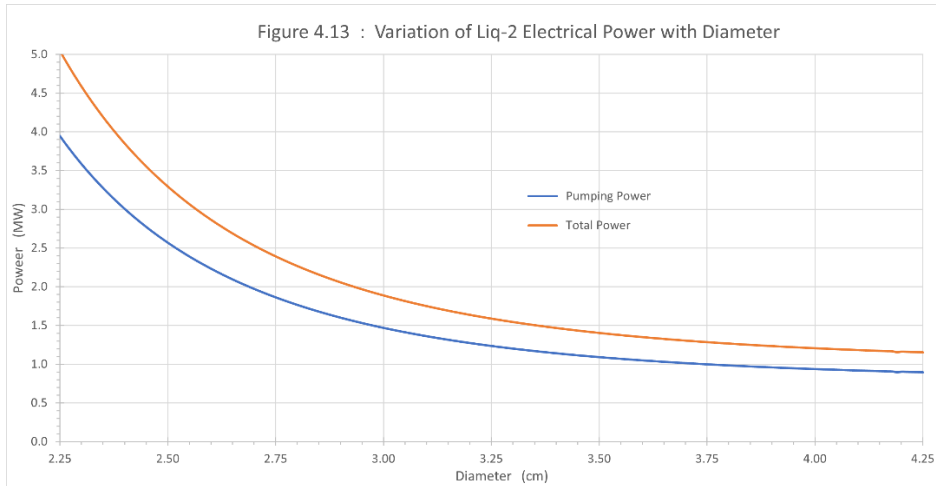

Another tradeoff is shown in Fig. 4.14, which displays the mass of hose components for the 10 pump case. As with the single-pump case (shown in Fig. 4.3) the mass of the $H_2S$ fluid climbs with bore size, while that of the hose itself drops rapidly and then flattens out for diameters from 3 to 4 cm. While these trends are similar, the details reveal major differences between single-pump and multi-pump designs. First, the total hardware weight drops from 10 tons in the single-pump case to under 8 tons with 10 pumps. Secondly, the mass of a single-pump hose was split between the hose itself (7 tons) and the tether (3 tons). With 10 pumps, the mass of these two components is cut four-fold, to 2.5 tons. Instead, most of the hose mass is devoted to two new components, the pumps themselves and the cables bringing electricity to them.

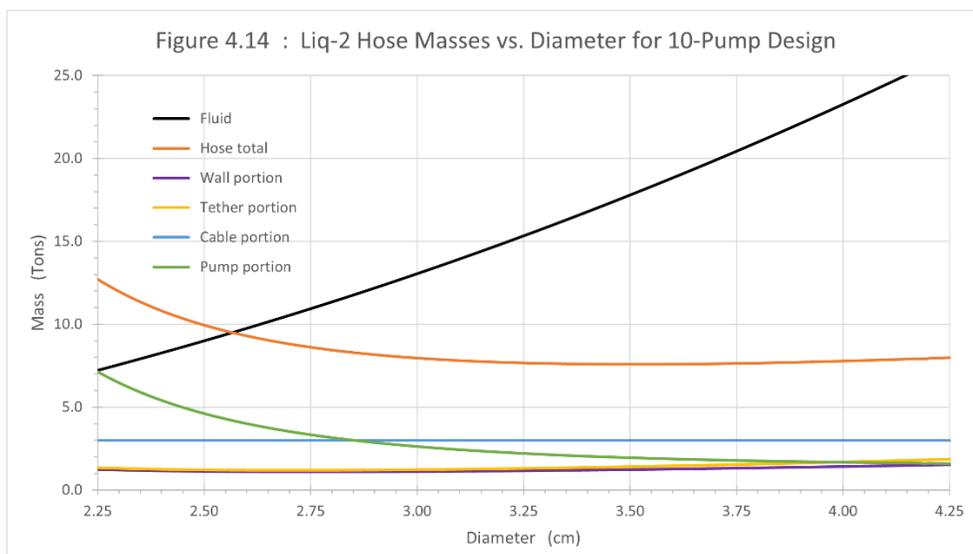

While the hardware mass of a ten-pump hose does not increase rapidly with bore diameter, the number of balloons required does. The reason for this, comes from the increase in the mass of the fluid, and the fact that its weight must be supported by the on-hose pumps, and ultimately by the balloons.

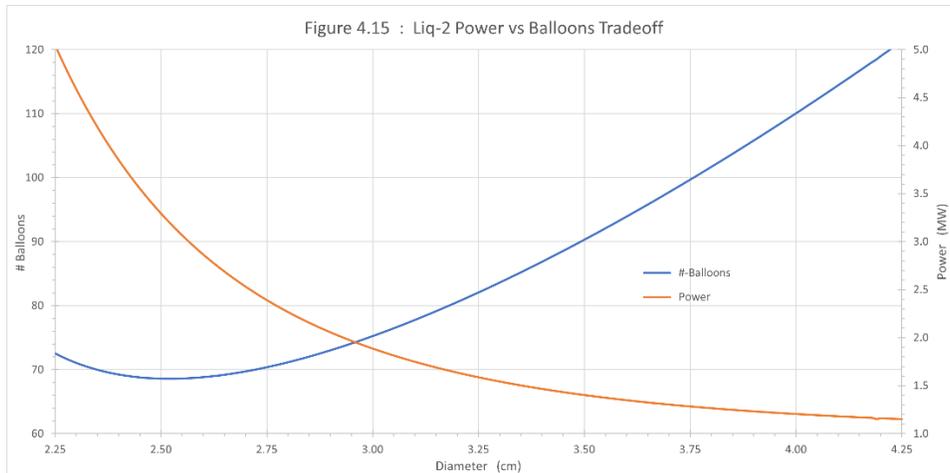

The decision on which bore size is best depends on priorities, electrical power versus number of balloons. In Fig. 4.15, we plot the variation of both with bore size. The optimum is likely between 2.9 and 3.3 cm; for present purposes let's settle on a bore size of 3.1 cm, the same value adopted for the single-pump design.

Using this value, let's look at the effect of cable area on the number of required balloons, and on the electrical power. Fatter cables are heavier, so will increase the number of balloons, but the larger conductor area will reduce resistive losses and hence the electrical power. The tradeoff between these two factors is shown in Fig. 4.16.

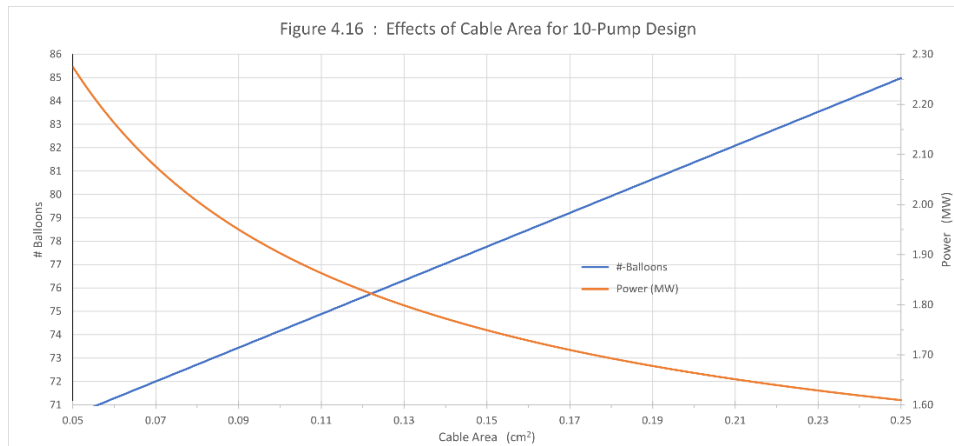

### 4.6 : Representative Designs

Let's present two representative hose designs; Liq-1 has a single pump mounted on the ground, and Liq-2 has ten pumps, nine of which are mounted on the hose itself.

| | Table 4.2 : Two Hose Designs | | |
|---|---|---|---|
| | Liq-1 | Liq-2 | Units |
| # Pumps | 1 | 10 | |
| Bore diameter | 3.10 | 3.10 | cm |
| # Sectors | 10 | 10 | Per pump |
| Peak pressure | 3.25 | 0.349 | Bars |
| # Balloons | 107 | 79 | |
| Mass - Fluid | 14.85 | 13.95 | Tons |
| Mass - Hardware | 9.74 | 7.81 | Tons |
|    Hose portion | 6.77 | 1.13 | Tons |
|    Tether portion | 2.97 | 1.24 | Tons |
|    Cable portion | | 3.00 | Tons |
|    Pump portion | | 2.44 | Tons |
| Electrical power | 1.290 | 1.749 | MW |
| Tension at base | 306.3 | 32.2 | kN |
| Peak excess-T | 12.07 | 15.22 | ºC |
| Average excess-T | 8.18 | 10.53 | ºC |

The two designs use the same bore diameters, which in both cases are larger than the values which would minimize the number of required balloons, trading the extra buoyancy needed, in exchange for lowering the power requirements, and the excess temperatures experienced. The most significant differences between them are probably qualitative; design Liq-1 keeps all its pumps on the ground, enhancing maintenance and reliability, but imposes almost ten times more pressure loads on its hoses. The most significant quantitative differences, are that design Liq-1 requires 35% more balloons, and almost ten times higher base tension, while design Liq-2 demands 42% more electrical power to operate.

Let's further illustrate these two designs, and the differences between them, by plotting pressure, tension, and temperature along the length of the hoses.

The pressure profiles are shown in Fig. 4.17. While very different, each drops with height at the rate required to support the weight of the liquid $H_2S$. The slopes are nearly constant, simply because the density of the liquid $H_2S$ is. For design Liq-1, since there is just a single pump at the ground, the pressure gradient must be maintained over the full length of the hose, hence the peak pressure is very large. The advantage that design Liq-2 gains by having multiple pumps, is that the pressure resets at each pump, so never gets nearly as large.

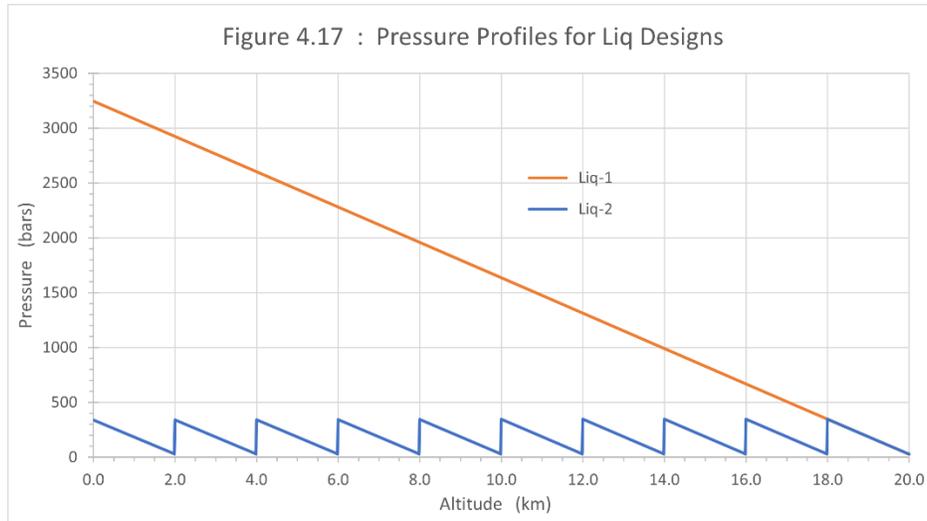

The tension profiles are displayed in Fig. 4.18. These require a bit more explanation than did the pressure.

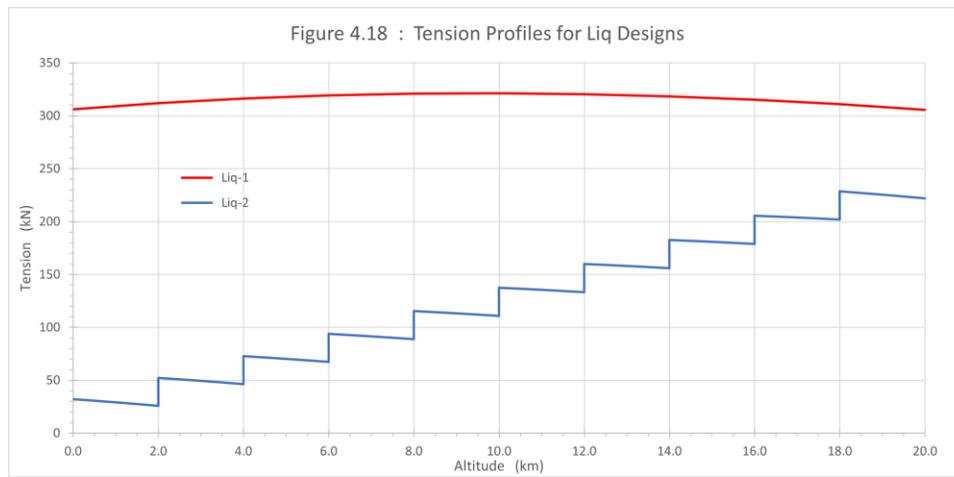

The tension in design Liq-1 starts at a very large value and remains fairly constant along the length of the hose, rising slightly for the lower portion and actually dropping towards the top of the hose. The large tension at the base is, as discussed previously, a consequence of the very large base pressure of a single-pump design. This generates a large compressive force along the hose which, by itself, would subject the hose to buckling; the large base tension is required in order to stabilize the hose against such deflections. The fact that this tension force doesn't increase much along the hose, is due to the fact that the weight of the hose itself (which would normally lead to such an increase) is almost completely balanced by the upward force from the $H_2S$'s flow friction. In the lower portions of the hose, where pressure is highest, the hose material is slightly heavier, requiring tension to increase. In contrast, the upper portions experience lower pressure, so can have thinner hose walls; here the frictional force dominates, causing tension levels to drop slightly.

The behavior of tension in design Liq-2 also requires explanation. It starts at a low base value, simply because the base pressure is much lower, and requires less tension to stabilize against buckling. The tension increases moving up the hose, but this occurs primarily by discrete jumps at each pump; in between pumps, tension actually drops slightly due to upward force from flow friction. The large jumps at each pump are due to two forces, weight of the pumps themselves, and that from the reaction force acting on the pump, as it supports the overlying $H_2S$.

Overall, the fact that the tension at the top of the Liq-1 hose is greater than that for design Liq-2, is why design Liq-1 requires more balloons; buoyancy from these balloons is what supplies this tension.

Temperature profiles for both designs, as well as that of the surrounding atmosphere, are displayed in Fig. 4.19. Atmospheric temperature declines throughout the troposphere, and then becomes basically constant until the top of the hose, at 20 km. Qualitatively, the temperature of the $H_2S$ fluid, in both designs, tracks this atmospheric profile. But, neither design is actually in thermal equilibrium with the atmosphere; both are consistently warmer. This excess temperature is, as discussed before, primarily due to the frictional heating resulting from the $H_2S$ being forced through the hose. Hose designs with smaller bores will have a greater thermal mismatch, while those with larger bores will be closer to equilibrium. One beneficial consequence for design Liq-1, of this extra warmth near the 10-12 km location, is less risk of the $H_2S$ solidifying. The temperature in design Liq-2 behaves similarly to that for design Liq-1, except it's always a bit warmer, and spikes at each pump. Both of these effects are due to the extra heat deposited in the flow by the pumps.

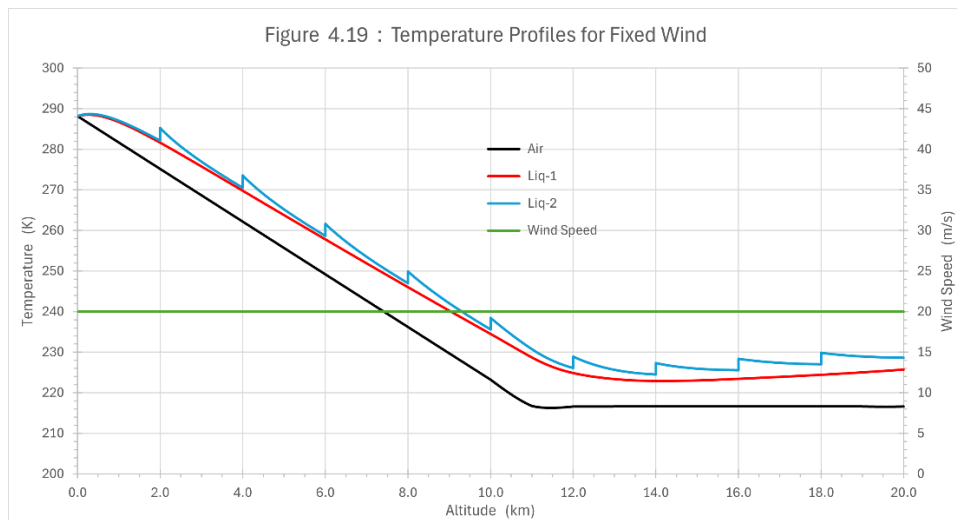

The temperature profiles shown in Fig. 4.19, were calculated using a fixed wind speed of 20 m/s. This profile has the advantage of being an unambiguous, clearcut, choice for the wind profile; but, of course, is not representative of real atmospheric wind profiles. In Fig. 4.20, we recalculate the temperature profiles of both designs using a more realistic wind profile. Here, the wind speed is relatively slow at low and high altitudes, but reaches much higher speeds in between. The $H_2S$ fluid temperatures are qualitatively similar to those found for the simple, fixed wind speed calculations; the primary difference is that the $H_2S$ sees lower warming in the middle portions of the hose, as a result of strong convection due to high winds within the jet stream.

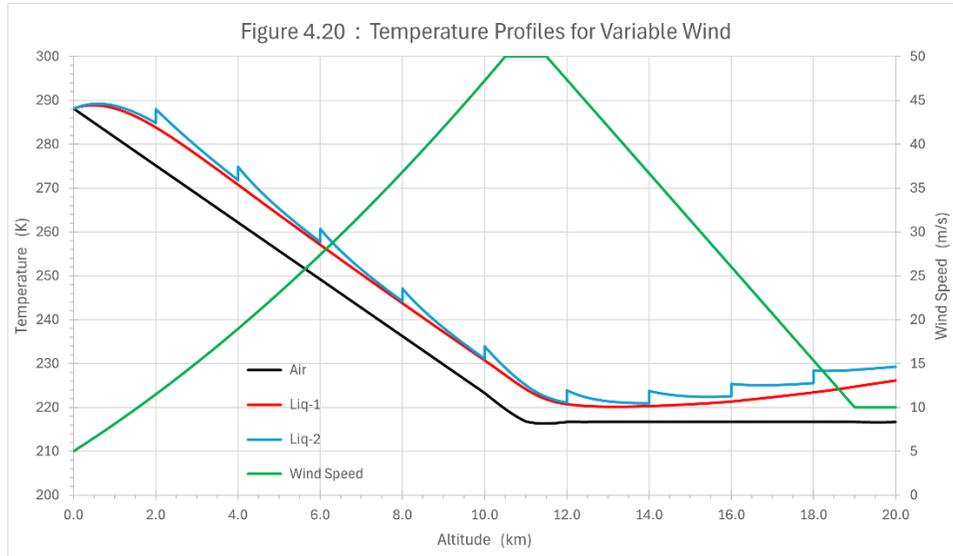

### 4.7 : Placing Balloons Along the Hose

Since the buoyancy of a balloon depends on the weight of air which it displaces, the lower in the atmosphere that balloons are placed, the greater their lifting power will be. Accordingly, it is more efficient to support the hose by placing a series of balloons along its length, than by locating them all at the top.

For simplicity, rather than trying to optimize where balloons should be located, we'll simply space them evenly along the hose. At each location, the number and size of the balloons are determined so as to support as much of the underlying weight as possible, while still retaining enough tension to stabilize the hose against pressure-induced buckling.

For the Liq-1 design, using mid-hose balloons cuts the number needed from 108 to 41. The reduction isn't quite as large for Liq-2, but is still more than a factor of 2, from 78 to 34.

| Table 4.3 : Support via Mid-Hose Balloons | | | | |
|---|---|---|---|---|
| | Liq-1 | | Liq-2 | |
| Altitude (km) | Number | Diameter (m) | Number | Diameter (m) |
| 4 | 3 | 18.7 | 2 | 17.8 |
| 8 | 4 | 19.1 | 3 | 18.2 |
| 12 | 6 | 19.3 | 4 | 20.0 |
| 16 | 10 | 19.5 | 8 | 19.8 |
| 20 | 18 | 19.7 | 17 | 19.6 |
| All Balloons at Top | | | | |
| 20 | 108 | 20.0 | 78 | 20.0 |

pg. 43

Clearly, using mid-hose balloons allows major reductions in the number required. The concern, as we'll see in Section 5.7, is that mid-hose balloons suffer more from wind, than do stratospheric ones.

# 5 : Wind

After being deployed, the hose and its supporting balloons will be pushed sideways by wind. This, unfortunately, is not a minor effect. The sideways forces are strong, and unless properly dealt with, can push the hose completely over, making the hose approach to geoengineering unworkable.

## 5.1 : The Wind Environment

Because the hose reaches up to the stratosphere, it is the high altitude winds which most affect it, not those near the surface. While these winds are steadier and more predictable than surface winds, they are by no means constant. The winds vary depending on site (e.g., latitude), with time (both seasonally and diurnally), and as a function of altitude. Just as crucially, they are not predictable, previous trends are merely guides to current and future behavior. While we can determine what the wind is most-likely to be at a given location and time, that's only an average value. Wind models often express results statistically, i.e., in a percentage likelihood that wind speed will be within a given range. This obviously complicates matters for the hose; the system can be designed for specified typical wind conditions, but must also be able to operate under more severe outlier situations.

For the hose, the two most important wind values are that at its top, i.e., 20 km altitude, and the jet stream value, which occurs near the midpoint of the hose and is generally the peak value the hose will experience. For mid-latitude locations, data from NASA's Horizontal Wind Model HWM14[43] indicates that typical peak values for these are about 20 m/s and 50 m/s respectively. Given the large variations to be expected in real-world wind conditions, we'll adopt a single, relatively simple W(z) profile (shown in Fig. 5.1) for design purposes, which corresponds to these two values. Then, to examine the hose's performance in greater or less severe wind conditions, we'll simply scale the entire magnitude of this wind profile accordingly.

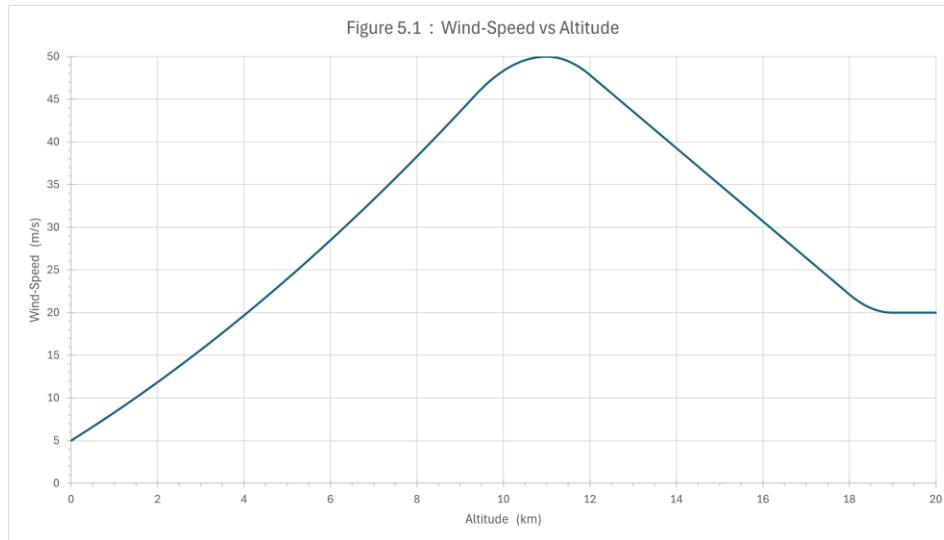

Figure 5.1 : Wind-Speed vs Altitude

## 5.2 : Behavior of a Tilted Hose

Wind forces push the hose and balloons sideways, deflecting them and tilting the hose away from vertical. In treating this behavior, we'll use a simplified model; assuming that motion lies in a 2-D plane, and is in static equilibrium, ignoring wind gusts, hose oscillations or inertial effects. This decision was made for simplicity, and because high-altitude winds are oriented predominantly in a single, east-west, direction, and vary relatively slowly in time. But, nevertheless, this model captures the primary role of steady-state, in-plane, tilt, in determining the response of the hose to forces from the wind.

The behavior of a tilted hose will be modeled sector by sector, treating discrete forces at the junction of two hose sectors, and continuous forces acting on the hose within each sector.

A hose sector which is tilted acts differently from a vertical one, even if there is no wind pushing on it. As shown in Fig. 5.2, there are 4 natural directions, two global ones (**x** sideways, and **z** vertical) and two local ones (**m** along the hose, and **n** perpendicular to it). Tension within the tether as well as flow-friction, act along the hose, while hose weights (of the fluid, the hose walls, the cables, and the tether) act vertically. The wind approaches sideways, along **x**, but there are two somewhat different ways to model the forces that it causes. In one model[44], the full wind velocity applies drag forces to the hose; this force acts globally along **x**, so splits locally into components along and perpendicular to the hose. In another model[45], only the perpendicular component of the wind applies drag to the hose, applying a purely perpendicular force. Both models generate the same perpendicular force, and hence induce the same local curvature, but the longitudinal force contribution in the first model tends to push upward on the hose, acting as does internal flow friction. As long as the hose tilts are kept small, the difference in these two approaches is unimportant.

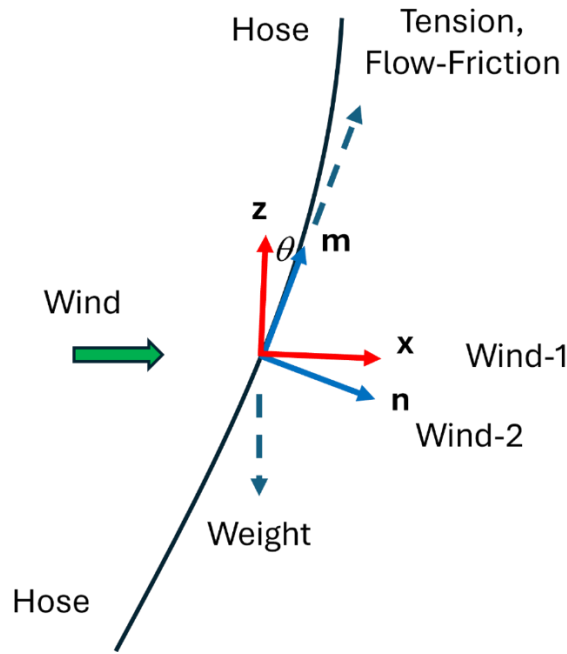

Figure 5.2 : Wind Forces on the Hose

The first of the two wind force models is:

$$dF_{w1} = \frac{1}{2}\rho_a W^2 C_D (D_{out} dz)\mathbf{x} \qquad (5.1)$$

When expanded into the local components, and using length ds along the hose, we get:

$$dF_{w1} = \frac{1}{2}\rho_a W^2 C_D (D_{out} ds)\cos\theta(\mathbf{n}\cos\theta + \mathbf{m}\sin\theta) \qquad (5.2)$$

The second of the two wind force models is:

$$dF_{w2} = \frac{1}{2}\rho_a (W\cos\theta)^2 C_D (D_{out} ds)\mathbf{n} \qquad (5.3)$$

As noted above, the **n** component of the two models is the same.

Once wind forces begin to tilt the hose, the fact that weight is no longer aligned with the hose accentuates the tilt. There are two components of weight, that of the hose hardware and of the fluid within it, and these act somewhat differently. The hardware (the hose itself, the tether, and any electrical cables) has a mass per length $m$; all of the resultant weight acts vertically, one component increasing tilt, and the other the tension. The effect of the fluid weight differs; the component perpendicular to the hose is unreacted and increases the hose's tilt, but the portion along the hose is supported by fluid pressure, not tension.

By expressing the forces in longitudinal and perpendicular components, we find equations for their effect on tension and tilt respectively:

$$\frac{dT}{ds} = gm\cos\theta + \left(\frac{\pi}{4}D^2\right)P'_F - \frac{1}{2}\rho_a W^2 C_D D_{out}\sin\theta\cos\theta \qquad (5.4)$$

Here, the last term is zero when using wind model 2, and the gradient of the frictional pressure, $P'_F$, is negative, thereby reducing tension. The tilt equation is:

$$-\left[T - \left(\frac{\pi}{4}D^2\right)(P - P_{air})\right]\frac{d\theta}{ds} = gm\sin\theta + \left(\frac{\pi}{4}D^2\right)g\rho_F\sin\theta + \frac{1}{2}\rho_a W^2 C_D D_{out}\cos^2\theta \qquad (5.5)$$

Here, the compressive force due to the fluid's net pressure, counteracts the effect of tension on the hose's stiffness; it's the net tension that stiffens the hose against deflections. Taking into account air pressure is, of course, inconsequential for hoses using liquid $H_2S$, but will matter in Chapter 6 when we consider gaseous $H_2S$.

Finally, the portion of fluid pressure representing the gravity-head is given by:

$$\frac{dP_g}{ds} = -g\rho_F\cos\theta \qquad (5.6)$$

The above equations describe the gradual changes in tension and tilt within a hose sector. But, at the junction of two sectors, tilt and tension can change abruptly. There are two types of junctions between sectors. In the simplest case, the transition is smooth, so that while the two sectors may have different properties, such as hose thickness or tether area, there are no objects at the joint; hence, both the tilt and the tension are continuous. But, the junction may also host discrete objects, such as pumps or balloons. This case is illustrated in Fig. 5.3, and results in abrupt changes in the hose's tilt and tension.

Figure 5.3 : Force Balance at Hose Joint

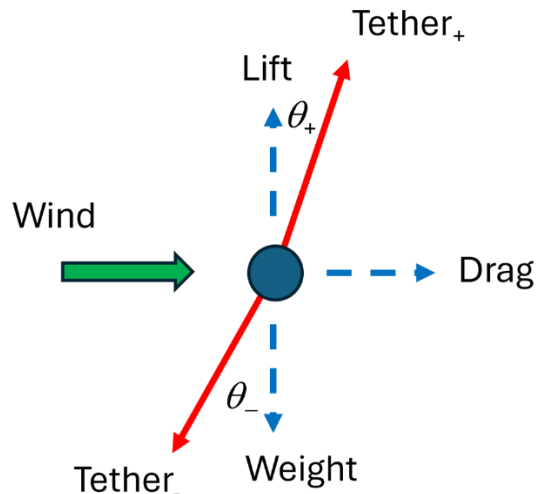

The terms involved in this force balance are clear. The weight is due to pumps and parasitic parts of balloon masses (walls and internal gas). Drag is due to aerodynamic drag from wind acting on balloon area. The lift comes from balloon buoyancy, as well as any aerodynamic lift provided by the wind.

$$\text{Aero}_{L,D} = \frac{1}{2} C_{L,D} \rho_a A_B W^2 \qquad (5.7)$$

A key factor here is $A_B$, the area presented to the wind by the balloons. This directly drives the aerodynamic forces, and increases with the buoyancy demanded from the balloons, i.e., with the tension carried by the tether. The details of this increase depend, however, on the configuration and design of the balloons. For the case we've been treating, where buoyancy is provided by $N_B$ separate balloons of diameter $D_B$, then

$$A_B = N_B \left(\frac{\pi}{4} D_B^2\right) \qquad (5.8)$$

In practice, however, the effective area of a group of many balloons will be significantly reduced by masking, as some balloons block wind exposure to others.

$$A_B = N_B^{2/3} \left(\frac{\pi}{4} D_B^2\right) \qquad (5.9)$$

This recovers the $A \sim V^{2/3}$ scaling presented by a single large balloon. Another way to greatly reduce the aerodynamic area of the balloons, is to use elongated shapes, rather than the quasi-spherical ones assumed above; these will be discussed in Section 5.4.

The tether forces are net values:

$$\text{Tether} = T - \left(\frac{\pi}{4} D^2\right)(P - P_{air}) \qquad (5.10)$$

Note, that the pressure portion of this force encompasses reaction loads on any pumps present. If there is a change in tilt at the junction, these pressure terms will produce a destabilizing lateral force, but this is not as strong as the stabilizing force due to the tension terms.

The buoyancy of the balloons (i.e., the lift term in Fig. 5.3) depends on the mass of the displaced air, so might be expected to change if the hose tilts, simply because the altitude of the balloon is now lower than when the hose was vertical, causing the density of the surrounding atmosphere to be higher. In an isobaric sealed balloon, this change does not actually happen; the balloon's volume shrinks if it drops to a lower altitude, and this balances the increased air density; the weight of displaced air (and hence the buoyant force) remains unchanged.

However, the wind forces on the balloon, given by Equation 5.7, will change with its altitude. The obvious reasons are that both the air density, $\rho_a$, and the wind speed, $W$, change. But, two other changes occur, one good (reducing drag) and the other bad (increasing it). As just noted, balloons shrink as they decrease in altitude; while this doesn't change their physical wall area, it will decrease $A_B$, the area directly exposed to the wind. A more significant change is that the drag coefficient, $C_D$, will almost certainly increase, as the balloon shrinks and its shape becomes less smooth. For spherical balloons, this may not matter much (to be blunt, $C_D$ is already about as bad as possible). But, when using streamlined balloons, as introduced in Section 5.4, $C_D$ is designed to be low; here, if the surface becomes wrinkly or irregular, $C_D$ will probably increase quite a bit.

The most effective way to prevent this, is to keep the balloon's shape taut and constant, even if it dips to a lower altitude as the hose tilts. This will require that the balloon be over-pressured enough to withstand the higher atmospheric pressures it will encounter at lower altitudes. This, in turn, demands a stronger, and hence heavier, balloon wall than needed at the balloon's intended altitude; which, of course, increases the incentive for us to keep the hose tilts small. In the deflection calculations that follow, we do not model changes in $C_D$ resulting from hose tilts. This neglect will have little effect as long as we can keep the tilts low (which is our goal), but means that the results with large tilts will actually be even worse than what will be presented below.

**5.3 : Effect of Wind on the Hose**

Using the above model, we can examine the effect of wind on the hose. Since the actual wind profile is unknown and ever-changing, we'll illustrate the behavior via a typical $W(z)$ variation in altitude, shown in Fig. 5.1, and examine different magnitudes by scaling the entire profile to match specified maximum velocities.

The shapes of the tilted hose (for different maximum wind speeds) are displayed in Fig. 5.4a for the Liq-1 design and in Fig. 5.4b for Liq-2. Both sets of calculations assumed that all the balloons are exposed to the wind (i.e., $A_B \sim N_B$), and have $C_D$ of 0.5 and zero $C_L$, (corresponding to spherical balloons) while the hose's $C_D$ is 1.0 (reflecting a circular cross-section).

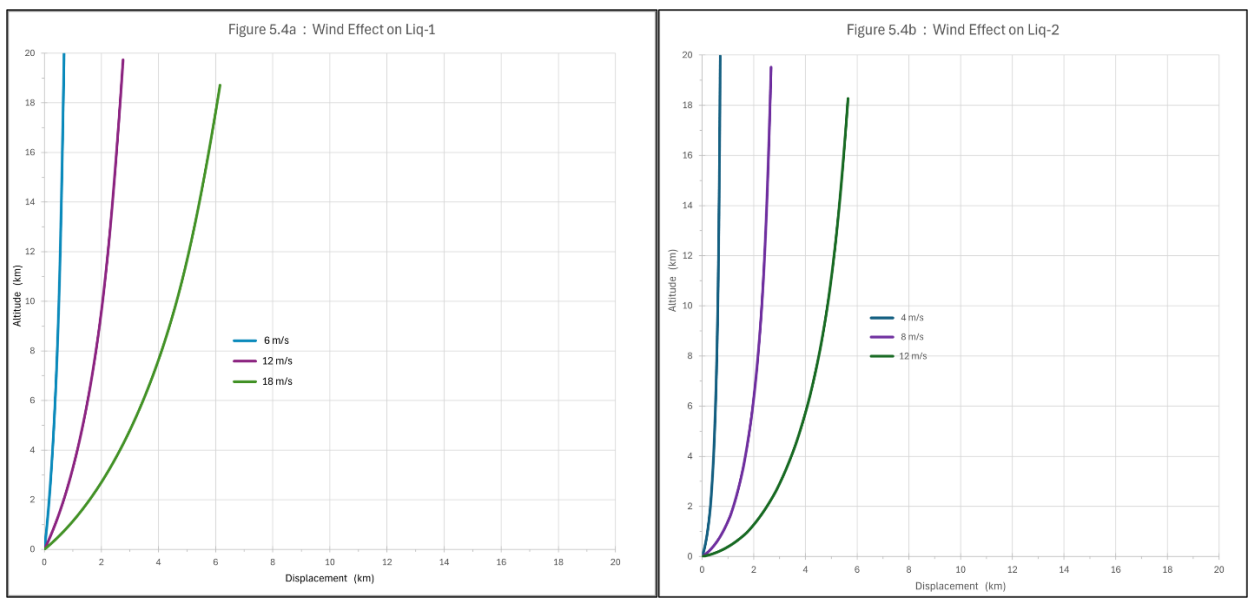

These calculations show that even modest winds, much weaker than the 50 m/s values frequently observed, cause large deflections of the hose. In each case, higher velocities, i.e., 19 m/s for Liq-1 and 13 m/s for Liq-2, can't be survived at all; the hoses collapse.

Note, that the Liq-1 design, the one with no on-hose pumps, can tolerate larger winds than Liq-2. This is because the higher tension values of this hose (required to stabilize it against the compression forces from its high internal pressures) make the hose stiffer. So, the lower tension in Liq-2, while beneficial in requiring fewer balloons for support, does come at a price; it's much more sensitive to wind-based deflections.

Nonetheless, while Liq-1 can tolerate stronger winds than Liq-2, neither design is remotely capable of operating in the 50 m/s winds they can be expected to frequently encounter. In such winds, forces on the Liq-1 design will be almost 8 times as high (and 17-fold for Liq-2) than what they can survive.

### 5.4 : Streamlining the Hose

Wind forces come from two places, the balloons at the top of the hose (as well as any positioned along the hose) and wind pushing on the hose itself.

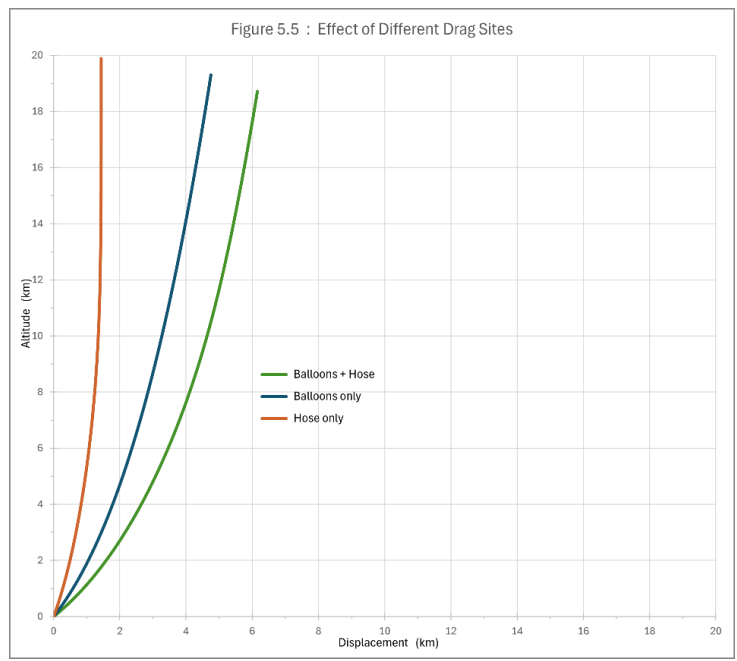

Their relative effects can be appreciated using Fig. 5.5. This shows the shapes of a Liq-1 hose in a wind with peak speed of 18 m/s, where forces act only on indicated portions of the system. The best case is where only the hose experiences forces; while the hose is curved, it has no tilt at the top, since there is no force on the balloons there. When the hose is not pushed, but the balloons are, the hose is tilted at the top in order to resist the balloon force; this tilt angle grows downwards towards the base. When the wind acts on both the hose and balloons, the hose is more strongly deflected, and near collapse.

While wind forces on the balloons are stronger than those on the hose, both cause deflections, and must be greatly reduced in order to field a survivable hose system.

The most effective way to obtain strong reductions in wind forces, is to streamline the shapes of both the balloons and the hose.

### *Reducing Balloon Drag*

The key to reducing balloon drag is to elongate them, adopting airship-type shapes. This has two advantages. The first, is that lengthening a balloon effectively masks most of its volume, achieving the same type of effect as stacking balloons behind each other. We therefore achieve the same buoyancy for much less wind-exposed area. The second, is that spheres are a remarkably high-drag shape, for the same frontal area, an elongated balloon will experience much lower drag forces.

There have been a number of papers discussing how to optimize the shape of an airship in order to minimize drag.[46][47][48] However, in the interest of fabricational simplicity, we'll focus on simple shapes, cylindrical balloons with ellipsoidal ends. The drag force on balloons with this type of shape was recently analyzed[49], and expressed in terms of a fineness-ratio dependent form-factor. Once the fineness ratio, $f$, (i.e., the length/diameter) of a body-of-revolution gets larger than about 4, then the drag forces on it are primarily due to skin-friction, and scale with the wetted surface area, not the frontal area. With increasing fineness, the shape-dependent pressure drag becomes ever less important. The total drag force can then be expressed as a multiplicative form-factor, times the drag from skin-friction.

$$F_D = FF(f) C_f S_{wet} \tag{5.11}$$

We'll use a skin-friction coefficient, $C_f$, from Eq. 26 in Chapter 2 of Hoerner[45], where $Re_L$ is a length-based Reynolds number.

$$C_f = 1/\{3.46 \log_{10}(Re_L) - 5.6\}^2 \tag{5.12}$$

For rotationally symmetric shapes, there are empirical expressions for the form-factor. We'll show two below, the first from Eq. 28 in Chapter 6 of Hoerner[45], and the second from Eqs. 12-15 of Götten[49]:

$$FF = 1 + 1.5 f^{-1.5} + 7 f^{-3} \tag{5.13}$$

$$FF = 1.09644 + 3.1742 f^{-2.6202} \tag{5.14}$$

While Götten's form does not properly asymptote to 1, we don't envision using large enough fineness ratios where this becomes a factor. In the regions of most interest to us, $f$~ 5-8, both expressions give similar values; we'll use Hoerner for historical reasons.

The shapes used by Götten had a 2:1 ellipsoid front, a 3:1 ellipsoid back, and a cylindrical middle section, so can be used for fineness ratios greater than 2.5. Accordingly, their surface area and volume are given by

$$S_{wet} = [3.4184 + 4(f - 2.5) + 4.9169]\pi R^2 = (f - 0.4162)\pi D^2 \quad (5.15)$$

$$Vol = \frac{2\pi}{3}R^3(2+3) + 2\pi R^3(f - 2.5) = \left(f - \frac{5}{6}\right)\frac{\pi D^3}{4} \quad (5.16)$$

To illustrate the reduction in drag possible with elongated balloons, compare the drag on the Liq-1 design between that with spherical balloons and that with elongated ones. For a 20 meter spherical balloon, the buoyancy is 372.4 kg which is reduced to 290.8 kg when accounting for the internal He and the material walls; supporting the hose requires 108 such balloons. Contrast this with a 5:1 elongated balloon, i.e., 20 meter diameter and 100 meter length. Each of these has more volume and surface area than the spherical balloons, so has a buoyancy of 2327 kg, which is reduced to 1868 kg when accounting for the weight of the gas and walls. Because of this higher lifting capacity, only 18 balloons are required. The reduction in the number of balloons shrinks the total frontal area accordingly, from 33930 m² to 5341 m². This sixfold gain (essentially from masking), is greatly magnified, when we consider the reduction in drag-coefficient due to the more streamlined shape of the elongated balloons. For a typical stratospheric wind speed of 10 m/s and a 100 meter balloon length, the Reynolds number will be 5,000,000, resulting in a skin-friction coefficient of 0.0032. At a 5:1 fineness ratio, the Hoerner friction-factor is 1.19, so in terms of the frontal-area, the drag coefficient is 0.071, another 7-fold improvement over spherical balloons. Overall, this switch to elongated balloons, reduces drag forces to only 2.2% of that using spherical balloons, a massive improvement.

Also, of course, the switch from 108 balloons to just 18 ones is a large practical advantage. This must be balanced against any increased fabricational and qualification difficulties, involved in lengthening the balloons from 20 meters to 100 meters. However, since the diameters remain the same, the balloons are still small enough to fit within many buildings, and the cylindrical midsection should be straightforward to incorporate; these challenges are well worth the payoff.

*Reducing Hose Drag*

It's clear from Fig. 5.5, that even without balloon drag, the wind forces on the hose itself are too strong to tolerate, and must also be greatly reduced. The solution is the same as for the balloons; the circular shape of the hose must be streamlined.

Unless the hose is strongly tilted (which, of course, the current drag reductions are meant to prevent) the wind flow is basically perpendicular to it and well described by 2D airfoil theory.

Airfoils have been designed and studied for many decades, and elongated teardrop-shaped ones feature much lower drag coefficients than circular cross-sections. Fig. 5.6 compares the shapes of two well characterized NACA airfoils to that of a circular hose.

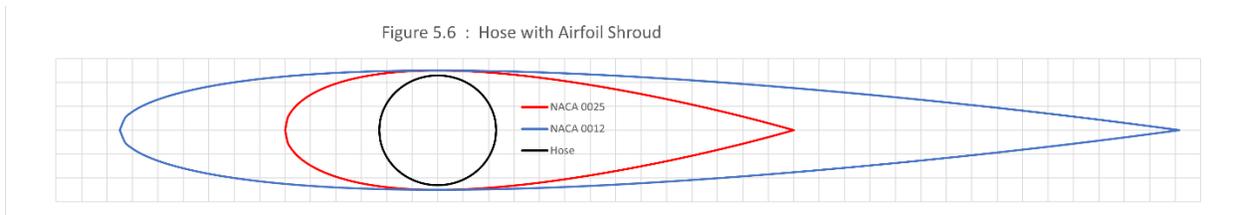

Figure 5.6 : Hose with Airfoil Shroud

Rather than trying to modify the shape of the hose itself to match that of such an airfoil, we'll adopt a simpler option, enclosing the hose within a hollow thin-walled shroud having the desired shape. The shroud will be free to rotate around the hose to align with the wind, and longitudinally segmented to adapt to variations in wind direction. As discussed elsewhere[28], the shroud can be kept passively aligned with the wind, by designing it with its center-of-mass in front of its center-of-pressure.

The drag coefficient of the shrouded hose, will depend on both its shape and the wind speed; the coefficient is lower, the more elongated the shape and the bigger the wind's Reynolds number. These trends are illustrated in Fig. 5.7 for data[50] calculated with the JavaFoil code. The NACA-0012 shape does have lower drag than does the NACA-0025 profile, but this benefit may not be worth the extra size; we'll adopt the NACA-0025 design.

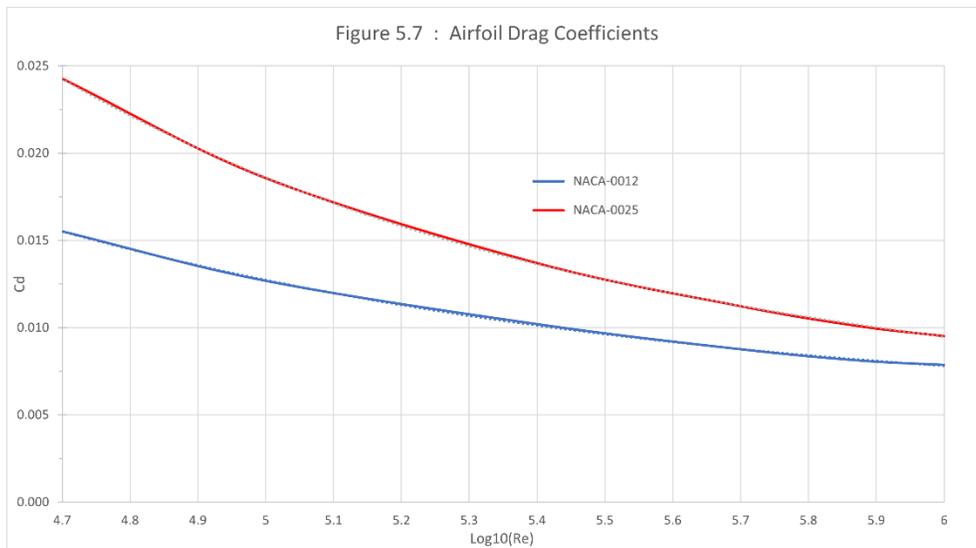

Figure 5.7 : Airfoil Drag Coefficients

The large drag reductions achievable by shrouding the hose, are accompanied by some costs. The shroud will slightly increase the frontal area exposed to the wind, and its weight will increase that of the hose. When calculating the performance of a shrouded hose, we'll assume a 0.02 gm/cc aerogel core with a 0.1 mm thick skin, locally sized to fit around the hose, and responding to the chord-based Reynolds number of the local wind. This has an areal mass of 0.8 gm/cm, i.e., a total mass of 1.6 tons. Given the low thermal conductivity of silica aerogel, its choice will greatly reduce atmospheric cooling of the hose; the H₂S will stay warm as it flows up to the stratosphere. This decoupling can be lessened, either by addition of high conductivity fillers, or by using a different aerogel material than silica.

Streamlining the balloons and hose, is very effective in reducing the deflection of the hose, and allowing it to operate during severe winds. This is illustrated for the Liq-1 design in Fig. 5.8a, and for Liq-2 in Fig. 5.8b.

Both hose designs operate comfortably in wind profiles featuring 50 m/s in the jet stream and 20 m/s at 20 km altitude; non-streamlined versions failed at less than 10% these dynamic pressures. In some circumstances, the hose may face even stronger winds; both designs can survive 50% faster winds (over double the dynamic pressure).

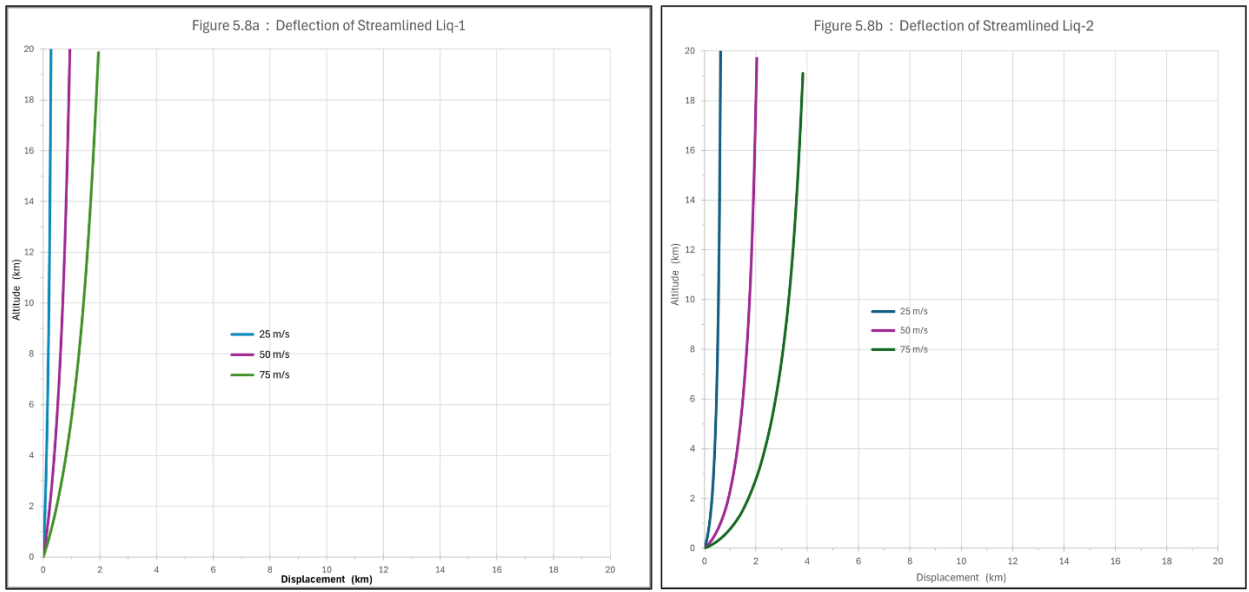

## 5.5 : Stiffening the Hose by Increasing Tension

While streamlining enables the Liq-2 hose to survive strong winds, it does not do so as robustly as Liq-1. This is a direct consequence to the fact, that its hose is under much lower amounts of tension, and thus deflects more under wind loading, than the high-tension Liq-1 hose.

A natural way to increase Liq-2's robustness, is to stiffen it, by artificially increasing its tension beyond the levels necessary to support its weight. This is achievable, simply by adding extra balloons at the top of the hose; these increase tension by pulling up more on the hose. The drawback, of course, is that adding extra balloons shrinks the advantage that Liq-2 had over Liq-1 in that regard; drag on them also increases the wind force on the top of the hose. However, the sequence of hose simulations displayed in Fig. 5.9, shows that significant extra stiffness is possible for the addition of only 1 or 2 extra balloons. But, improving Liq-2 to fully match the wind performance of Liq-1, will require just about the same number of balloons employed by it.

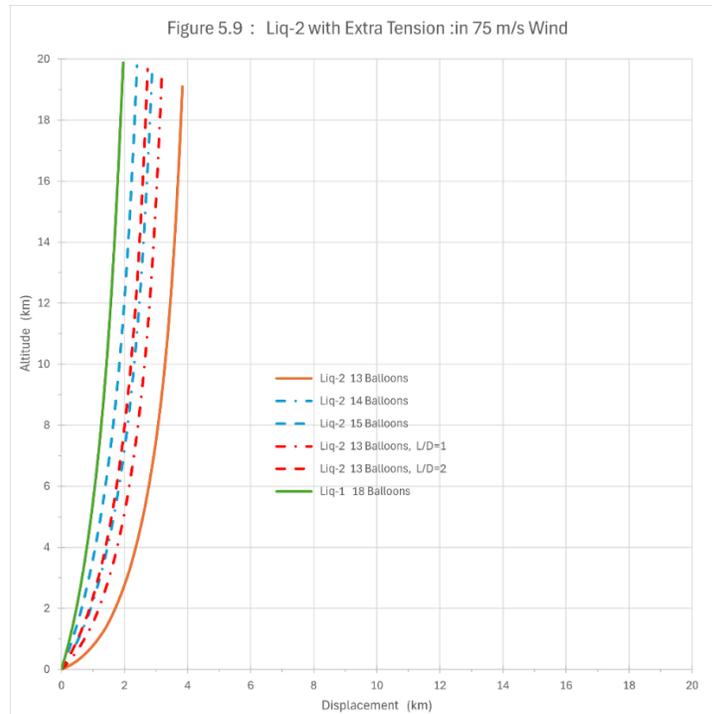

Figure 5.9 : Liq-2 with Extra Tension :in 75 m/s Wind

There is another way of increasing hose tension, which does not require adding any extra balloons, and which applies ever more tension the stronger the wind becomes. This approach, also illustrated in Fig. 5.9, is to shape the balloons so that they provide aerodynamic lift, not just drag.

In principle, such balloons could derive much of their overall lift capability from wind-based aerodynamic lift rather than buoyancy, allowing the hose to use fewer and/or smaller balloons to support itself. While potentially attractive, the large time-dependent variations in winds (and hence the upward force) would greatly complicate design and operation of the hose system. Active control of the balloon dynamics would probably be necessary in order to keep the uplift applied to the hose steady, as well as large enough to always support it. To avoid committing to the time and expense involved in developing such a capability, we'll instead consider a much more conservative use of aerodynamic lift. Here, we'll continue to size the balloons so that buoyancy is large enough to fully support the hose. Aerodynamic lift is then used purely to increase hose tension above what's necessary to support the hose, providing extra stiffness which increases and decreases with the wind's dynamic pressure, hence in direct proportion to the lateral forces acting on the hose. The one structural change the hose must make to accommodate this option, is to increase the tether's area to be able to handle this additional tension. The performance of this approach is displayed in Fig. 5.9.

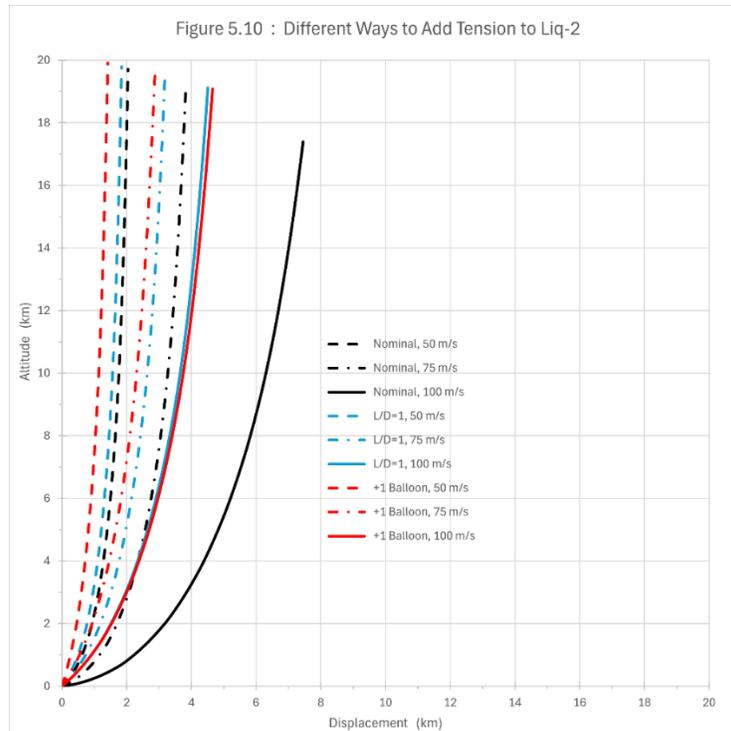

Figure 5.10 : Different Ways to Add Tension to Liq-2

In Fig. 5.10, we display the ability of the hose to handle very high winds; focusing on Liq-2, as it is most affected by wind. In these simulations, jet stream winds vary from 50 to 100 m/s, while the corresponding stratospheric winds are 20 to 40 m/s. The nominal low-drag hose, using streamlined balloons and hose-shrouds, does well in winds up to 75 m/s, but is just about to fail at 100 m/s. To handle such intense winds, artificially increasing tension is essential. Both approaches, adding extra balloons, or using aerodynamic lift, are equally effective in 100 m/s winds, although lift is relatively less helpful in lighter winds.

Comparing the two approaches, simply providing 1 or 2 extra balloons seems to be a less complicated, and more reliable, option than having to design, field, and operate lifting-body balloons.

**5.6 : Surviving Winds by Emptying the Hose**

Occasionally, very severe winds do happen. While winds in the jet stream and lower stratosphere are generally steadier, and more predictable, than those in the troposphere and near the surface, intense variations do sometimes happen. It is hard to place guaranteed upper limits on the winds that the hose will face.

This places a premium on designing as wind-resistant a hose as possible; insuring that the hose can handle, not just the most severe winds already documented at its site, but also providing extra margin for outlying events.

We've discussed methods which should be sufficient to do so. Streamlining both the balloons and the hose, is essential just to handle predictable winds, ones which the hose will be expected to face over the course of a year. But, for stronger winds, this may not be good enough; the hose can be made stiffer and more robust by increasing tension. This "wind insurance", however, comes at a cost, requiring the addition of extra balloons, and potentially foregoing the balloon reductions possible by distributing pumps or balloons along the hose. Furthermore, of course, we have to accurately predict the upper wind limit the hose will face, and design and build accordingly. If more severe conditions do happen, the hose will fail, if they never do, then we may have overbuilt or forewent more efficient designs.

There is, however, another way to survive the occasional occurrence of intense winds; empty the $H_2S$ from the hose, ride out the event, then refill and resume operations once the severe winds recede.

Operationally, this is acceptable because the hose's job is to deliver an average flux of sulfur to the stratosphere; occasional cessations of this flow are fine, as long as the proper aggregate amount gets delivered. Outages should be kept to a minimum, since the hose itself must be designed to handle its maximum flowrate, but rare outages are tolerable since they won't increase this value much.

An empty hose experiences the same wind forces that a filled one does, but is more wind resistant for two reasons. The first, is because it's lighter. When a hose is tilted, a portion of its weight acts sideways (rather than along the hose), and this force acts to further increase the tilt. The less weight the hose has, the less severe this destabilizing effect is.

The other improvement from emptying the hose, is that it becomes stiffer. This is not because the tension increases (in fact, average tension actually drops), but because what counts in resisting hose deflection is net tension, and this increases significantly once the compressive force from fluid pressure goes away.

The effect of emptying a hose supported only by top-mounted balloons is shown in Fig. 5.11a for the Liq-1 version, and in Fig. 5.11b for Liq-2 with its on-hose pumps. The improvement is clear even for the higher tension, Liq-1 hose, but is particularly dramatic for the inherently less stiff Liq-2 case. Once emptied of fluid, both hose designs can readily survive in 100 m/s winds, and likely even stronger ones.

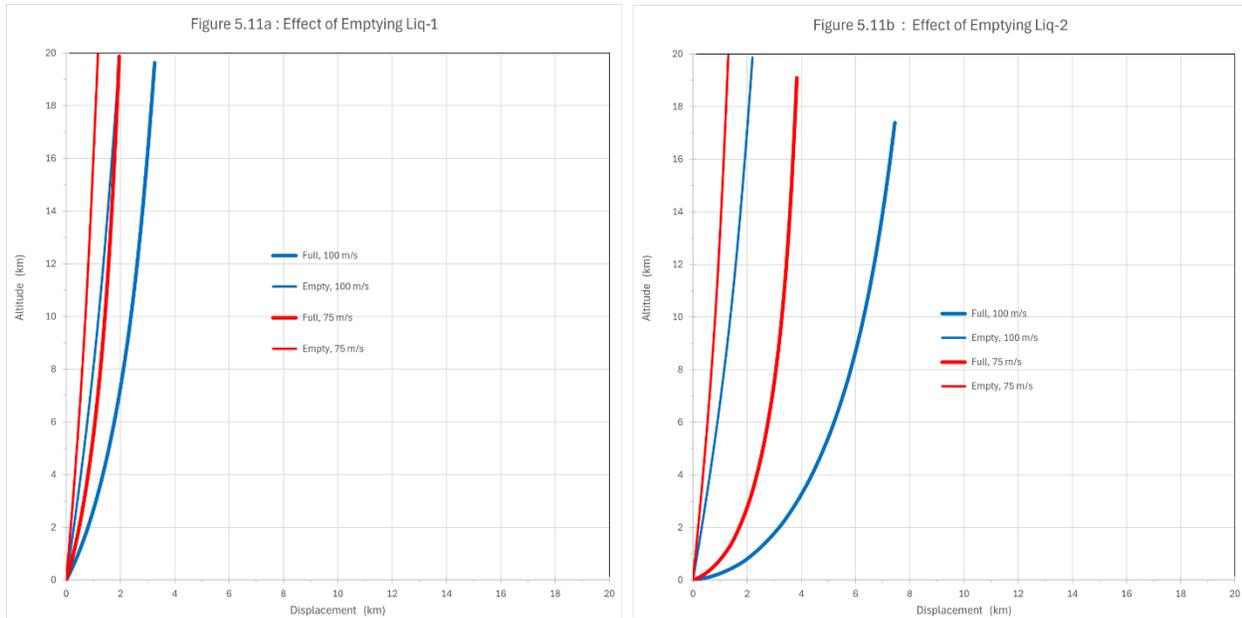

In practice, we'll probably adopt a two-tiered strategy towards survival in high winds. The hose will be designed to live, and operate, in winds of a designated strength and probability. But, if the hose starts to encounter stronger winds, it will be emptied, temporarily suspending sulfur delivery until the winds drop again, after which it is refilled and resumes operation. What this threshold condition is, depends on anticipated wind statistics, how much passive protection (e.g., extra balloons) one is willing to incorporate, and how often you're willing to suspend operations.

## 5.7 : Wind and Mid-Hose Balloons

We previously showed, in Section 4.7, that placing some of the balloons along the length of the hose, was more efficient than putting them all at the top.

Now, we're in a position to discuss the problem with this approach; the hose is now much more susceptible to wind. Given how serious wind is for hoses where all the balloons are at the top, we will need to use all the approaches developed for that case, when attempting to use mid-hose balloons.

The first, most essential, technique, is to use streamlined elongated balloons, rather than spherical ones. This, of course, reduces the number of balloons required; updating Table 4.3 (for spherical balloons) to Table 5.1 (for elongated ones).

| Table 5.1 : Support via Mid-Hose Balloons ||||||
| Liq-1 ||| Liq-2 ||
| Altitude (km) | Number | Diameter (m) | Number | Diameter (m) |
| --- | --- | --- | --- | --- |
| 4 | 1 | 14.9 | 1 | 12.4 |
| 8 | 1 | 16.7 | 1 | 14.6 |
| 12 | 1 | 19.3 | 1 | 17.5 |
| 16 | 2 | 18.4 | 2 | 17.4 |
| 20 | 3 | 19.6 | 3 | 19.2 |
| All Balloons at Top |||||
| 20 | 18 | 19.8 | 13 | 19.9 |

Despite converting to elongated balloons, designs with mid-hose balloons will still suffer more from wind than designs with only stratospheric ones. This is due to two effects: The first problem, is that the wind pushing on mid-hose balloons is stronger than that at the top of the hose. Not only are lower altitude wind speeds (e.g., jet stream ones) generally faster, but they occur at higher air densities; accordingly, they deliver much larger dynamic pressure to the balloons. This more than offsets the smaller aggregate balloon area possible with mid-hose siting. This problem can be illustrated, by comparing the total lateral force applied to the vertical hoses (i.e., before deflections) by the reference wind profile of Fig. 5.1. For Liq-1, the total lateral force on the balloons increases from 6.2 kN to 13.3 kN, and for Liq-2, it climbs from 4.5 kN to 11.3 kN. Obviously, this increased force is likely to lead to more deflection.

However, not only are these hoses pushed more, they are also more susceptible to being deflected when pushed. This is because using mid-hose balloons to help support the hose, does so by periodically reducing tension, and this inherently makes the hose less stiff and less capable of resisting lateral forces.

These dual effects, more lateral force acting on a more flexible hose, inevitably cause hoses supported by mid-hose balloons, to suffer greater wind-caused deflection than designs with only stratospheric balloons.

This can be demonstrated in Fig. 5.12a (for Liq-1) and Fig. 5.12b (for Liq-2). Here, we compare deflection of hoses with mid-hose balloons, to that where they're all at the top. Clearly, use of mid-hose balloons leads to greater deflections. If the winds are weak enough, this is acceptable. The Liq-1 design can readily handle winds peaking at 50 m/s, probably deal with 75 m/s ones, and may actually survive 100 m/s ones; but in all cases, with much greater deflections than designs with only stratospheric balloons. But, as seen before, Liq-2 designs are inherently less stiff. Here, the hose will probably fail in 75 m/s winds, and certainly does so at 100 m/s. In 50 m/s winds, the mid-hose design still undergoes large deflections, of the same size that the top-supported design does at 100 m/s.

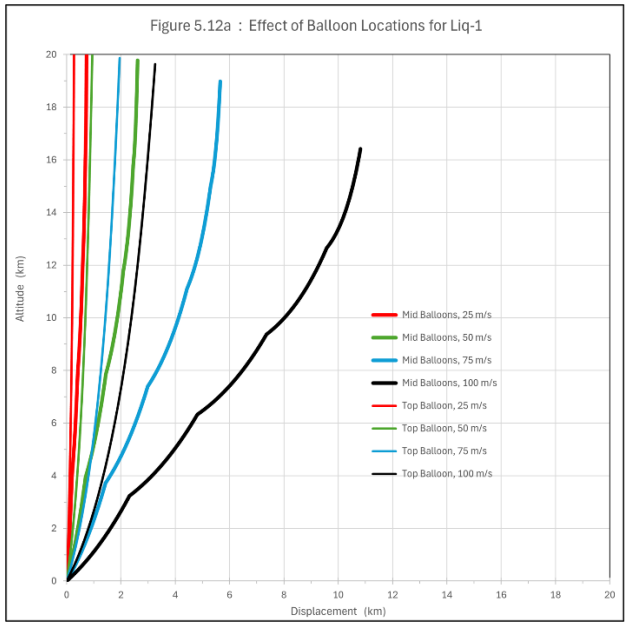
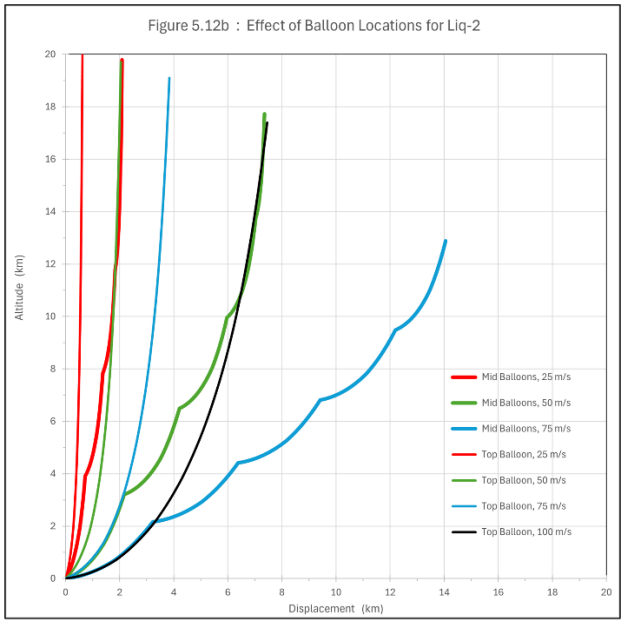

Previously, for top-mounted balloon designs, we found that adding extra stratospheric balloons was an effective way to decrease deflections. The same will be true here, although this will necessarily lessen the balloon savings of mid-hose sitings.

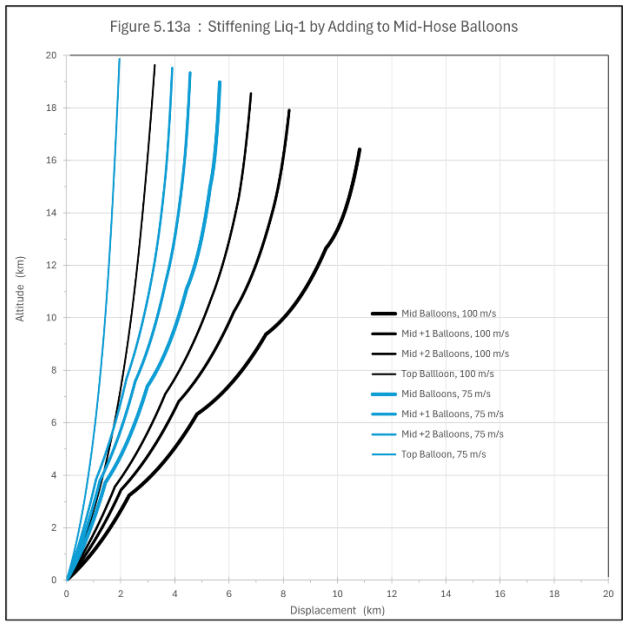
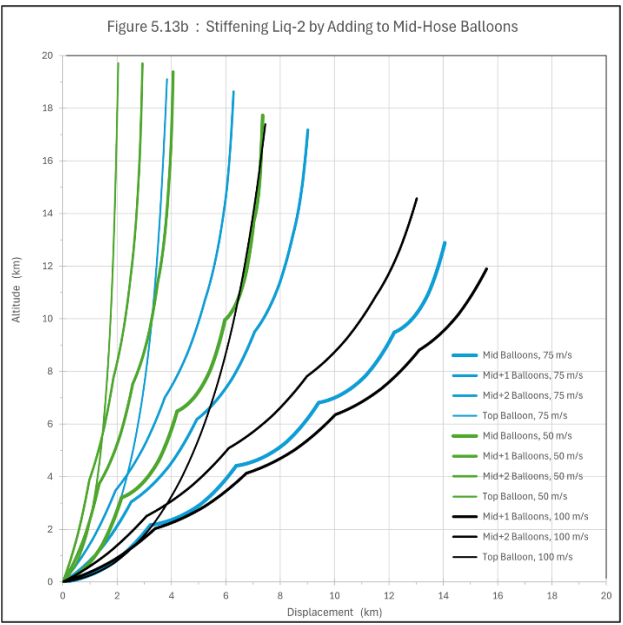

In Fig. 5.13a, we show the effect of adding different numbers of extra stratospheric balloons, to a Liq-1 hose which also employs mid-hose balloons; this progression is compared to the design with all stratospheric balloons. It's clear that the addition of 1 or 2 extra balloons (bringing the total to 9 or 10 respectively) does significantly reduce deflections; 75 m/s winds are now clearly tolerable, while 100 m/s ones may be so. But, it's also clear, that these augmented mid-hose designs are still much less rigid, than the baseline all-stratospheric ones.

Similar comparisons are shown in Fig. 5.13b for the Liq-2 hose. Because this design inherently has less tension, the addition of extra stratospheric balloons has greater beneficial effect, but, of course, its starting from an even more deflected state. One extra balloon is sufficient for 50 m/s winds, two are needed at 75 m/s, and even three will probably not allow 100 m/s to be survived; unfortunately, such balloon counts are now approaching the 13 used in all-stratospheric support.

Ultimately, the most effective way to protect these hoses from extreme winds is, as discussed previously, to temporarily empty them.

The efficacy of this is shown in Fig. 5.14a for Liq-1 designs, and in Fig. 5.14b for Liq-2 ones. In these plots, we show not only the deflections for full and empty hoses, but also the improvement proffered by our earlier strategy of adding extra balloons to a full hose. Clearly, emptying the fluid from the hose, reduces deflections much more than is possible by adding balloons. This gain is particularly apparent for the Liq-2 design; even with two extra balloons (raising its total to 10, compared to 13 for the top-mounted system) it could probably not survive 100 m/s winds. But, the pure 8 balloon system can, if emptied, handily survive these winds.

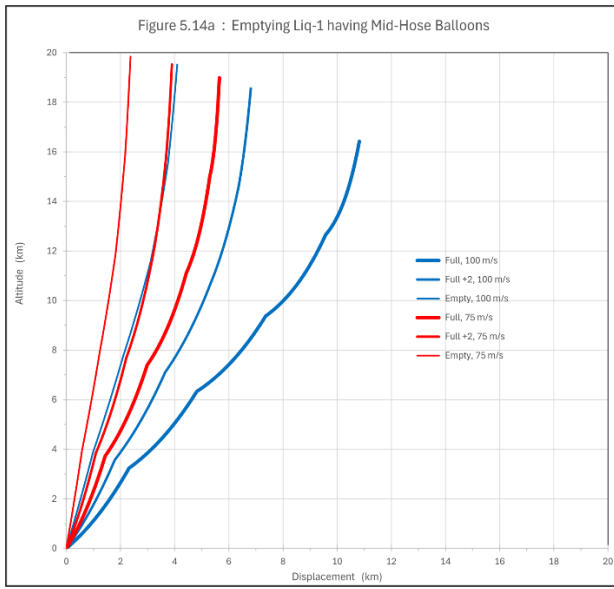
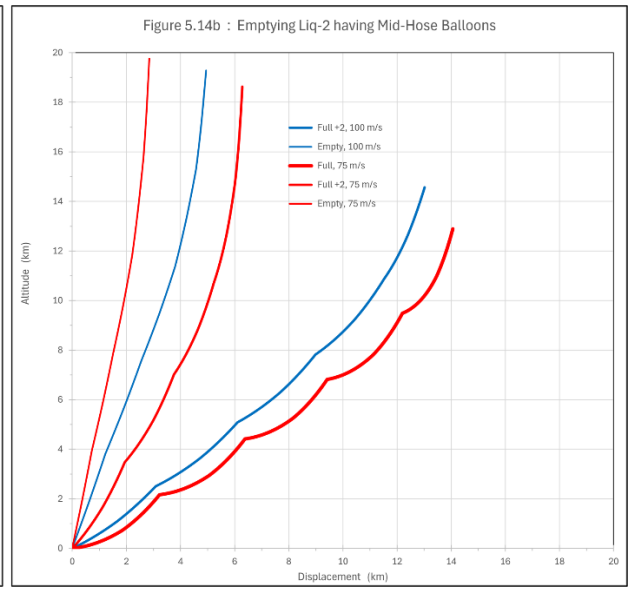

Overall, placing balloons midway along the hose will certainly reduce the number of them needed. It will also make the hose more susceptible to winds. Whether this tradeoff is acceptable or not, will probably only become apparent with operational experience.

# 6 : Gas-based Hose (Gas-1)

So far in this study, we've assumed that the sulfur-bearing fluid pumped up the hose was a liquid. Delivering the material as a gas instead, discussed in Section 6.1, has two main advantages, simplicity and much lower pressures.

However, the drawback to using gas, is that it requires a much fatter hose than liquids do, and hence is even more susceptible to wind. Accordingly, using an airfoil-shaped shroud to reduce wind forces is essential; and, as discussed in Section 6.2, can reduce deflections to manageable levels.

Because of its large size, however, making the shroud at an acceptable weight is difficult. Since the solid shrouds used for liquid-based hoses, are too heavy for these much larger ones, we'll focus on thin-walled hollow shrouds. This provides a volume, which can either be left empty, or can be filled with gas, so as to perform useful functions. One possibility, having the shroud act as the $H_2S$ conduit, is treated in Section 6.4. Another, filling the shroud with helium and using it for buoyancy, is treated in Section 6.5.

In any event, hollowing out the shroud does reduce mass, but its walls become a crucial concern. They must be strong enough to maintain the shroud's airfoil-type shape, yet remain lightweight enough for use in the hose. Use of membranes, thin shells, or more heterogeneous wall designs, will be discussed in Section 6.6.

## 6.1 : Nominal Performance

Let's first discuss a gas-based hose in an ideal, windless, atmosphere.

The first question, is which gas to use. Referring back to Table 2.1 for the options and material properties, hydrogen sulfide ($H_2S$) is again the preferred choice; for the same basic reasons as before. First, $H_2S$ is (from a mass standpoint) almost all sulfur; it imposes virtually no mass penalty through delivering extraneous elements. Secondly, while we've already explored operation of hoses using liquid $H_2S$ at atmospheric temperatures, $H_2S$ can alternatively be operated as a gas, simply by keeping pressures low enough.

A gas-based hose is particularly simple to model. The pressures, set by gravitational head and fluid-friction, will be low enough so that $H_2S$ can be treated as a perfect gas. Likewise, we'll assume that the gas is in thermal equilibrium with the surrounding atmosphere. There would certainly be advantages to operating at elevated temperatures, but thermal coupling to the atmosphere is strong enough, so that this does not appear feasible. The difference here, from liquid-based hoses, is that gas-based ones will be much wider, with thinner walls; and hence couple even more strongly to the surrounding atmosphere.

Given these two facts, perfect gas law and defined temperature profiles, modeling flow up the hose is straightforward. Unlike the liquid system, where the fluid density was nearly constant along the hose, here the gas density varies with altitude, due to changes in both the gas's pressure and the atmospheric temperature. The mass flowrate is set by the same delivery requirements as for liquid-based systems, so its value is the same for a gas-based hose. This makes the $\rho v$ product constant along the hose; but, since the density varies, so will the flow velocity.

$$\rho = m_w \frac{P}{R_* T} \tag{6.1}$$

$$\rho v = \Lambda \dot{M} / \left(\frac{\pi}{4} D^2\right) \tag{6.2}$$

The pressure of the $H_2S$ gas drops along the hose, due to both the weight of the gas, and the flow-friction encountered when pumping it. Here, the friction factor, $f_D$, is evaluated with the same Reynolds number based Serghides[34] correlation employed in Section 3.3; albeit now using the temperature dependent viscosity for gaseous $H_2S$[51].

$$P' = -\rho g - \frac{f_D}{2} \frac{\rho v^2}{D} \tag{6.3}$$

For the liquid $H_2S$ system, both of these gradients were essentially constant along the hose. With gas, both terms vary; the first as $\rho$ and the second approximately as $1/\rho$. Nonetheless, it is straightforward to numerically solve the equations and find the pressure variation along the hose. As for liquid-based systems, the most important design parameter is the hose diameter; the smaller it is, the greater the flow friction, and the more pressure is required to pump $H_2S$ up the hose.

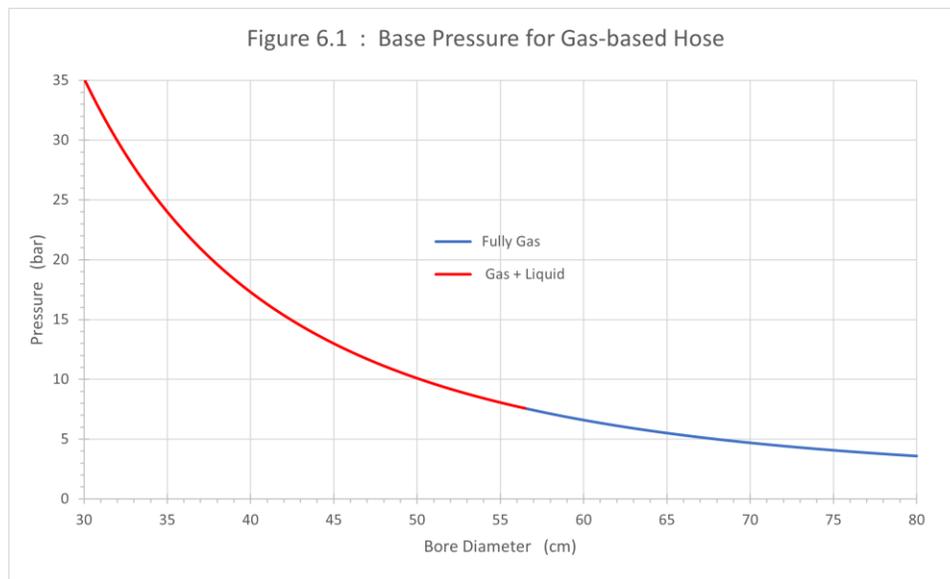

Figure 6.1 : Base Pressure for Gas-based Hose

This effect is demonstrated in Fig. 6.1, where we plot base pressure versus hose diameter, assuming the pressure at the top of the hose is that of the local atmosphere.

It's clear that the pressures involved in gas-based hoses are much smaller than for liquid ones. By adopting large diameters, these pressures can be reduced to several bars.

The reason for preferring to keep pressures this low, is to insure that the $H_2S$ in a gaseous state; if the pressure is too high, the $H_2S$ will tend to liquify. This threshold is indicated by color in Fig. 6.1; only within the low pressure, large diameter portion of the curve, is the $H_2S$ always a gas. At smaller hose diameters, pressures in parts of the hose exceed $H_2S$'s vapor pressure, which varies along the hose as the gas temperature changes. This situation is illustrated in Fig. 6.2, where the variation of the $H_2S$ vapor pressure along the hose, is compared to that of the gas pressure in hoses of different diameters.

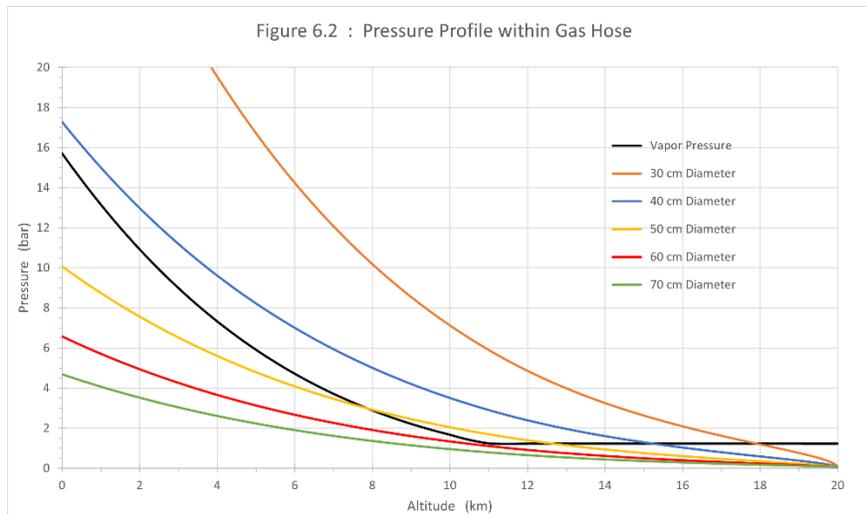

For large hose diameters, the gas pressure remains below the vapor pressure along the entire length of the hose; for these systems, flow is entirely gaseous. But, smaller diameter hoses require larger pressures, and eventually these exceed $H_2S$'s vapor pressure. The details of what happens once this occurs are unclear (the pressure curves in Fig. 6.2 cannot be relied upon once liquid appears), but the result is clear; a gas-based hose requires large diameters.

Structurally, a gas-based hose can be analyzed in the same way as liquid ones were. Since pressures are low, there is no need for locating some pumps along the hose; all pumping will be performed at the base. Thus, the hose design closely resembles the Liq-1 system; the hose consists of only a tube to contain the gas and its pressure, and a tether to carry forces along it. These two functions could be combined, using the tube walls to carry axial load, but to limit design and implementational interactions, we'll keep them separated into two different components. Using the same material properties for the tube wall and the tether as for the liquid-based hose (as detailed in Table 4.1), we'll plot their masses, as well as that of the gas, in Fig. 6.3.

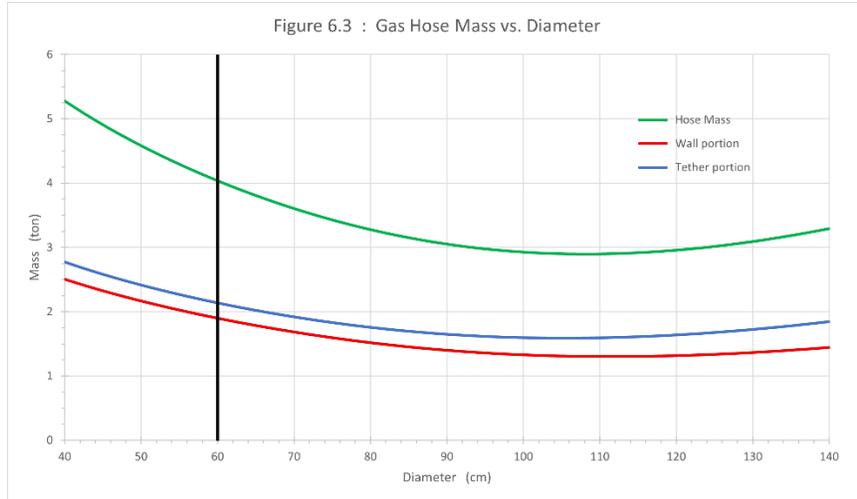

The mass of the hose and its components first drops with diameter, before reaching a gentle minimum at about 105 cm. This behavior is largely determined by that of the compressive force, $F_c$, acting along the hose.

$$F_c = \frac{\pi}{4}(P - P_{air})D^2 \tag{6.4}$$

This term is what determines the amount of tension the hose needs to prevent buckling, and hence the number of balloons (spherical ones in Fig. 6.4) required to supply this tension. All three of these quantities are plotted in Fig. 6.4, and mirror the hose masses shown in Fig. 6.3.

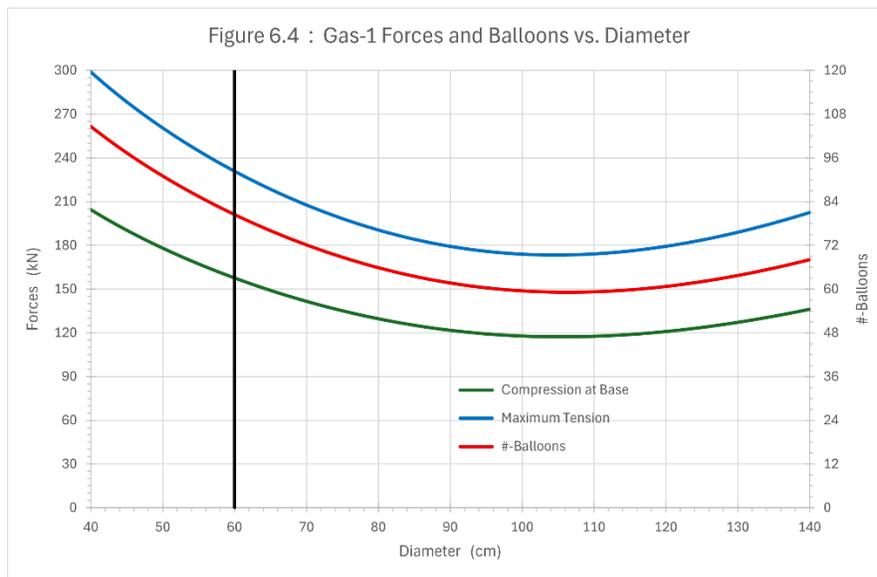

While the mass and balloon values suggest that the hose should have a diameter of about 1 meter, we will actually baseline a value of 60 cm, just above the liquid transition shown in Fig. 6.1. The reason, is wind. The larger the diameter, the greater the wind force pushing on the hose, and the harder it is to deal with it. These issues (as we'll see) are the biggest problem facing a gas-based hose, and strongly favor keeping its diameter as small as possible; 60 cm is much preferable to 1 meter.

Simulations, summarized in Table 6.1, demonstrate key differences and similarities between gas-based and Liq-1 liquid-based hoses. Both are exceedingly simple systems, using a pump at the ground to push fluid up a tubular hose. Their primary differences are that liquid hoses require much larger internal pressures, whereas gas hoses pay for their lower pressures through the need for much larger bore diameters.

| Table 6.1 : Gas vs Liquid Comparison | | | |
|---|---|---|---|
|  | Liq-1 | Gas-1 | Units |
| Bore diameter | 3.10 | 60.0 | cm |
| Peak pressure | 3,250 | 6.59 | Bars |
| # Balloons (spheres) | 108 | 80 |  |
| # Balloons (elongated) | 18 | 13 |  |
| Mass - Fluid | 14.8 | 18.1 | Tons |
| Mass - Hardware | 9.8 | 4.0 | Tons |
|    Hose portion | 6.8 | 1.9 | Tons |
|    Tether portion | 3.0 | 2.1 | Tons |
| ΔPressure*Area | 2.45 | 1.58 | Bar*m$^2$ |
| Tension at base | 306 | 198 | kN |

However, the effects of these differences largely cancel out at the system level. Both designs use nearly the same number of balloons, so need almost the same top tension. The pressure-area product, which dictates the buckling-limited tension at the ground, is surprisingly close, despite the gas system's much lower pressure. Accordingly, they both operate at similar average tension (gas a bit lower), so will have similar tether mass and stiffness. The actual mass of fluid within the hoses is also close, despite the large difference between gas and liquid densities.

The largest component difference shown in Table 6.1, is that of the heavier walls for the liquid hose; but much of their weight is supported by that system's greater flow-friction. However, there's an even bigger mass difference between gas and liquid hoses, which is not yet shown in Table 6.1, namely the mass of the aero-shroud needed to offset most of the wind forces on gas-based hoses. This mass is likely to dominate that of the Gas-1 hose, and Sections 6.3 – 6.6 will be focused on dealing with it.

**6.2 : Dealing with Wind**

It is their larger diameter that causes gas-based hoses the most problems. These come both from the increased wind force acting on it, and also from the greater shroud mass needed in order to limit the force increase.

In principle, wind force on the hose scales with its diameter. So, a 60 cm gas-based hose should experience 17 times the force seen by the 3.5 cm (outer diameter) liquid-based one. In practice, the fact that flow around the larger hose occurs with a greater Reynolds number, does help by reducing the drag coefficient, but the force on the hose is still 10 times larger.

Obviously, the same streamlining steps that were essential for liquid-based hoses, are available for gas-based ones. Using elongated balloons, instead of spherical ones, is straightforward, and will be just as effective in this case; ballon drag is similar for both types of hose. So, while the increase in hose size shifts the main drag source, from being the balloons to the hose, the overall force only increases by about three-fold.

Unfortunately, this increase in wind force is accompanied by the fact that, with lower tension, the gas hose has less stiffness. This combination means that the gas hose is deflected much more by the wind than a liquid one.

Wind deflections (at a peak wind speed of 50 m/s) are compared in Fig. 6.5 for gas and liquid hoses; both use the same streamlined balloons and shrouds which are essential to handling wind forces. The Gas-1 hose is clearly affected much more by wind than the Liq-1 one, and fails completely for stronger winds (i.e., 60 m/s and above).

The simplest way to improve the gas-based hose's performance, is to increase its stiffness, i.e., to increase tension by adding more balloons at the top. This is shown in Fig. 6.5, for additions of one to five extra balloons (above the nominal value of 13 shown in Table 6.1). These additions help considerably, but even when adding 5 balloons (bringing the Gas-1 allotment up to the same value of 18 as Liq-1), the gas-based hose deflects more; reflecting the fact that its hose is wider, and is hence pushed more.

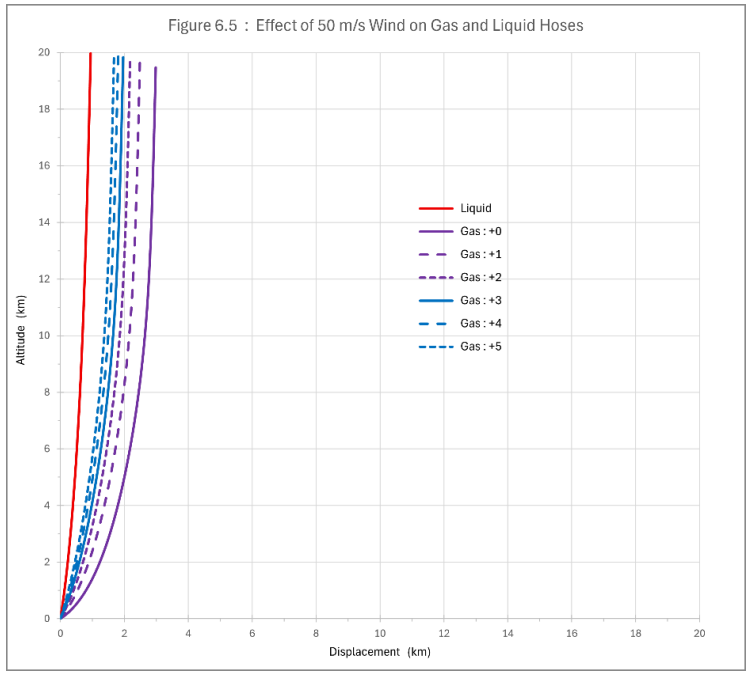

Since it's apparent, that the helpfulness of balloons is diminishing as we add more of them, we'll baseline the addition of 3 balloons rather than 5, raising the Gas-1 total to 16, which adds 48 kN to its tension. The effect of this at different wind speeds is seen in Fig. 6.6, where we compare the Liq-1 and Gas-1 hoses.

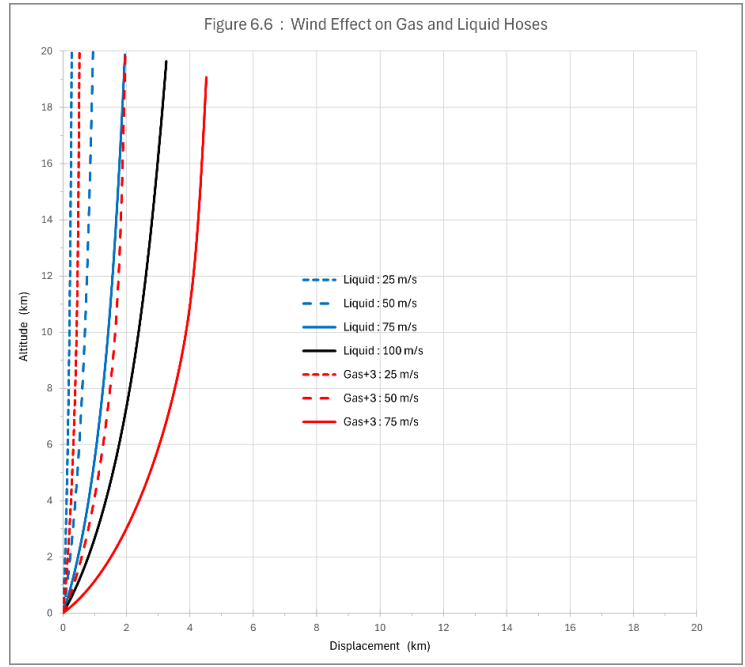

The increased vulnerability of Gas-1 is particularly apparent, the stronger the winds it's exposed to. The deflections of Gas-1 in a wind peaking at 75 m/s, are considerably worse than Liq-1 at 100 m/s; and the Gas-1 hose cannot survive 100 m/s winds at all.

We could, of course, try to improve Gas-1's extreme-wind performance by further increasing its number of balloons. But, rather than penalize it in this way, we will handle rare and extreme winds in the same way discussed for liquid hoses, by temporarily emptying the hose. The effect of this procedure is displayed in Fig. 6.7. Note, that while 100 m/s winds could not be survived when the hose was full, Gas-1 can probably deal with them once emptied.

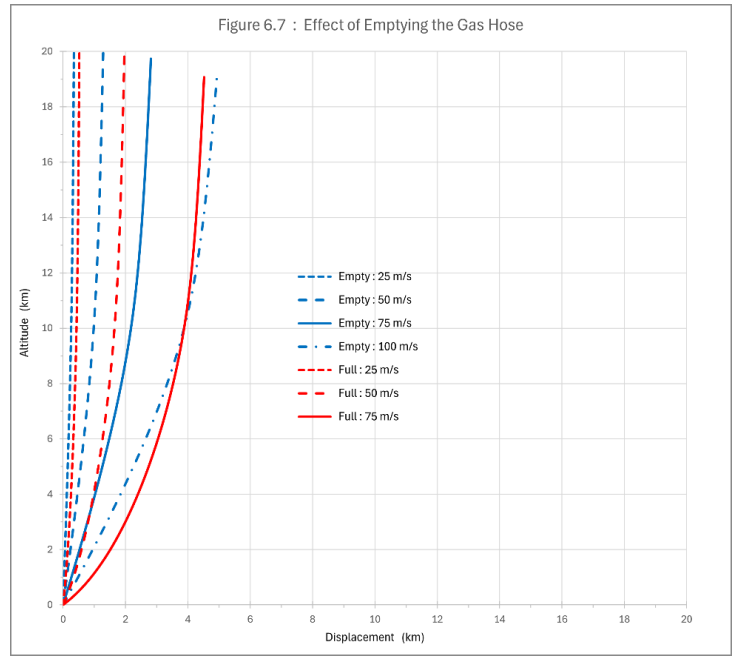

Figure 6.7 : Effect of Emptying the Gas Hose

So, the direct effects of wind on large diameter gas-based hoses are troublesome, but tolerable.

**6.3 : The Challenge of Shroud Mass**

The indirect effect of wind, that on the hose's mass, is much more serious. The problem is that, because the gas hose is so much wider than a liquid one, the aerodynamic shroud, which is absolutely essential in reducing wind forces, is correspondingly larger and heavier. The same aerogel-and-skin shrouds, which imposed only modest weight additions for the liquid hose, would completely dominate the weight of a gas-based hose.

First, consider how heavy a shroud we can afford: The hose is 20 km long, so every 1 gm/cm of lineal weight, adds 2 tons to the hose. Given that the lifting power of each elongated balloon (20 meters wide, 100 meters long) is about 1.9 tons, we see that each 1 gm/cm of shroud weight requires the hose to have one additional balloon. Compare these numbers to those of the gas hose itself (shown in Table 6.1); the hose mass is only 4.0 tons (i.e., just 2 gm/cm), and utilizes 16 ballons (mostly to increase the tension enough to limit deflections). So, even a 10 gm/cm shroud (i.e., 1 kg/m) would completely dominate the mass and balloon complement of the hose; smaller values are strongly preferred, larger ones become prohibitive.

Next, how challenging will it be to make a shroud this light?

As a reference point, the liquid hose's shroud used a 0.1 mm skin around a fairly sturdy (0.02 gm/cc) aerogel core; together they had a mass of 0.8 gm/cm. The shroud used for the gas hose is large, 2 feet wide and 8 feet long; it encloses an area, beyond that of the hose itself, of 7000 cm$^2$ with a perimeter of 510 cm. If it used the same material composition as the liquid hose's shroud, the skin itself would contribute 5 gm/cm; but the aerogel core would require 140 gm/cm. Clearly, this is not viable.

One option, is to use a much lower density aerogel, While the earliest aerogels were made from silica, they have since been made from many other materials, and can be both lighter and stronger. Given the large area required for the hose's shroud, a suitable aerogel should have density of 1 mg/cc or lower. While materials such as graphene can be used to make aerogels with such densities, and intrinsically offer better strength and elasticity than silica, these properties drop very rapidly with density[52]. Beyond material properties, a practical challenge with this approach is simply the amount of aerogel required. Making aerogel-filled shrouds for the hose will require 14,000 m$^3$ of material, formed into stiff large-scale sections. For these reasons, aerogel-filled shrouds are not suitable for near-term gas hoses.

**6.4 : Using the Shroud as the H$_2$S Conduit**

An alternative approach is to fill the shroud with gas, rather than a solid material. We'll discuss two interesting options; one is the use of the shroud itself as the hose's gas conduit, the other is to make the shroud buoyant, by filling it with helium.

One advantage of using an elongated, airfoil-type, shape for the gas conduit is that it requires less width for the same flow area. We can model flow in this-shaped conduit by knowing its flow area and perimeter, and from these, its hydraulic diameter. In terms of maximum width $D_w$, these are respectively 2.723$D_w^2$, 8.517$D_w$, and 1.280$D_w$ for the NACA0025 shape.

Because of its smaller width per area, an airfoil-shaped conduit can be thinner than a circular one while still avoiding H$_2$S liquification; 35 cm rather than 60 cm. This smaller size, of course, reduces wind forces, and should help lower the conduit-shroud's mass.

The effect of the smaller wind forces is illustrated in Fig. 6.8, which compares deflections of a gas hose with the traditional circular conduit, to one using an airfoil-shaped conduit. To reiterate, both hoses use airfoil-shaped envelopes to drastically reduce wind forces, but in one case, this is an external shroud surrounding a circular conduit, while in the currently discussed version, the airfoil itself is the conduit. The improvement from using the smaller airfoil-shaped conduit is clear; for instance, it can now handle 100 m/s winds that the circular conduit could not.

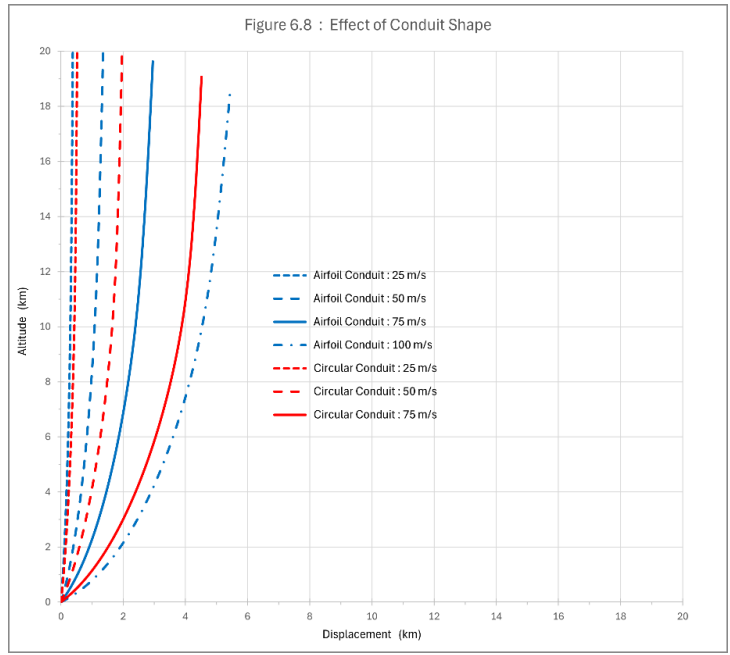

Figure 6.8 : Effect of Conduit Shape

However, while allowing the shroud to double up as the gas conduit, is an attractive way to improve the hose and its wind resistance, it does little to actually address the airfoil's mass. In fact, although its size reduction should help, this is offset by the fact that now, the airfoil must withstand a significant pressure differential, between its internal $H_2S$ and the external atmosphere. The airfoil conduit calculations, shown in Fig. 6.8, were not based on accurate calculations for its mass, but rather, an extremely optimistic adoption of a fixed skin thickness, similar to that of an equal-area circular conduit. This led to a value of 6 gm/cm, smaller than our 10 gm/cm target, but still dominating the overall mass and balloon budget. Circular conduits are uniquely well shaped to handle pressure loads, but elongated shapes are not; an airfoil conduit will actually be structurally much heavier than an equal-area circular one. This will be demonstrated in Section 6.6, where we consider the shell-type walls needed for airfoil-shaped pressure vessels. For weight reasons, we will abandon this approach.

**6.5 : Making the Shroud Buoyant**

One way to minimize the effect of the shroud mass, is to make it buoyant. Filling the shroud with helium will do this, while also providing enough internal pressure to help keep its skin taut. At the same time, this internal pressure need only be slightly higher than that of the local atmosphere, so will not impose nearly as large a pressure load as using the airfoil for the main $H_2S$ conduit.

The first hurdle for this concept is the numbers; does the shroud contain enough volume to provide the buoyancy the hose needs? It does, but barely: The buoyancy of a 20 km column of helium is given by the air's top-to-bottom pressure differential, derated slightly by the ratio of air-to-helium molecular weights. This buoyancy is 0.96 bars, which when cut by 14% due to the helium, becomes 0.82 bars. The shroud encloses a column of 6970 $cm^2$, so, acting as a balloon it provides a buoyancy of 57.5 kN, enough to support 5.86 tons. While this is a significant value, it only corresponds to an average hose mass of 3 gm/cm, so if it's the only support for the hose, mass will be have to be very strictly controlled.

But, while a buoyant shroud might provide enough force to support the hose's mass, it doesn't provide enough to stabilize against buckling. From Table 6.1, the conduit's pressure-area product is 1.58 bar-m$^2$, i.e., 158 kN. To stabilize this, with a 25% margin, requires a tension force of 198 kN at the base. Since the conduit's buoyancy of 57.5 kN cannot provide this, balloons are required as well. In principle, we could artificially scale-up the shroud volume four-fold, in order to supply enough buoyancy for this tension. But the shroud would now be twice the width needed to encompass the gas conduit, 4 feet wide and 16 feet long; far larger than needed for its primary job, and totally dwarfing the 1.5 inch wide shroud sufficient for liquid hoses.

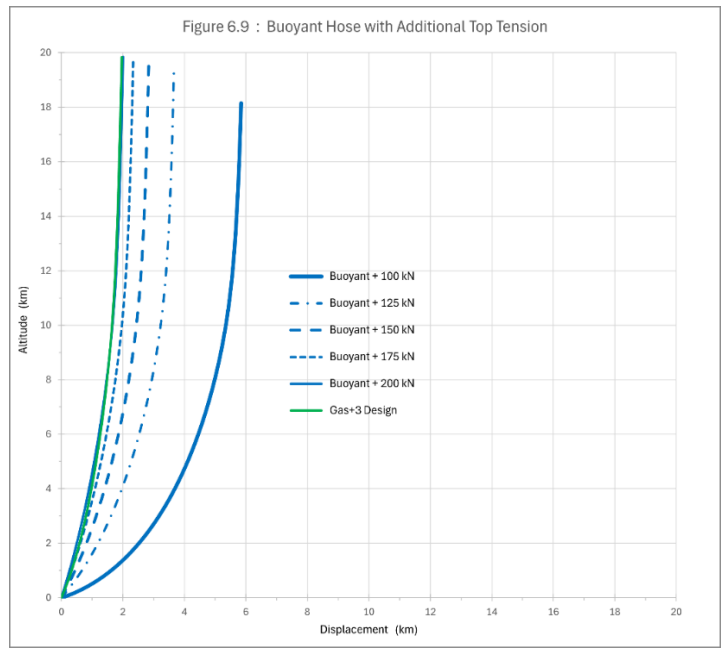

The other reason, that the hose will not be able to rely upon the shroud for all of its buoyancy, is that, even if we could ignore the buckling instability, it still needs additional balloons at the top, to apply extra tension to help stiffen the hose; without these, the hose cannot resist the sideways wind forces. This is illustrated in Fig. 6.9, where we show deflections in a 50 m/s wind, for different amounts of upward force applied to the top of the hose. No cases are shown for tensions below 100 kN, since hoses with 75 kN or less collapse. Only with the addition of 200 kN, about 12 balloons, does this buoyant hose reach the same performance as a standard, fully balloon-supported, hose.

So, a helium-filled shroud simply cannot provide enough buoyancy to be worth the effort: It can only lift about 3 gm/cm (probably not even its own weight), and still needs additional balloons at the top of the hose to stabilize it against buckling and wind forces.

## 6.6 : The Shroud Wall

The largest drawback to a gas-based hose, is the mass of its aerodynamic shroud. Its large conduit size dictates that it must have a streamlined shroud in order to reduce wind forces, yet the shroud's large size drives up its own weight. Unfortunately, this weight is not a minor concern, if more than 5-10 gm/cm it will dominate the weight of the entire hose, making it increasingly uncompetitive with liquid-based alternatives.

Solid shrouds, even if made from aerogel, are too heavy and impractical to field. The shrouds must be hollow, either empty or filled with gas. We've already discussed a couple of interesting ways to make use of gas-filled shrouds. Regardless of what (if anything) is inside the shroud, its walls are crucial. These have to define and maintain the shroud's aerodynamic shape, contain any internal gas, and yet satisfy the stringent weight constraints.

### *Membranes*

The simplest, and probably lightest, approach would be to use a thin membrane for the wall. Membrane walls can be considered as pure 2-D surfaces, carrying tensile loads along their surface. They are attractive because they can be made from very high strength fibers; because of applications such as sailcloth for racing yachts, very lightweight fabrics are commercially available. The problem, is that membranes are flexible, and can only maintain shapes for which all loads lie along their surface. Unfortunately, the only long tubular shape for which this is true, is a simple, constant diameter, cylinder. This is why the gas hose's conduit is circular, and can be so lightweight. Pressurized shapes like the NACA airfoils, cannot be made from simple membranes; such a tube would promptly deform into a circular cylinder.

It is possible, that more complexly-shaped shrouds can be designed, allowing one to take advantage of membrane materials. For instance, the shroud could use internal fibers to tie opposing sites on the surface together, forming a surface defined by a series of intersecting circular arcs. Or, an array of reinforced webs can be built into the membrane, forming a scalloped 3-D surface. The challenge to these types of approaches, is that their shrouds no longer have aerodynamically smooth surfaces; drag forces will increase. Optimal design will require tradeoffs; are the weight savings from use of high strength fabrics, worth the increased wind forces from poorer aerodynamic shapes?

### *Shells*

Since membrane walls cannot hold an airfoil shape, we could adopt thicker and stiffer walls. These will be inherently heavier than simple membrane fabrics, and yet must still meet the aforementioned weight target, of 10 gm/cm or less. A shroud based on the NACA0025 airfoil has a perimeter of just over 500 cm, so in order to limit the wall's mass to 10 gm/cm, its local mass cannot be more than 0.02 gm/cm$^2$, equivalent, for instance, to a 1 centimeter thick layer of 0.02 gm/cc aerogel.

The strength which the wall will need, to maintain its shape, can be found via shell theory, in which a wall, while thin, carries loads by shear and moments, in addition to the in-plane force, which is all that a membrane is capable of. Since the shroud's shape does not vary axially, the behavior of its shell can be analyzed with the same force-and-moment methods used in beam theory. Using $F$ for in-line force, $S$ for shear force, and $M$ for moments, the equilibrium equations are:

$$\frac{dF}{d\ell} = -\frac{S}{\mathcal{R}} \tag{6.5}$$

$$\frac{dS}{d\ell} = \frac{F}{\mathcal{R}} - p \tag{6.6}$$

$$\frac{dM}{d\ell} = -S \tag{6.7}$$

Here, $\ell$ is length along the surface, $p$ is internal pressure, and $\mathcal{R}$ is the local radius-of-curvature. These reduce to the standard beam equations when $\mathcal{R}$ goes to ∞. However, for a complexly curved surface, where $\mathcal{R}$ is not constant as in a cylinder, the in-line force $F$ exists and varies along the shape, giving rise to shear and moments. For curved shapes, the first two coupled equations can be combined into a simple second order differential equation, in terms of the angular slope, $\phi$:

$$\frac{d^2 F}{d^2 \phi} + F = p\mathcal{R} \tag{6.8}$$

For a circular tube, $\mathcal{R}$ is a constant (equal to the tube's radius), and we recover the classic solution, $F = p\mathcal{R}$. But, for general shapes, such as our airfoil one, $\mathcal{R}$ varies along the surface, and hence so does $F$. This equation can be solved numerically, using two boundary conditions:

$$\text{Nose}: \frac{dF}{d\phi} = 0 \tag{6.9}$$

$$\text{Tip}: F \cos\phi + \frac{dF}{d\phi} \sin\phi = 0 \tag{6.10}$$

Here, $\phi$ is measured clockwise from horizontal, and the derivative $dF/d\phi$ is $S$, the shear force.

We have solved the force and moment equations for a shroud shaped like the NACA0025 airfoil (as was displayed in Fig. 5.6), and their values are plotted in Fig. 6.10. The values were calculated for $p=1$, while expressing $\mathcal{R}$ in terms of $D$, the airfoil's width (i.e., the diameter of the enclosed $H_2S$ conduit). Hence, $F$ is in terms of $pD$, while $M$ is in terms of $pD^2$.

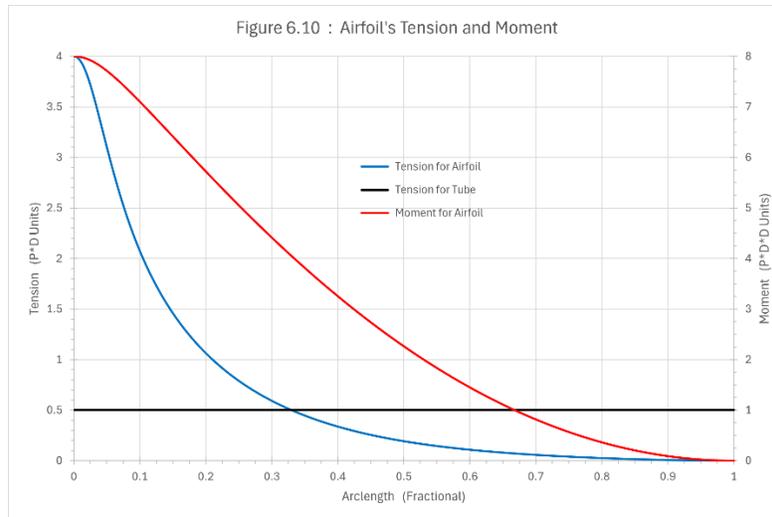

Physically, these solutions are not at all surprising. Both *F* and *M* peak at the nose of the airfoil, and drop to zero at the tip. Mathematically, however, the fact that their peak values are such precise numbers is more so. This turns out to be a general property of smooth mirrored shapes; holding for other teardrop shapes than just the NACA0025 one, as well as for rounded shapes such as ellipses. They each have $F_{max} = (L/W)$, and $M_{max} = (L/W)^2/2$.

Of more immediate concern, is the size of the values. The peak force value, needed for the NACA0025 shape, is 8 times that required for the same pressure, if carried by the gas conduit itself; since it occupies 2.5 times the perimeter it inherently would, this incurs a twenty-fold mass penalty. In principle, most of this could be avoided, by varying the thickness of the tension-carrying part of the shell with position, in order to match the local load. If this were done, while avoiding minimum-gauge practicalities, the penalty would be reduced to just under 3.5X. However, the need to handle bending moments is an even greater concern. Shell walls usually resist moments via a sandwich or truss arrangement, carrying tension and compression in skins on opposite sides of a spacer. The challenge faced by a thin shell, is the large ratio between the moment's lever arm, scaling as *D*, to the skin thickness, which in the above mass-budget strawman was 1 cm. This will multiply the required stress values by a factor of 60.

Clearly, an airfoil-shaped shell cannot affordably carry gas at any significant pressure. This prevents it from being used as the $H_2S$ conduit (as was considered in Section 6.4). While a helium-filled shell would nominally be pressure matched to the local atmosphere, in practice it must be designed to handle some pressure mismatch, and an airfoil-shaped shell is not a mass efficient way to do this. The most plausible use of such a shell is for it to be empty, presenting no pressure barrier to the surrounding atmosphere. In this case, the primary concern will be damage due to concentrated handling-loads. Here, fully flexible structures are more forgiving than shells, and hence preferrable.

*Rib-and-Skin*

A third approach, to making a lightweight shroud that can maintain its airfoil-type shape, is to use a rib-and-skin approach, analogous to aircraft wings. The ability to do so, is greatly aided by the fact that the hose's airfoil is vertical, rather than horizontal, and that it does not have to structurally collect and transfer aerodynamic forces to a distant airframe.

This type of shroud, illustrated in Fig. 6.11, will consist of a series of airfoil-shaped ribs, separated vertically from each other along the hose. The airfoil's surface is formed by high-strength fabric, which is stretched vertically between the ribs. The fabric is kept in its proper shape, not by lateral forces or shell-type rigidity, but rather by the shape of the ribs, which it is axially stretched between. Ultimately, it is gravity which acts to keep the fabric taut, as the weight of lower ribs and fabric, holds the fabric above them in tension.

Figure 6.11 : Vertical Ribbed Airfoil

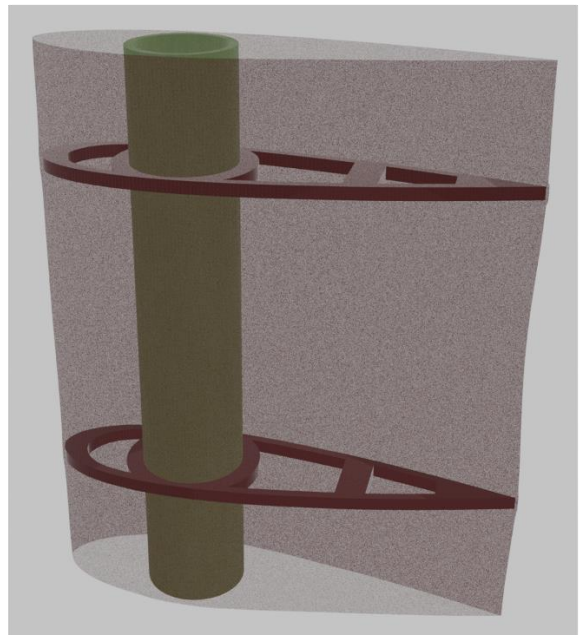

While it's not certain that this type of shroud will be lightweight enough for the gas hose, neither of its two components need be heavy. The fabric skin of the shroud, carries loads purely in longitudinal tension, so can use high-strength fiber material. Commercial aircraft fabric, which actually provides 2-D strength, is available with masses of about 2 - 3 $Oz/yd^2$. Since a NACA0025 shroud has a perimeter of 5 meters, this will contribute a mass of 4 - 5 gm/cm. Individual ribs are rigid objects, so will be relatively heavy; but with composite materials and efficient, hollowed out, structural design (e.g., each rib containing 10 lineal meters of 1 $gm/cm^2$ struts), a 1 kg weight seems attainable. The contribution of this to the overall shroud mass, depends on how closely ribs need to be spaced. Given the shroud's 2 foot by 8 foot size, we'll target a rib spacing of about 6 feet; hence, they'll contribute 5.5 gm/cm of weight. Overall then, a shroud weight of 10 gm/cm (i.e., 1 kg/m) is plausible; more detailed design should be able to reduce this further.

**6.7 : Summary**

We can use hoses to deliver sulfur up to the stratosphere as a gas, instead of as a liquid. A gas-based hose operates at much lower pressure, but in exchange requires a much fatter conduit.

This larger size, makes the hose more susceptible to wind, and hence even more reliant on having a streamlined shroud to reduce aerodynamic forces. Such shrouds are effective enough at this task, so that gas hoses can be operated successfully; wind deflects them more than liquid-based hoses, but not enough to prevent their use.

The shroud's size does, however, raise its mass; and this may be a serious enough problem, to make gas hoses uncompetitive with liquid-based alternatives.

To keep the shroud's mass affordable, it must be hollowed out. In principle, this empty volume can be taken advantage of by filling it with gas, using the shroud, either as the main gas conduit for flowing $H_2S$, or for buoyancy by filling it with helium. In practice, neither of these gas-filled shroud options look worthwhile. Both of them require strong enough walls to contain their gas, and the structural inefficiency of their airfoil-type shapes, makes these walls impractically heavy. In addition, the shroud does not provide enough buoyant volume to support the hose, or to provide the tension needed to keep it stable; so this approach does not eliminate the need for conventional balloons.

Fortunately, it does seem feasible to field lightweight shrouds, by foregoing the need for gas containment, and using a variant of rib-and-skin type aircraft wing construction.

While the large aero-shroud, and its associated mass, will always be a concern for gas-based hoses, their extreme simplicity, and low pressure operation, makes them a worthwhile alternative to liquid-based hoses.

# 7 : Design Selection

In this report we've discussed many aspects of designing stratospheric hoses. These have ranged from choosing the proper fluid, requirements for pumping it to the stratosphere, making the hose from strong yet lightweight components, using balloons to suspend it, learning how to deal with strong winds, and, of course, properly modeling the interactions and performance of the hose.

At this point, it's time to boil this discussion down. We'll summarize three different hose options, first tabulating and comparing them, and then recommending which are best to pursue.

## 7.1 : Three Hose Options

We've presented 3 designs for sulfate-delivery hoses, two liquid-based (Liq-1 and Liq-2) and one delivering gas (Gas-1). All of these will work, and can be developed and built quickly to test geoengineering, and if desired can be replicated and operated to actually perform it.

The three designs do differ however, each having advantages and disadvantages relative to the others.

The Liq-1 design uses a very high pressure pump on the ground to send liquid $H_2S$ up the hose. The fact that pumps stay on the ground allows routine maintenance, helping to assure the hose's reliability. Because the $H_2S$ is liquid, the hose is narrow, which inherently reduces wind forces on it (and the mass of any aerodynamic shroud). It's the high pressure that drives Liq-1's challenges. The direct concern, is simply fielding a hose which can handle this pressure, and for as low a weight as possible. But, the pressure also poses an indirect cost, as the hose needs to supply enough tension to prevent pressure-induced buckling. Both this tension, and support for the weight of the hose, have to be supplied by balloons; so the Liq-1 design tends to require more balloons than its Liq-2 alternative.

The Liq-2 design also uses liquid $H_2S$, so shares the same compactness and hence inherent wind tolerance as Liq-1. It's difference, is that it uses a series of small pumps mounted on the hose, rather than a single large one on the ground. This means that the hose requires much less pressure, thereby making its design more straightforward, and reducing its weight. The lower pressure also means that tension levels can be lower, since buckling is less of a concern. Both the lighter hose and the lower tension levels reduce the number of balloons that Liq-2 needs. These advantages, however, come at a cost. First, the pumps can't be serviced easily, which will lower hose reliability, and likely up-time. Secondly, the hose needs to carry not just the weight of the pumps, but also the electrical cables bringing power to them. Finally, most of weight of Liq-2's liquid $H_2S$ is now carried by the hose itself (and hence by its balloons) rather than being supported from the ground as with Liq-1. These weight increases drive up the number of balloons; Liq-2 still needs less balloons than Liq-1, but this cuts its advantage.

Gas-1 is, of course, a much different type of hose than Liq-1 or Liq-2. Because it pumps gas instead of liquid, its pressure values are miniscule compared to the two liquid-based hoses. And, since its pump is on the ground, it shares the simplicity and reliability virtues of Liq-1. The problem with Gas-1, is that a much fatter hose is required to transport gas than liquid. The penalty from this larger size is not directly due to the weight of the hose itself (the greater wall area is offset by it needing less thickness), but comes indirectly via effects of the wind. Gas-1's large diameter causes it to intercept much more wind than a liquid-based hose; an aerodynamic shroud to reduce wind forces is absolutely essential. But, the hose's large diameter makes such a shroud inherently large and heavy; whereas the 20 km length of the shroud means that very little cross-sectional weight can be tolerated. The challenge facing Gas-1 therefore, is making its aerodynamic shroud light enough so that Gas-1's balloon requirements remain competitive with Liq-1 and Liq-2.

In Table 7.1 we detail and compare numerical properties of the three designs, while in Fig. 7.1, we show their deflections in our baseline, Fig. 5.1, wind (50 m/s in the jet-stream, 20 m/s at the top).

| Table 7.1 : Hose Designs | | | | |
|---|---|---|---|---|
| | Liq-1 | Liq-2 | Gas-1 | |
| # Pumps | 1 | 10 | 1 | |
| Bore Diameter | 3.1 | 3.1 | 60.0 | Centimeters |
| Peak Pressure | 3250 | 350 | 6.6 | Bars |
| Tension at Base | 306 | 65.1 | 246 | kN |
|    Anti-Buckling Value | 306 | 32.2 | 197 | kN |
| Fluid Mass | 14.86 | 13.99 | 18.05 | Tons |
| Hose Mass | 11.53 | 7.07 | 25.57 | Tons |
|    Conduit | 6.78 | 1.13 | 1.90 | Tons |
|    Tether | 3.17 | 1.62 | 3.67 | Tons |
|    Pumps, Cables | | 3.00 | | Tons |
|    Aero-Shroud | 1.58 | 1.32 | 20.0 | Tons |
| Buoyant Lift | 32.82 | 27.72 | 49.96 | Tons |
|    # Balloons | 18 | 15 | 27 | |
|    Balloon Wall Area | 10.2 | 8.6 | 15.4 | Hectares |

As expected, Liq-1 has by far the highest pressure, while Gas-1 has by far the lowest. The need to avoid pressure-induced buckling dictates that Liq-1 must have a large tension, much higher than its hose weight would imply. All three designs have comparable fluid masses, but different hose masses; Gas-1 is the heaviest, and its weight is completely dominated by the aerodynamic shroud, which is included here at a value of 10 gm/cm, i.e., 1 kg/m.

The Liq-2 and Gas-1 designs are more susceptible to wind than Liq-1. To reduce their deflections, the designs shown in Table 7.1 use higher tension values than those needed to prevent pressure-induced buckling; this, of course, has to be obtained by adding extra balloons (incorporated in the numbers displayed in Table 7.1).

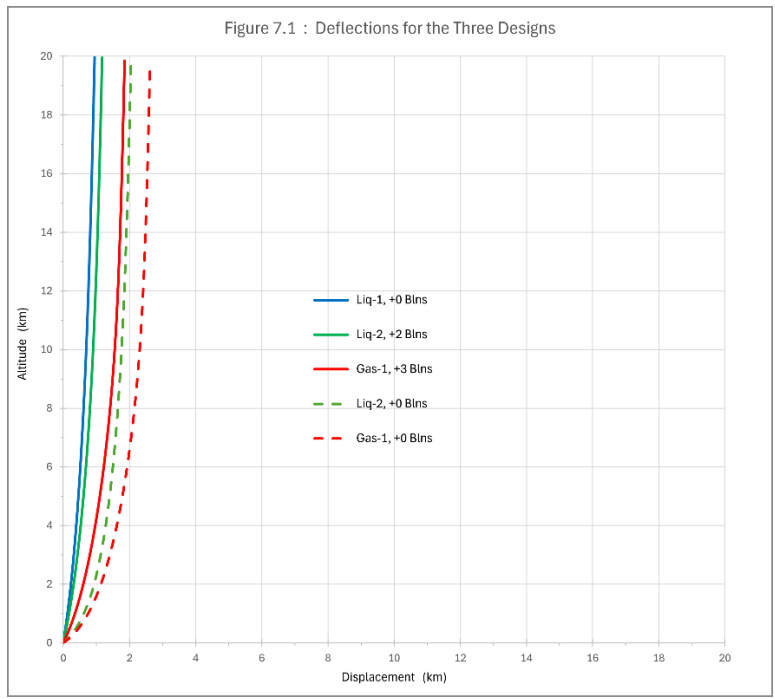

The deflections shown in Fig. 7.1, illustrate both the nominal deflections (those where only enough tension is used to prevent buckling), as well as the deflections corresponding to the Table 7.1 designs (where both Liq-2 and Gas-1 have added extra balloons in order to increase tension). Note, that the extra tension has more influence on Liq-2 (due to its very low inherent tension levels) than on Gas-1, where the extra tension has less relative effect on the total value.

**7.2 : Recommendations**

Let's use 3 criteria to judge the performance of the hose, as a hose. These are size, reliability, and wind handling.

Size does not necessarily mean weight or physical dimensions, instead it refers to the number of the balloons needed to support the hose. Here, Liq-2 is the best, Liq-1 only slightly behind, and Gas-1 is clearly the worst.

Reliability and simplicity are important to keep the hose operating as much as possible with the smallest effort (and support staff). Both Liq-1 and Gas-1 should do well here, since their pumps are all located at the ground where they can be easily maintained and, if necessary, replaced; Liq-2, with on-hose pumps is clearly a more complex and less reliable choice.

The Liq-1 design is substantially better at handling wind than the others. This is partly because of its small cross-section, but also due to stiffness resulting from its high tension. Gas-1 has the most trouble as a result of its large cross-section.

But, how well a hose performs is only part of what matters: The hose is just a means to an end, namely delivery of sulfates to the stratosphere in order to lower temperature rises from global warming. Accordingly, the goal is not necessarily to develop the "best" hose, but rather one which delivers relief from global warming as early, effectively, and certainly as possible. Accordingly, development time and risk are at least as important criteria as the hose's actual performance. It's certainly important to develop a hose that works well, but fielding a "good enough" hose soon is better than an "optimum" one years later.

By this metric, two hoses (Liq-1 and Gas-1) are most worth pursuing. Either one of these could turn out to be quick and easy to develop, or, in contrast, turn out to be too slow or risky. At the moment, we don't know which.

In principle, Liq-1 should be straightforward to develop. It has only one unique make-or-break development issue. This, of course, is the ability to make a lightweight hose which is capable of handling its large fluid pressure. The reasons that this can be done, are that the strength and weight of existing fiber-based materials are sufficient for Liq-1's hose, as well as the fact that such materials are already in use in existing (albeit weaker and heavier than we need) hoses.

Its risk, is not feasibility, but one of near-term availability. Right now, we can't simply buy hoses with the pressure and weight values that Liq-1 needs. As we'll discuss in Section 8.1, either existing products will have to be upgraded, or we'll have to develop and source the hoses ourselves. Both routes are doable, and would ultimately succeed, but how long they'll take is unknown.

As with Liq-1, the Gas-1 development hangs on one make-or-break issue; in this case it's the need for a lightweight aerodynamic shroud. There are two issues here, designing a shroud that's light enough, and then demonstrating that it actually works. The first challenge is extremely straightforward, easily addressed by existing structural design tools. We should know very quickly whether a 1 kg/m rib-and-skin shroud can be built, and will probably be able to design even lighter versions. If, of course, the answer is bad, and the shroud will be too heavy, we'll know this very early, without investing much time or effort into Gas-1.

The second challenge, demonstrating that the shroud works aerodynamically, is less clearcut. This also can be addressed computationally, using existing software. Again, the answer may be negative, e.g., the shroud undergoes too much flutter to work; if so, we'll know this early. But, if the computational answers are ambiguous, or even if they're positive, experimental verification will be necessary. This will lengthen the duration, and increase the risk and uncertainty, of developing Gas-1.

So, Liq-1 and Gas-1 involve different types of development, with different advantages and risks. For Liq-1, the risk is finding someone to make the high-pressure hose; but, once made, there is little doubt that the hose will work. For Gas-1, if a lightweight shroud can be designed, finding companies to build it will be easy; however, until tested, there will always be some doubt that the shroud will work correctly.

In contrast, Liq-2, is not as attractive; its drawback is the reliance of hose-mounted pumps. The presence of active components on the hose (pumps and electrical equipment) makes operation of the hose less reliable and predictable than either Liq-1 or Gas-1. On-hose events such as pump failures or lightning strikes are inevitable; while they can be planned for, these are complications not faced by Liq-1 or Gas-1. Dealing with these issues, not any structural or aerodynamical ones, will be the focus of Liq-2's development process. Both matters (reliable, light-weight pumps, and lightning resistant electrical systems) can be addressed by careful design. However, the success of such designs is not as clear-cut to evaluate as the structural issues faced by Liq-1 or Gas-1; here, it will only be established after long-term operational experience. This inherent uncertainty is the strongest reason not to adopt Liq-2; it might turn out to be reliable and work very well, but only operational experience will be able to prove this.

One other design decision, common to all three hose options, is whether to attach some balloons along the hose, or to keep them all at the top.

This matter was discussed previously in Sections 4.7 and 5.7; because mid-hose balloons have more buoyancy than stratospheric ones, they reduce the number of balloons needed by the hose. But, as demonstrated there, mid-hose balloons increase wind deflections; both by increasing wind forces, and by decreasing hose tension and hence its stiffness. Wind performance can be partially, but not fully, recovered by placing additional balloons at the top of the hose to improve its stiffness.

There is, however, another price to be paid for using mid-hose balloons, beyond the poorer wind performance. Mid-hose balloons will experience a more varied, and severe, weather environment than stratospheric ones. They also require the development of a different harness design; the hose will now extend past and above the balloons, rather than always staying below them.

Whether reducing the number of balloons is worth the extra design effort and operational risk needed to deal with their mid-hose placement is a judgement call. On balance, I would prefer using only stratospheric balloons, but this is not a clear-cut choice.

# 8 : Implementation

This report has outlined what is required for a hose to deliver sulfates to the stratosphere, in order to perform geoengineering. Now, it's time to focus on how to make this happen.

We're not in a position to simply purchase Liq-1 or Gas-1 hoses, or even to buy, and then assemble, all of their components. While some parts do exist, or can be readily adapted, there are others which must still be developed. While the need to do this is disappointing, the developments themselves are straightforward engineering projects, there are no scientific unknowns to making the hose.

In the next sections, we'll focus on what has to be done; on the challenges we'll need to address, and the steps that will let us solve them.

There are two basic subsystems we need, the hose and the balloons; these can (and should) be developed in parallel.

## 8.1 : Developing the Hose

The main parts that have to be developed for the hoses are, the Liq-1 high pressure conduit, the Gas-1 aero-shroud, the tethers, and the aerosol sprayer. Components such as the pumps, sensors, and control electronics are important and will need to be properly specified, acquired, and perhaps customized, but these don't require major design efforts.

### *Liq-1 Conduit*

The most challenging item that has to be developed for the Liq-1 system, is the hose conduit itself; this is unique to Liq-1, and its development is a make-or-break endeavor.

The Liq-1 design calls for a hose capable of transporting liquid $H_2S$ at a pressure of 3250 bars, through a conduit with a 3.1 cm bore. The weight estimates we've employed were based on walls formed by using Kevlar fibers, with strength derated by a 2:1 safety margin. This is a realistic target, but hoses with this size, strength, and weight aren't off-the-shelf; they'll have to be developed.

The preference, of course, is to work in concert with existing companies, finding ones with the right expertise, and paying for them to develop a hose with these properties; and then to manufacture sufficient quantities of it. Given the commercial landscape, there are two parallel routes to pursue; one is with existing hose manufacturers, and the other is with fiber wrapping experts.

There are a number of companies who sell hoses (for hydraulics) which handle these pressures. But these are usually only about a third of the necessary diameter, and use steel for their strength, so are far too heavy. For our application, such manufacturers would have to switch to using fiber reinforcement; this would be a big change. It's probably only a relevant avenue for us, if we can identify firms already planning for this upgrade.

There are, however, some firms who already do use Kevlar to make high pressure hoses. Some of these hoses are the right diameter, and, naturally, are much lighter than steel hoses. The problem, is that they are aimed at lower pressures than we need. So, they are not strong enough, reaching only about one third of the pressures we require. However, upgrading strength should be achievable by increasing the thickness of the hose's Kevlar walls, a more straightforward process than fully switching from steel to Kevlar.

The other route, is to use fiber-wrapping methods to make the hose. This is attractive because the fiber choice, the wall thickness, and the fiber orientation are all controllable, and by manufacturers used to such customization. The main challenge here, is making long hoses; the commercial market is focused on fairly short tubes. While attaching tubes together every few meters would be possible, the joints would have to be strong enough to handle the pressure, and the hose erection process would be complex, resembling that used for underground drilling. The same fiber-winding techniques used for tubes are applicable to hoses, as long as the bare hose can be gradually translated through the wrapping device (or vice versa). Our challenge would be to induce a company to do so.

Hopefully, this would just require finding a company interested in expanding their market from tubes to hoses, and then helping with this development, via hardware design and/or financing. In this case, our primary in-house tasks will be coordinating with the client companies to insure a suitable product, and to perform testing of hose samples with flowing $H_2S$, at the low temperatures the hose will encounter.

However, if we can't find an outside contractor to develop the conduit, we would have to do this ourselves, or abandon Liq-1 in favor of Gas-1 (or even Liq-2). Any internal effort should be focused on the fiber-wrapping approach, and requires us to do two things; the fiber wrapping itself, and performing it on a hose. We can certainly buy fiber wrapping tools, and can either hire or license the expertise to use them. Applying this to a hose, is the new skill that we'd prefer a company to undertake. But, if we were doing it, we'd have to choose whether to translate the hose underneath the tool, or move the tool along the hose; the latter seems easier.

*Gas-1 Aero-shroud*

The only make-or-break item unique to Gas-1 is, as discussed before, its aero-shroud. The primary need here, is to design and make an aero-shroud lightweight enough to make the Gas-1 hose attractive. If a rib-and-skin approach is chosen, several issues are important; choosing the overall airfoil shape, designing a stiff yet lightweight rib, picking the skin material and thickness, and determining the right spacing between ribs, as well as whether rigid stringers are necessary between them.

The skin will probably be made from advanced sailcloth; this is commercially available with excellent strength-to-weight properties. Since the stiffness and weight of the ribs is so crucial, these will likely be made from carbon-fiber-based material, and designed with modern commercial structural software.

The ribs, their spacing, and the airfoil shape form a coupled system, which can be optimized by combining aerodynamics codes with the structural ones. Since the software tools necessary to do this are readily available, and optimization techniques are well understood; this design process will be performed quickly. The resulting computational predictions for the aero-shroud's static properties (such as weight, strength, stiffness, and drag) will be accurate enough to tell us whether the Gas-1 approach is *not* feasible, and hence should be dropped.

However, a computational answer that Gas-1 is not infeasible, doesn't actually mean that it *is* feasible. The problem, is that dynamic properties, such as flutter, or other aero-driven oscillations, of this highly flexible structure, are harder to be sure of than static ones. Software tools can be used here, to predict whether serious oscillations will exist, and indicate steps to lessen them, but experimental confirmation will be important. Two testing methods can be used, wind tunnels or in-air tests, as part of the balloon tether development discussed below.

Unfortunately, while these tests are straightforward, the need for them complicates, and lengthens, the Gas-1 development process; and, of course, any decision on whether Gas-1 is the right approach to pursue.

If it is pursued, the actual fabrication of the Gas-1 aero-shroud involves buying long rolls of high-strength fabric, outsourcing the production of some 10,000 ribs, and then joining them together with a conduit, into the flexible tube and aero-shroud assemblage.

### *Gas-1 Conduit*

The gas conduit needed for Gas-1 is, of course, larger than that of Liq-1, and operates at very low pressure. While similar strength-to-weight performance is required as for Liq-1, the Gas-1 tube can be fabricated out of several layers of commercially available fabric, e.g., high strength sailcloth. This can be designed in-house, sample tube sections produced and tested, and then out-sourced for full fabrication.

### *Liq-1 Aero-shroud*

The Liq-1 aero-shroud is much smaller than the one needed by Gas-1. It is just as necessary, but its lightweight fabrication should be straightforward enough, that it's not the make-or-break issue it will be for Gas-1. In this report, we've designed aero-shrouds using the NACA0025 airfoil. The availability of data made this convenient, but there is nothing sacred about this shape. Since the computational tools for combined aerodynamic and structural optimization are readily

pg. 86

available, this choice should be revisited and likely improved upon. Beyond designing a good aero-shroud, our main job will be making sure it is lightweight enough, and is easy to fabricate and deploy. In Section 5.4, we modeled the aero-shroud as a solid block of aerogel, covered by a thin skin. This gave an acceptable mass, but a largely-hollow plastic construction may be stiffer and easier to produce. Trial configurations can be computationally designed, then made and evaluated via in-house additive manufacturing. The preferred design can then be out-sourced for manufacturing (e.g., injection-molded as a large number of identical axial sections).

*Other Components*

The other components for Liq-1 and Gas-1 are similar to each other, and can share the same development and fabrication processes.

We expect the tether to be made from commercially available high strength fibers; it may require some customization from existing products, but nothing beyond what commercial firms are used to providing. Our main task, will be deciding how to interface the tether with the hose. The most efficient solution would be to combine its function, with that of the hose, as both are formed from high-strength fibers. However, their structural responsibilities are different; the hose needs azimuthal strength, and the tether longitudinal. In the interest of design simplicity, and fast implementation, it is best to keep these two missions separate and non-interactive.

Likewise, nozzles capable of delivering sprays of finely sized aerosols are commercially available. The main design responsibility is ensuring that the droplets remain separated, and don't coalesce into larger drops. This will probably require an array of nozzles, spaced relatively far apart. Designing this, in a fashion that can be delivered in place, and suspended at the top of the hose, is the main task here. It's not clear what amount of testing will be necessary. The spraying and coalescence issue can be evaluated in a low-pressure chamber, but this will probably not have the extent needed to evaluate the actual evolution of $H_2S$ droplets to $H_2SO_4$ ones.

The above developmental programs are straightforward engineering. When they're done, we will be able to make a 20 km long hose, capable of delivering 100,000 tons/yr of sulfur up to the stratosphere; either as a liquid with Liq-1 or as a gas with Gas-1.

## 8.2 : Developing the Balloons

But the hose is only half the job, actually fielding it requires enough large balloons to lift and support it.

*Balloons*

Actually designing and making stratospheric balloons is not a major challenge. This has been done for decades, and there are commercial firms which can perform this service. The biggest decision here, is whether to field a single very large balloon, or a number of smaller ones. In this document, we've baselined the multi-balloon approach, believing that this will cut the design time, make fabrication and testing easier, and accommodate changes in hose weight simply by varying the number of balloons; it will also likely widen the pool of potential contractors. But, a single balloon does reduce wall area and wind-drag, and simplifies the connection to the hose; so this decision may be revisited when the project is undertaken. In any event, most of the balloon work should be commercially outsourced, our task will be limited to specifying the overall size and shape of the balloon(s), their acceptable weight, and to design issues related to connecting to the hose, spray equipment, and other payloads.

*Tethering the Balloons*

The one aspect of the balloons that is a make-or-break issue, common to all the hose options, is anchoring them to the ground. While there's a long history of tethering low altitude airships, there is virtually no experience in doing so for stratospheric balloons. Besides their value in keeping a balloon anchored to a fixed location (valuable for many communications and surveillance applications), tethers also solve the slow-leakage concerns facing long-duration balloons; they can include a small tube to deliver makeup gas.

The main challenge to tethered ballons is wind, both to the balloons (which no longer simply drift along with it), and to the hose itself. We've already shown that these are severe problems, and plan to deal with them by streamlining both the hose and the balloons.

Unless these steps work, the stratospheric hose approach to geoengineering will not be practical. And, although modeling and numerical simulation indicate that streamlined systems will survive wind forces, this has to be experimentally demonstrated.

Since experimentation rarely goes as smoothly as planned, this real-world experience should begin from the start, in parallel with hose development. It is probably not a task that can be out-sourced to an external company, but will require a dedicated project by our own team. Note, that there is interest in tethered stratospheric balloons for other applications than geoengineering (e.g., communications or surveillance platforms); so some design or testing partnerships may be convenient.

We do not have to work at full-scale (or even full altitude), to begin to gain valuable experience with tethered balloons. Experiments can, and should, begin with single balloons anchored by pure tethers; multiple balloons lifting an actual hose are not needed. In fact, the earliest tethers will probably not be streamlined, although the balloon should be.

Some of the most important early lessons, will come from multi-day flight experience with a wide range of different wind conditions. We'll plan to attach wind and temperature sensors along the tether to collect long duration, time resolved, data. In addition to wind information, this flight experience will teach us about the dynamics of the balloon, the tether, and their dependence on the coupling between them.

It'll also be important to gain experience with managing the ascent of a tethered balloon. In principle this is simple; as long as the balloon is buoyant, it will rise, pulling the tether up with it, and remaining anchored to the launch site by it. Due to wind, the reality will be more complicated{53}. One reason, is that the wind forces encountered during ascent, will be larger than stratospheric ones (given higher wind speeds and gas density). Another issue, is that, unless the balloon is fully inflated, it will not have the streamlined shape necessary for giving it a low drag coefficient. To ensure the full inflation needed for this, we'll want to fill the balloon at launch with much more He than it will need in the stratosphere, and then gradually bleed this off, to avoid catastrophic over-pressure as the balloon rises. All of these factors can be computationally modeled beforehand, but plans must be verified by experience.

As time and our experience proceeds, experiments will become more advanced and directed towards the eventual hose. Altitude will have to be extended to the 20 km stratospheric target; this, of course, will also enhance our suite of weather data.

*Other Work*

Another point to learn, is how to harness multiple balloons to the tether. A key aspect of this, is how closely they can be spaced; how does this affect their dynamics, and how does it change their drag? While spherical balloons could be arranged in a dense 3-D clump, harnessed by simple tethers, this is probably not appropriate for elongated balloons, since extreme proximity may increase their drag. Arranging them in a 2-D array yields a fairly compact envelope, but maintaining spacing requires a harness with compression members, not just tethers. The simplest technique, requiring only flexible tethers, would be to arrange the balloons in a vertical, one-dimensional stack, extending above the hose. While a 1-D stack is simple, it will require an increase in the number of balloons, as higher altitude ones provide less buoyancy.

At this point in the effort, we should also start to experiment with aero-shrouds. The early ones will naturally be smaller than the Liq-1 shroud, let alone that for Gas-1, so variations will be easy to fabricate and test; nor do early test aero-shrouds have to extend the full length of the tether.

Eventually, we will have learned enough to begin to test components of the actual Liq-1 or Gas-1 systems; these may begin with sub-scale versions, or may immediately use full-scale ones. One natural thing to test will be the balloons themselves, e.g., adding single copies of the final design to arrays of previously verified balloons. Other tests can be focused on adding sections of the hose (conduit, tether, and/or aero-shroud), to an existing test-tether.

In this way, we'll gradually learn how to operate tethered hoses, gaining the experience needed to implement a full-scale hose.

**8.3 : Required Effort**

How much will developing the hose and balloons cost?

The unsatisfying, but true, answer is that it's hard to know. Cost estimates for new R&D efforts are difficult to make, and generally inaccurate. At this stage, it's more useful to understand the level of effort required.

Here, I'll be drawing on my personal experience, which is limited to working (over the course now of 50 years) with teams of highly skilled scientists, engineers, and technicians in the San Francisco Bay area. However, as will be clear from Section 8.4, the hose may well be developed under very different conditions; outside the United States, and using different pools of researchers. Accordingly, one should base estimates of time, personnel, and hence cost, on local conditions; what's constant is the work that has to be performed.

We've already laid out this work in Sections 8.1 and 8.2. Some of this can be done internally, by dedicated staff and facilities. The rest will be performed externally, by companies acting either as partners or vendors.

*External Work*

The two most crucial and likely expensive aspects of this external work are the development and production of the Liq_1 conduit, and the balloons used to support both the Liq-1 or Gas-1 hose. While we could attempt to perform this work internally, it makes more sense to leverage the capabilities and expertise of existing companies. How much this will cost, is unclear. For instance, if carbon-wrapping firms are enthused by the prospect of expanding into hose manufacture, they will likely invest their own resources as well as ours. The order for 20 or so large stratospheric balloons should also be attractive, while it won't provide manufacturers with fundamental new capabilities, it does represent a substantial amount of business; with the promise of much more, if this approach to geoengineering is pursued.

The resources needed for our internal effort are determined by two main factors: How many people will we need? And, how long will it take?

*Internal Work*

The largest, most intensive, efforts will be creating the high-pressure Liq-1 conduit, and flight experiments with tethered balloons.

Hopefully, the former will largely be done by outside firms. In that case, our team's responsibilities will be both to support the contractors (e.g., to identify and help solve problems beyond their experience), and testing hose sections at low temperature and with $H_2S$. These tasks should require a team of about 4 people, and continue until the conduit development is complete. How long this takes depends on the contractor's progress, although if their work extends beyond 2-3 years, then we're probably using the wrong approach or contractors.

As discussed before, if we can't find contractors interested, or capable of, developing a high pressure conduit, then we'll have to either abandon the Liq-1 approach, or do this work ourselves. In the latter case, we'll need more in-house personnel, both for applying the wrapping and doing so on a long hose; let's target 7 people for 3 years.

Since we're not making the balloons themselves, our team will just be responsible for tethering and flying them. This work will likely have to be done in a sparsely populated area, so its team will probably not be collocated with the people working on the hose itself. Most of this effort will be preparation for each new experiment, and handling the ascents and descents; the flights themselves will usually last weeks or months, with automated monitoring and data collection. Given the need to perform long-duration flight operations in a remote location, this will be fairly people intensive; we'll plan for 7-8 people in an experimental program lasting 3 years.

While the aero-shroud is a make-or-break issue for Gas-1, developing it will not require as much effort as the Liq-1 conduit. The most immediate issue, whether a rib-and-skin shroud will be lightweight enough, can be quickly determined computationally, taking two people 6 months or less. While we'd prefer the weight values to be low (5 gm/cm or less), unless they are extremely excessive (e.g., 20 gm/cm or higher), it'll be prudent to continue working on Gas-1, as insurance in case the Liq-1 conduit turns out to be impractical.

This additional Gas-1 work has two components. The first, is to get prototypes of the ribs made, tested, and then mated with skin fabric for further testing. As discussed in Section 8.1, finding contractors to make the ribs will be easy. Their testing, and connection with the skin will probably be done in-house; but this process is also straightforward, so let's plan on 3 people for 6 months.

The other task, aerodynamic testing of the aero-shroud, is more complicated. Computational analysis will presumably have continued after the initial structural design (using the same personnel), so the physical testing will start with wind-tunnels. Rather than devoting the time and personnel needed to make our own wind-tunnels, it'll be preferable to lease wind-tunnel time from external facilities. Given the large size of the Gas-1 aero-shroud, this testing will probably be done at sub-scale. Our responsibilities will be, defining the test conditions, making the aero-shrouds to be tested, and then evaluating the results. Allowing for multiple experiments, and for coordinating schedules with the wind-tunnel facilities, this process should require 3 people and 1 year duration. After this round of wind-tunnel experimentation, future aero-shroud testing will be in the atmosphere, attached to the test tether as part of the tethered-balloon development laid out before.

The other hose developments, the tether, the small Liq-1 aero-shroud, the aerosol sprayer, and balloon harness are smaller efforts and can be performed by 1-2 people each.

*Organization*

Overall then, I expect that both hoses can be developed by a research team of 20-25 people, requiring 45-65 man-years of effort. Ideally, these should be as driven, innovative, and hardworking people as possible. The nature of the project will help here; while geoengineering is controversial, being part of as important and pioneering an effort as this, will appeal to many of the type of people we need,.

Traditionally, an organization's research team is augmented by a large support staff, handling overall management, purchasing, human relations, facilities, public relations, etc. But, in our case, a large support staff should be avoided; in my experience, research teams work best with minimal structure, performing many jobs, such as technically-sensitive purchasing, by themselves. No more than 5 people should be allotted for support functions.

Accordingly, the overall effort need require no more than 25-30 people and 60-80 man-years.

*Duration*

More important than this, is the question of how long the development will take.

Obviously, knowing the duration, helps determine the cost of the effort. But, minimizing time is much more important, in order to cut the real "cost", that of global warming. The longer it takes to develop the hose, and to start geoengineering, the more harm global warming will do in the meantime, and the greater the risk of reaching climatic tipping points will become. So, any opportunity to trade people for time should be taken.

Predicting how much time developing the hose will take, is hard to do. Part of the reason, is that lightweight high-pressure hoses, and tethered stratospheric balloons, don't exist yet. That said, the challenges involved in both are understood, and approaches to solve them have been identified. In the discussion above, I've tried to estimate how long these developments will take, including allowance for some of the anticipated problems and contingencies; this leads to a prediction of 3 years.

But, even under ideal conditions, things take longer than just what's technically required. One delay comes from assembling the right team; the highest quality people are not fungible. Another problem, is that specialized facilities, or long-lead-time equipment, can take a long time to acquire. Fortunately, I don't expect this to be much of a factor; e.g., the balloon vendors presumably have big enough hangers for construction and pressurization tests. One issue that might arise, is obtaining permits for testing tethered stratospheric balloons; this may well dictate where we do, or do not, choose to test. These issues can easily add another year to the project's duration. However, given the urgency, we should be ruthless against accepting such delays, and inventive in working around them. By holding the line, we should be able to keep the length of the development program to 3-3.5 years.

*Fabrication*

But, learning how to make a hose is not sufficient, we have to actually field one. Unlike development, this requires making many copies of our designed components; and this fabrication will take time. The Liq-1 conduit may be considered successfully developed once we've demonstrated the process with a 20 meter section, but we need 20,000 meters for Liq-1, 1000 such sections. Likewise, making and successfully flying a 20 meter wide, 100 meter long stratospheric balloon is nice, but we'll probably need 20-25 of these for the hose. And, as promising as making a lightweight Gas-1 rib will be, the aero-shroud will likely need about 10,000 of them. So, development demonstrates that we know how to build the hose, but the job isn't done until we've actually done so.

Clearly, this replication and assembly of components, and then the fielding of an operational hose, is going to take time, as well as money. How much of each is unknown. Since the actual amounts of materials involved are small, the money will not be large. Given the financial benefits of averting global warming, once successful development proves that we know how to build a hose, the money to do so will exist.

As with development, the most valuable coin is time. Here, what counts is not fabrication time per-se, but the date-of-completion, when the hose is finally erected. Hence, we should start replicating each type of hose component as soon as its development is done, not wait until the entire development process is complete.

Development and fabrication should be done in parallel, not series. One example of this could be the Gas-1 aero-shroud; by the end of year 1 we will have acquired, and structurally tested, prototypes of both the ribs and skin. At this point, if making enough of them for the 20,000 meter hose looks to be a long process, we could start production; not waiting to know how successful the Liq-1 conduit development, or even Gas-1's aerodynamic testing, will be. Similarly, the contractors will probably have made and tested the first full-scale balloon by the end of year 2; making 20 copies could start then. Depending on the speed of fiber-wrapping machines, forming 20,000 meters of the Liq-1 conduit may turn out to be the rate-limiting step. If so, the answer may be to replicate the wrapping tools themselves, e.g., have 5 machines making 4,000 meters each.

Clearly, fabrication is crucial and should be planned for from the start. It will likely be just as important a criteria for choosing contractors, as their technical prowess.

Given all the uncertainties involved both in the development of the hose, and fabricating it, any prediction for how long it will take to field the first hose is, at best, a well-informed guess. If development takes longer than expected, or if fabrication is sloppily handled, then we should be done in 5 years. But, with strong development, and careful preplanning for fabrication, then I anticipate a 4 year completion.

**8.4 : The Future**

Technical feasibility, of course, doesn't insure that hose-based geoengineering will ever occur; this will depend on real-world considerations. There are strong reasons for proceeding with geoengineering, but also continuing resistance to it.

The difference that the hose makes to this continuing dispute, is that it's so inexpensive; hose-based geoengineering can be pursued by a great number of people, in many different ways. This "entropy", while disturbing, should prove decisive.

***The Impetus***

Global warming keeps getting worse; global temperatures and $CO_2$ concentrations are all higher than they have ever been, and they continue to rise. Our current strategy to fight global warming, counting on transitioning from hydrocarbons to clean energy, is not working; certainly not fast enough to prevent the warming from continuing, and from growing worse.

Geoengineering, via stratospheric aerosols, offers a quick and inexpensive way to stop the progression of global warming, giving the time needed for more permanent solutions to work. Yet, despite debate and considerable theoretical research, it has never been attempted, nor even experimented on.

There are several reasons for this; including active opposition by some, disinterest and lack of urgency by most, no proof that it will actually work, and no easy way to test it.

At present, things are starting to change. While active opposition to geoengineering remains strong, there is growing frustration with our current failure to stop global warming, along with an increasing willingness to consider other approaches, even geoengineering.

The significance of the hose in this environment, is that it makes geoengineering quick and inexpensive to try; and if desired, to then implement.

While we've focused in this report on the details involved in fielding a stratospheric hose, fundamentally, it's just a strong lightweight tube, held up by balloons. To be effective, the hose needs to deliver little more flow than a typical garden hose. Despite being 20 km long, the Liq-1 hose is only a few inches wide, making it no bigger than the many thousands of underground pipes or cables routinely installed throughout the world. By far, the largest part of the hose is the suite of balloons needed to hold it up; but although big, each is little more than a hollow cigar-shaped plastic bag filled with helium.

Even more telling, is the effect the hose can have. Each Liq-1 hose weighs less than 20 tons, yet delivers over 5000 times this much stratospheric coolant each year. A collection of 20 such hoses can offset the global warming caused by the world's yearly emission of 40,000,000,000 tons of $CO_2$. This gives hose-based geoengineering an Archimedean "lever arm" of 100 million; each year, the hoses have a cooling effect 100,000,000 times their own size. No other tool in our arsenal has, even remotely, such an outsize effect on global warming.

Of course, this hose system doesn't exist yet. But, developing it, despite the issues we've described, is a straightforward engineering job; cheap in time and in people, and hence, also in money.

*Many Routes Forward*

This means that hose-based geoengineering, with all of its potential, is within the reach of many (possibly most) governments, and of hundreds of companies, organizations, and even rich individuals. For better or for worse, the major national governments will no longer be the sole gatekeepers to geoengineering.

This possibility is not a new concern; it's long been feared by opponents (as well as some advocates) of geoengineering. In practice, of course, almost none of the groups which could attempt geoengineering, will actually try to do so. Most are occupied with their own concerns, and have no interest at all in geoengineering. It's the large size of this pool, resulting from the extremely low cost of hose-based geoengineering, that raises the odds that one or more agents will actually attempt to act.

But, "action" does not have to mean carrying out a full geoengineering operation, encompassing all steps from design of the hose, construction of the first one, use of it for scientific testing, multi-site replication, and then operation of global geoengineering. In fact, while this full-stack approach is possible for governments or a few large organizations, it's not at all necessary that one agent perform all these functions. Design, prototyping, scientific testing, replication, and operation, can well be performed separately by different groups with different motivations, budgets, and risk tolerances.

As just one example of this, a first group could decide to kick-start progression towards geoengineering by designing and prototyping a stratospheric hose, and then make the hose available to a second, independent, group for scientific evaluation.

The effort required from the first group is small enough, some tens of millions of dollars, that a huge number of countries or other organizations could easily undertake it. They might be motivated by any number of factors, ranging from altruistic (to help find out if geoengineering works), to less so (being hydrocarbon proponents). This action would let them achieve a large effect for little cost; and with less risk of retribution than actually performing geoengineering would entail, since they are not operating the hose, simply making it available for testing.

The actual scientific evaluation of the hose could then be performed by one or more independent groups, such as universities, international organizations, government agencies, or the like. There is no shortage of groups capable of, and interested in, performing such testing.

And, this scenario itself may be more structured than what will turn out to happen. For example, many of the developmental tasks detailed in Section 8.1 represent separate bite-sized efforts, that can be undertaken by small teams or even motivated individuals, they need not be undertaken by a single common organization. Likewise, the tethered stratospheric balloons of Section 8.2 are of interest for a number of commercial applications, not just geoengineering; interests and developments may overlap, but organizations don't have to. Similarly, the testing may be split into separate tasks and performed by multiple teams; there may not wind up being consensus on all results, and no single yea/nay assessment of geoengineering may be requested or delivered.

Once a prototype hose has been built and tested, then the information learned will enable other groups to proceed with the replication and operation needed for global geoengineering.

It's certainly possible that this process will be well-organized and controlled, e.g., performed by major governments, acting through some international governance regimen. But, there are also many, more chaotic, scenarios, e.g., where one or more groups reach their own conclusions about the test results, and start performing geoengineering by themselves.

For instance, individual organizations or countries might decide (e.g., for local political reasons, impatience with existing failures to stop global warming, in order to offset their own carbon usage, etc.) to establish and operate their own hose at a local site. They would not necessarily have to build this hose themselves, but might be able to contract out the fabrication of it to independent contractors, i.e., companies established for making and erecting such hoses.

Or, in another example, hose-based geoengineering might serve as the basis for a world-wide carbon-offset effort. This might be done by a commercial entity, selling carbon offsets to companies, small countries, or individuals, across the globe. Or, it could be done by oil producers, publicly offsetting the global warming caused by their products, by applying a matching amount of geoengineering cooling. In these cases, the practitioners would probably build and operate hoses themselves, at multiple sites across the globe.

With all these possibilities, is hose-based geoengineering inevitable?

*Resistance*

Not necessarily; the opposition to geoengineering is intense, and will surely grow even stronger, the closer geoengineering gets to occurring.

However, the profusion of participants and procedures, allowed by hose-based geoengineering, will make efforts to stop it much more difficult than has been the case to date.

The international scope of participants makes simple national edicts ineffective. A country can simply forbid its own citizens from practicing geoengineering, but it's much harder to prevent foreigners (possibly other governments) from doing so.

The most likely attempt to stop it will be an international treaty (or some less formal accord) proclaiming geoengineering forbidden. Such edicts are relatively easy to issue; enforcement, of course, is less simple. Success will depend on who is being told to stop, and from what activity.

If a country, or group of them, decided to pursue hose-based geoengineering, it would be difficult to force them to stop. Similarly, the profusion of separate steps, from first making an inactive hose, then scientifically testing it, before finally attempting to operate it, makes preventing this progression difficult; particularly if each action is performed by different agents, for different reasons. And, once the means to perform geoengineering exist, and have been successfully demonstrated, it will be very hard to enforce any prohibition against employing it, particularly if the counter-argument is framed as, "forbidding geoengineering means permitting global warming".

In practice, the most effective response will probably be to try to control and regulate geoengineering, not to forbid it outright; i.e., to install an international governance regimen, and do so soon, before hose-based geoengineering becomes a reality.

***The Entropic Future***

Ultimately, it is the fact that hose-based geoengineering is so cheap, and can take effect so rapidly, that makes it a compelling tool against global warming. The inexpensiveness of this approach allows large numbers of potential actors to take part; and to do so in multiple ways, either as a single coordinated process, or in a multitude of separate independent steps.

Because of this, implementation is likely to be much more complex than traditional, governmentally-controlled, approaches to geoengineering, but this chaos also makes adoption of geoengineering, for better or worse, that much more assured.

# References

{1} International Energy Agency (2024). World Energy Investment 2024, https://www.iea.org/reports/world-energy-investment-2024

{2} Hannah Ritchie, Pablo Rosado, and Max Roser (2020) - "Energy Production and Consumption" Published online at OurWorldinData.org. Retrieved from: 'https://ourworldindata.org/energy-production-consumption' [Online Resource]

{3} Hannah Ritchie and Max Roser (2020) - "$CO_2$ emissions" Published online at OurWorldinData.org. Retrieved from: 'https://ourworldindata.org/co2-emissions' [Online Resource]

{4} Hannah Ritchie, Pablo Rosado, and Veronika Samborska (2024) - "Climate Change" Published online at OurWorldinData.org. Retrieved from: 'https://ourworldindata.org/climate-change' [Online Resource]

{5} United Nations Environment Programme. (2023). *Emissions Gap Report 2023*. https://doi.org/10.59117/20.500.11822/43922 (Figure 2.2)

{6} Herzog, Howard J. *Carbon Capture*. Cambridge, Massachusetts, The MIT Press, 2018. ISBN 978-0262348874

{7} Williamson, P., Wallace, D. W. R., Law, C. S., Boyd, P. W., Collos, Y., Croot, P., Denman, K., Riebesell, U., Takeda, S., & Vivian, C. (2012). Ocean fertilization for geoengineering: A review of effectiveness, environmental impacts and emerging governance. *Process Safety and Environmental Protection*, *90*(6), 475–488. https://doi.org/10.1016/j.psep.2012.10.007

{8} Teller E, Wood L, Hyde R. (1997). Global warming and ice ages: I. Prospects for physics-based modulation of global change. *UCRL-JC-128715*. Presented at 22nd Int. Sem. Planet. Emerg., Aug. 20–23, 1997, Erice, Italy

{9} Crutzen, P. J. (2006). Albedo Enhancement by Stratospheric Sulfur Injections: A Contribution to Resolve a Policy Dilemma? *Climatic Change*, *77*(3-4), 211–220. https://doi.org/10.1007/s10584-006-9101-y

{10} Caldeira, K., Bala, G., & Cao, L. (2013). The Science of Geoengineering. *Annual Review of Earth and Planetary Sciences*, *41*(1), 231–256. https://doi.org/10.1146/annurev-earth-042711-105548

{11} National Academies of Sciences, Engineering, and Medicine. 2021. Reflecting Sunlight: Recommendations for Solar Geoengineering Research and Research Governance. Washington, DC: The National Academies Press. https://doi.org/10.17226/25762

{12} Keith, David W. *A Case for Climate Engineering*. Cambridge, M Ass., The Mit Press, 2013. ISBN 978-0262019828

{13} Wagner, Gernot. *Geoengineering : The Gamble*. John Wiley & Sons, 8 Sept. 2021. ISBN 978-1509543069

{14} MacMartin, D. G., & Kravitz, B. (2019). The Engineering of Climate Engineering. *Annual Review of Control, Robotics, and Autonomous Systems*, *2*(1), 445–467. https://doi.org/10.1146/annurev-control-053018-023725

{15} Parker, D. E., Wilson, H., Jones, P. D., Christy, J. R., & Folland, C. K. (1996). The Impact of Mount Pinatubo on World-Wide Temperatures. *International Journal of Climatology*, *16*(5), 487–497. https://doi.org/10.1002/(sici)1097-0088(199605)16:5%3C487::aid-joc39%3E3.0.co;2-j

{16} Dykema, J. A., Keith, D. W., & Keutsch, F. N. (2016). Improved aerosol radiative properties as a foundation for solar geoengineering risk assessment. *Geophysical Research Letters*, *43*(14), 7758–7766. https://doi.org/10.1002/2016gl069258

{17} Govindasamy, B., & Caldeira, K. (2000). Geoengineering Earth's radiation balance to mitigate CO2-induced climate change. *Geophysical Research Letters*, *27*(14), 2141–2144. https://doi.org/10.1029/1999gl006086

{18} Ban-Weiss, G. A., & Caldeira, K. (2010). Geoengineering as an optimization problem. *Environmental Research Letters*, *5*(3), 034009. https://doi.org/10.1088/1748-9326/5/3/034009

{19} Visioni, D., Pitari, G., & Aquila, V. (2017). Sulfate geoengineering: a review of the factors controlling the needed injection of sulfur dioxide. *Atmospheric Chemistry and Physics*, *17*(6), 3879–3889. https://doi.org/10.5194/acp-17-3879-2017

{20} McClellan, J., Keith, D. W., & Apt, J. (2012). Cost analysis of stratospheric albedo modification delivery systems. *Environmental Research Letters*, *7*(3), 034019. https://doi.org/10.1088/1748-9326/7/3/034019

{21} Lockley, A., MacMartin, D., & Hunt, H. (2020). An update on engineering issues concerning stratospheric aerosol injection for geoengineering. *Environmental Research Communications*, *2*(8), 082001. https://doi.org/10.1088/2515-7620/aba944

{22} McClellin, J. (2010). Geoengineering Cost Analysis, *Aurora Flight Sciences*, AR10-182.

{23} Smith, W., Bhattarai, U., Bingaman, D. C., Mace, J. L., & Rice, C. V. (2022). Review of possible very high-altitude platforms for stratospheric aerosol injection. *Environmental Research Communications*, *4*(3), 031002. https://doi.org/10.1088/2515-7620/ac4f5d

{24} J. J. Blackstock, D. S. Battisti, K. Caldeira, D. M. Eardley, J. I. Katz, D. W. Keith, A. A. N. Patrinos, D. P. Schrag, R. H. Socolow and S. E. Koonin, Climate Engineering Responses to Climate Emergencies (Novim, 2009), archived online at: http://arxiv.org/pdf/0907.5140

{25} The Stratospheric Shield. (2009) https://climateviewer.com/downloads/Stratoshield-white-paper-300dpi.pdf

{26} Chan, A., Hyde, R., Myhrvold, N., Clarence, T., & Wood, L. (2012). *High Altitude Structure for Expelling a Fluid Stream Through an Annular Space* (United States of America Patent 8,166,710).

{27} Davidson, P., Burgoyne, C., Hunt, H., & Causier, M. (2012). Lifting options for stratospheric aerosol geoengineering: advantages of tethered balloon systems. *Philosophical Transactions of the Royal Society A: Mathematical, Physical and Engineering Sciences*, *370*(1974), 4263–4300. https://doi.org/10.1098/rsta.2011.0639

{28} Davidson, P., Hunt, H., & Burgoyne, C. (2016). *Atmospheric Delivery System* (United States of America Patent 9,363,954).

{29} SPICE (2025). http://spice.ac.uk/about-us/aims-and-background/

{30} L. Ruffine, Mougin, P., & Barreau, A. (2006). How To Represent Hydrogen Sulfide within the CPA Equation of State. *Industrial & Engineering Chemistry Research*, *45*(22), 7688–7699. https://doi.org/10.1021/ie0603417

{31} Gosselin, F. P., & Païdoussis, M. P. (2014). Dynamical stability analysis of a hose to the sky. *Journal of Fluids and Structures*, *44*, 226–234. https://doi.org/10.1016/j.jfluidstructs.2013.11.003

{32} NASA (1976). U.S. Standard Atmosphere, *NASA-TM-X-74335, N77-16482,* Table I.

{33} Mo, C., Zhang, G., Zhang, Z., Yan, D., & Yang, S. (2022). A Modified Solid–Liquid–Gas Phase Equation of State. *ACS Omega*, *7*(11), 9322–9332. https://doi.org/10.1021/acsomega.1c06142

{34} Serghides, T. K. (1984). Estimate Friction Factor Accurately. *Chemical Engineering Journal, 91*, 63-64.

{35} Fan, T.-B., & Wang, L.-S. (2006). A viscosity model based on Peng–Robinson equation of state for light hydrocarbon liquids and gases. *Fluid Phase Equilibria*, *247*(1-2), 59–69. https://doi.org/10.1016/j.fluid.2006.06.008

{36} Babb, S. E. (1969). New Phases in H2S and COS. *The Journal of Chemical Physics*, *51*(2), 847–848. https://doi.org/10.1063/1.1672085

{37} Saito, Y., Iijima, I., Y. Matsuzaka, Matsushima, K., Tanaka, S., Kajiwara, K., & S. Shimadu. (2014). Development of a super-pressure balloon with a diamond-shaped net. *Advances in Space Research*, *54*(8), 1525–1529. https://doi.org/10.1016/j.asr.2014.06.043

{38} Sweeney, A. J., & Fu, Q. (2021). Diurnal Cycles of Synthetic Microwave Sounding Lower-Stratospheric Temperatures from Radio Occultation Observations, Reanalysis, and Model Simulations. *Journal of Atmospheric and Oceanic Technology*, *38*(12), 2045–2059. https://doi.org/10.1175/jtech-d-21-0071.1

{39} Fleming, E. L., Chandra, S., Barnett, J. J., & Corney, M. (1990). Zonal mean temperature, pressure, zonal wind and geopotential height as functions of latitude. *Advances in Space Research*, *10*(12), 11–59. https://doi.org/10.1016/0273-1177(90)90386-e

{40} Cole, A. & Kantor, A. (1978) Air Force Reference Atmospheres, AFGL-TR-78-0051

{41} Giauque, W. F., & Blue, R. W. (1936). Hydrogen Sulfide. The Heat Capacity and Vapor Pressure of Solid and Liquid. The Heat of Vaporization. A Comparison of Thermodynamic and Spectroscopic Values of the Entropy. *Journal of the American Chemical Society*, *58*(5), 831–837. https://doi.org/10.1021/ja01296a045

{42} Churchill, S. W., & Bernstein, M. (1977). A Correlating Equation for Forced Convection From Gases and Liquids to a Circular Cylinder in Crossflow. *Journal of Heat Transfer*, *99*(2), 300–306 (Equation 7). https://doi.org/10.1115/1.3450685

{43} Drob, D. P., Emmert, J. T., Meriwether, J. W., Makela, J. J., Doornbos, E., Conde, M., Hernandez, G., Noto, J., Zawdie, K. A., McDonald, S. E., Huba, J. D., & Klenzing, J. H. (2015). An update to the Horizontal Wind Model (HWM): The quiet time thermosphere. *Earth and Space Science*, *2*(7), 301–319. https://doi.org/10.1002/2014ea000089

{44} Chiba, Kazuhisa, et al. (2015). Feasibility Studies on a High-Altitude Captive Lighter-Than-Air Platform System. *14th AIAA Aviation Technology, Integration, and Operations Conference*, 18 June 2015, AIAA 2015-3378. https://arc.aiaa.org/doi/10.2514/6.2015-3378

{45} Hoerner, S. F. (1965). *Fluid-dynamic drag : practical information on aerodynamic drag and hydrodynamic resistance*. Hoerner Fluid Dynamics. https://archive.org/details/FluidDynamicDragHoerner1965

{46} Lutz, T., & Wagner, S. (1998). Drag Reduction and Shape Optimization of Airship Bodies. *Journal of Aircraft*, *35*(3), 345–351. https://doi.org/10.2514/2.2313

{47} García-Gutiérrez, A., Gonzalo, J., Domínguez, D., & López, D. (2022). Stochastic optimization of high-altitude airship envelopes based on kriging method. *Aerospace Science and Technology*, *120*, 107251. https://doi.org/10.1016/j.ast.2021.107251

{48} Wang, X.-L., & Shan, X.-X. (2006). Shape Optimization of Stratosphere Airship. *Journal of Aircraft*, *43*(1), 283–286. https://doi.org/10.2514/1.18295

{49} Götten, F., Havermann, M., Braun, C., Marino, M., & Bil, C. (2021). Improved Form Factor for Drag Estimation of Fuselages with Various Cross Sections. *Journal of Aircraft*, *58*(3), 549–561. https://doi.org/10.2514/1.c036032

{50} *NACA0012*. (2021). Bigfoil.com. https://bigfoil.com/ed677818-5a94-4df6-8006-2eebed47df40_info.php

*NACA0025*. (2021). Bigfoil.com. https://bigfoil.com/e6ee7e4f-5145-4f61-806c-5f25109b5779_info.php

{51} Huber, M., & Harvey, A. (n.d.). *Viscosity of gases*. https://tsapps.nist.gov/publication/get_pdf.cfm?pub_id=907539

{52} Qin, Z., Jung, G. S., Kang, M. J., & Buehler, M. J. (2017). The mechanics and design of a lightweight three-dimensional graphene assembly. *Science Advances*, *3*(1), e1601536. https://doi.org/10.1126/sciadv.1601536

{53} Grant, D,& Rand, J. (1996) Dynamic analysis of an ascending high altitude tethered balloon, *AIAA paper 96-0578*, presented at AIAA 34th Aerospace Sciences Meeting and Exhibit, Reno, NV Jan 15-18, 1996. https://arc.aiaa.org/doi/10.2514/6.1996-578